\documentclass[a4paper,11pt]{article}
\pdfoutput=1 

\usepackage{jheppub}
\usepackage[T1]{fontenc} 
\usepackage{lmodern}
\usepackage{amsfonts}
\usepackage{graphicx}
\usepackage{amsthm}
\usepackage{amsmath}
\usepackage{amssymb}
\usepackage{mathrsfs}
\usepackage{bbold}
\usepackage{mathtools}
\usepackage{slashed}
\usepackage{float}
\usepackage{epstopdf}
\usepackage{etoolbox}
\usepackage{xspace}
\usepackage{tabularx}
\usepackage{amsbsy}
\usepackage{array}
\usepackage{bm}
\usepackage{cases}
\usepackage{lscape}
\usepackage{empheq}
\usepackage{longtable}
\usepackage{cancel}
\usepackage{footmisc}

\usepackage{multirow}

\makeatletter
\newcommand{\linethrough}{\mathpalette\@thickbar}
\newcommand{\@thickbar}[2]{{#1\mkern0mu\vbox{
    \sbox\z@{$#1#2\mkern-1.5mu$}%
    \dimen@=\dimexpr\ht\tw@-\ht\z@+2\p@\relax 
    \hrule\@height0.5\p@ 
    \vskip\dimen@
    \box\z@}}
}
\makeatother



\usepackage{subcaption}
\usepackage[labelformat=simple]{subcaption}

\newcolumntype{L}[1]{>{\raggedright\let\newline\\\arraybackslash\hspace{0pt}}m{#1}}
\newcolumntype{C}[1]{>{\centering\let\newline\\\arraybackslash\hspace{0pt}}m{#1}}
\newcolumntype{R}[1]{>{\raggedleft\let\newline\\\arraybackslash\hspace{0pt}}m{#1}}

\newcommand{\mathematica}{\textsc{Mathematica}\xspace}
\newcommand{\feyncalc}{\textsc{FeynCalc}\xspace}

\newcommand{\jaxodraw}{\textsc{JaxoDraw}\xspace}

\allowdisplaybreaks[2]

\usepackage{subcaption}
\usepackage[labelformat=simple]{subcaption}

\title{\boldmath On the three-point order parameters of chiral symmetry breaking}

\author{Tom\'{a}\v{s} Kadav\'{y},}
\author{Karol Kampf}
\author{and Ji\v{r}\'{i} Novotn\'{y}}

\affiliation{Institute of Particle and Nuclear Physics, Charles University\\V Hole\v{s}ovi\v{c}k\'{a}ch 2, 180 00 Prague 8, Czech Republic}

\emailAdd{\{kadavy, kampf, novotny\}@ipnp.mff.cuni.cz}

\abstract{The nonlocal order parameters of the QCD chiral symmetry breaking, namely the nonanomalous three-point Green functions of the quark bilinears belonging to the odd intrinsic parity sector, are studied within the framework of the Resonance chiral theory. The general form of these correlators is then matched with various high energy constraints: the leading and sub-leading OPE for all three momenta simultaneously large, the leading OPE for two momenta large and the leading and sub-leading Brodsky-Lepage limit for the corresponding transition form factor. In order to satisfy these constraints, the Resonance chiral theory has to be enlarged with additional resonance multiplets and with higher derivative operators as well. We discuss a minimal extension of this kind in detail and present corresponding correlators in a closed form. The remaining free parameters are then constrained from lattice data. As a phenomenological application, we discuss the pion-pole contribution to the muon $g-2$ and the decay $\pi^0\to e^+e^-$.}

\begin{document}

\maketitle

\flushbottom


\section{Introduction}\label{sec:intro}
We believe that Quantum chromodynamics (QCD) is a fundamental theory of strong interactions between quarks and gluons. However, its nonabelian nature is the reason for the unique properties that QCD inherits --- the asymptotic freedom and confinement. Consequently, the perturbative approach fails in the low-energy region. In order to be able to describe theoretically the region where QCD becomes nonperturbative, one is required to use an effective field theory that would be formulated in terms of the relevant degrees of freedom --- the mesons and baryons. Such a theory, however, is not known from the first principles in the full-energy region.

For energies typically less than $M_{\rho}$, i.e.~the mass of the $\rho(770)$ meson, there exists an effective field theory of QCD --- the Chiral perturbation theory (ChPT) \cite{Coleman:1969sm,Weinberg:1978kz,Gasser:1983yg,Gasser:1984gg}.

In an intermediate energy region, i.e.~for $M_{\rho}\leq E\leq 2\,\mathrm{GeV}$, one relies on the phenomenological theory inspired by the large-$N_{c}$ counting, in which the resonances are explicit degrees of freedom. We will refer to the theory as the Resonance chiral theory (RChT). Its Lagrangian was introduced in ref.~\cite{Ecker:1988te} and then further developed and enlarged both for the even-parity sector and the odd-parity sector e.g. in refs.~\cite{Cirigliano:2006hb,Kampf:2006yf,Masjuan:2007ay,Geng:2008ag,Jiang:2009uf,Kampf:2011ty,Nieves:2011gb,Nugent:2013hxa,Roig:2013baa,Czyz:2017veo,Guevara:2018rhj}. The concept of the renormalization within RChT was recently studied for example in~\cite{Rosell:2009yb,Sanz-Cillero:2009bcv,Kampf:2009jh,Pich:2010sm,Terschlusen:2013iqa,Bruns:2013tja,Terschlusen:2016kje}.
The RChT includes a finite number of massive $\mathrm{U}(3)$ multiplets of vector $V(1^{--})$, axial-vector $A(1^{++})$, scalar $S(0^{++})$ and pseudoscalar $P(0^{-+})$ resonances, due to which the number of degrees of freedom increases. Note that, however, in contrast with ChPT, the RChT with only a finite number of resonance multiplets cannot grasp all the features of large $N_C$ QCD. Other problems are the missing mass gap between the degrees of freedom involved and the rest of the hadronic spectrum and a lack of systematic and physically motivated treatment of the higher loop corrections. Therefore RChT should be regarded as a QCD-motivated phenomenological model rather than an effective theory.

Interactions of these resonances can be studied with the help of the Green functions of the chiral currents and densities that allow us to comfortably obtain physical observables of the processes studied within this context. In these studies, the odd-intrinsic parity sector of QCD plays a significant role. Here, we restrict ourselves purely to the three-point Green functions of chiral currents and densities, which can be easily associated with two-body decays of various mesons. In total, there are five such correlators. We will discuss only the order parameters of the chiral symmetry breaking in the chiral limit: $\langle VVP\rangle$, $\langle VAS\rangle$ and $\langle AAP\rangle$. 

The $\langle VVP\rangle$ Green function represents probably the most important phenomenological object within the set of correlators of the odd sector. It has already been studied extensively, see e.g.~\cite{Moussallam:1994xp,Knecht:2001xc,RuizFemenia:2003hm}. In ref.~\cite{Kampf:2011ty}, the NLO resonance contribution has been presented, and some phenomenological aspects have been studied as well, such as the decays $\rho\rightarrow\pi\gamma$, $\pi(1300)\rightarrow\gamma\gamma$, $\pi(1300)\rightarrow\rho\gamma$ or the pion-pole contribution to the muon $g-2$ factor.
Therein, also the $\langle VAS\rangle$ Green function has been studied, and that is only within the context of its resonance contribution. As it seems, a detailed phenomenological study of such a correlator is still missing. Similarly, a systematical study of the resonance contribution to the $\langle AAP\rangle$ Green function is still unknown, although some theoretical aspects have been analyzed in the past, namely in \cite{Knecht:2001xc}.

The motivation behind this paper is to provide a matching of the operator product expansion (OPE) of the Green functions with their resonance contributions, including the subleading OPE terms recently studied in~\cite{Kadavy:2020hox}. To this end, we evaluate the contributing Feynman diagrams in RChT to the $\langle VVP\rangle$, $\langle VAS\rangle$ and $\langle AAP\rangle$  correlators and analyze its matching onto the high-energy behaviour obtained within the framework of OPE in the chiral limit. 
Although it is not automatically guaranteed that this approach provides a trustworthy interpolation between the low energy and the asymptotic energy regions and that all the UV constraints are satisfied (see e.g.~\cite{Peris:1998nj,Bijnens:2003rc}), it allows for reducing the number of free parameters of the phenomenological model using the first principle constraints derived directly from QCD.

The paper is organized as follows. After a short introduction in section \ref{sec:intro}, we briefly cover the necessary preliminaries on the chiral perturbation theory and resonance chiral theory in section~\ref{sec:QCD}. Section~\ref{sec:GF} consists of a description of the Green functions of chiral currents and densities in the odd-intrinsic parity sector of QCD, their high-energy behaviour within OPE and a discussion on a general form of respective resonance contributions, as well as on a procedure of duplicating the resonance multiplets. Section \ref{sec:two_multiplets} presents the analysis of the $\langle VVP\rangle$, $\langle VAS\rangle$ and $\langle AAP\rangle$ Green functions with two resonance multiplets taken into account in each channel. Section \ref{sec:three_multiplets} introduces a discussion on the $\langle VVP\rangle$ ansatz, for which three vector and three pseudoscalar resonance multiplets are considered. We also discuss in detail the connected pion transition form factor. Several phenomenological examples are studied, such as the decays $\rho^{+}\to\pi^{+}\gamma$, $\omega\to\pi^{0}\gamma$, $\omega\to\pi^{0}e^{+}e^{-}$, $\omega\to\pi^{0}\mu^{+}\mu^{-}$ and the pion-pole contribution to the muon $g-2$ factor. In appendix \ref{sec:struktura}, we justify in detail our approach regarding the duplication of the resonance multiplets. Finally, appendix \ref{sec:lowest_resonances} contains details on RChT--OPE matching in the special case when only the contributions of the lowest resonance multiplets to the respective Green functions are taken into account.


\section{Framework of low-energy QCD}\label{sec:QCD}

\subsection{Chiral perturbation theory}\label{ssec:chpt}
In the chiral limit, spontaneous breaking of the chiral $\mathrm{SU}(3)_{L}\times\mathrm{SU}(3)_{R}$ symmetry down to $\mathrm{SU}(3)_{V}$ in QCD leads to the presence of the Goldstone bosons. They can be identified with the octet of pseudoscalar mesons $(\pi,K,\eta)$, the lightest hadronic observable states. The effective Lagrangian, expressed in terms of the mentioned hadronic degrees of freedom, represents the so-called Chiral perturbation theory (ChPT) ~\cite{Weinberg:1978kz,Gasser:1983yg,Gasser:1984gg}. For recent constructions, see also~\cite{Kampf:2021jvf}. The basic building block of ChPT is
\begin{equation}
u=\exp\bigg(\frac{i\phi}{\sqrt{2}F}\bigg)\,,\label{eq:u}
\end{equation}
where the low-energy parameter $F$ is related to the pion decay constant in the chiral limit, $\phi=\sqrt{2}\phi_{a}T^{a}$
is the matrix of the octet of pseudoscalar meson fields, and $T^{a}$ are the $\mathrm{SU}(3)$ generators, defined as the halves of the Gell-Mann matrices, normalized as $\mathrm{Tr}\,(T^{a}T^{b})=\frac{1}{2}\delta^{ab}$.

The monomials of mesonic chiral Lagrangian are constructed by taking traces of products of suitable operators made of chiral tensors. Such tensors are defined as follows:
\begin{align}
u_{\mu}&=i\big[u^{\dagger}(\partial_{\mu}-i r_{\mu})u-u(\partial_{\mu}-i\ell_{\mu})u^{\dagger}\big]\,,\notag\\
\chi_{\pm}&=u^{\dagger}\chi u^{\dagger}\pm u\chi^{\dagger}u\,,\notag\\
f_{\pm}^{\mu\nu}&=u F_{L}^{\mu\nu}u^{\dagger}\pm u^{\dagger}F_{R}^{\mu\nu}u\,,\notag\\
h_{\mu\nu}&=\nabla_{\mu}u_{\nu}+\nabla_{\nu}u_{\mu}\,,\label{eq:h}
\end{align}
where
\begin{align*}
r_{\mu}&=v_{\mu}+a_{\mu}\,,\\
\ell_{\mu}&=v_{\mu}-a_{\mu}
\end{align*}
are the right and left external sources, respectively, defined as the combinations of the vector $v_{\mu}$ and axial-vector $a_{\mu}$ external sources. Furthermore,
\begin{equation}
\chi=2B_{0}(s+ip)\,,\label{eq:chi}
\end{equation}
with $s$ and $p$ being the scalar and pseudoscalar external sources, respectively, and $B_{0}$ is a constant not restricted by chiral symmetry and related to the quark condensate. The external sources in \eqref{eq:chi} can be decomposed as $v_{\mu}=v_{\mu}^{a}T^{a}$, $a_{\mu}=a_{\mu}^{a}T^{a}$, $s=s^{a}T^{a}$ and $p=p^{a}T^{a}$.
We have also denoted
\begin{align}
F_{L}^{\mu\nu}&=\partial^{\mu}\ell^{\nu}-\partial^{\nu}\ell^{\mu}-i[\ell^{\mu},\ell^{\nu}]\,,\notag\\
F_{R}^{\mu\nu}&=\partial^{\mu}r^{\nu}-\partial^{\nu}r^{\mu}-i[r^{\mu},r^{\nu}]\label{eq:velkeFR}
\end{align}
as the left and right nonabelian field-strength tensors. The covariant derivative $\nabla_{\mu}$ of an arbitrary operator and the chiral connection $\Gamma_{\mu}$ are defined as
\begin{align}
\nabla_{\mu}\,\bullet=\partial_{\mu}\bullet+[\Gamma_{\mu},\bullet]\,, \qquad
\Gamma_{\mu}=\frac{1}{2}\big[u^{\dagger}(\partial_{\mu}-ir_{\mu})u+u(\partial_{\mu}-i\ell_{\mu})u^{\dagger}\big]\,.\label{eq:gamma}
\end{align}

The ChPT Lagrangian is organized as a series of terms with an increasing number of derivatives and
quark masses. Up to $\mathcal{O}(p^{6})$, such Lagrangian can be written down in the form
\begin{equation}
\mathcal{L}_{\mathrm{ChPT}}=\mathcal{L}_{\mathrm{ChPT}}^{(2)}+\mathcal{L}_{\mathrm{ChPT}}^{(4)}+\mathcal{L}_{\mathrm{WZW}}^{(4)}+\mathcal{L}_{\mathrm{ChPT}}^{(6)}+\mathcal{L}_{\mathrm{WZW}}^{(6)}+\ldots\,.\label{eq:chiralLagrangian}
\end{equation}
For our purposes, only the terms discussed briefly below are relevant.

\begin{itemize}
\item[a)] The ChPT Lagrangian of the lowest order in the chiral limit \cite{Gasser:1983yg,Gasser:1984gg},
\begin{align}
\mathcal{L}_{\mathrm{ChPT}}^{(2)}=\frac{F^{2}}{4}\langle u_{\mu}u^{\mu}+\chi_{+}\rangle\,,\label{eq:lagrangiangb}
\end{align}
depends only on two parameters: the constant $F$ and the parameter $B_{0}$, related to the quark condensate.

\item[b)] The leading order of the pure Goldstone-boson part of the odd-intrinsic parity sector starts at $\mathcal{O}(p^{4})$ and the parameters are set entirely by the chiral anomaly. The relevant part of the respective Wess--Zumino--Witten Lagrangian reads\footnote{The action that gives rise to the anomaly can be found in ref.~\cite{Bardeen:1969md} and \cite{Wess:1971yu}. We also note that all the terms that do not contribute to the three-point correlators have been omitted.} \cite{Witten:1983tw} 
\begin{align}
\mathcal{L}_{\mathrm{WZW}}^{(4)}\ni\frac{i N_{c}}{48\pi^{2}}\varepsilon^{\mu\nu\alpha\beta}\Big(\big\langle\mathcal{W}_{\mu\nu\alpha\beta}(u,\ell,r)\big\rangle-\big\langle\mathcal{W}_{\mu\nu\alpha\beta}(1,\ell,r)\big\rangle\Big)\,,\label{eq:wzwlagrangian}
\end{align}
where
\begin{align*}
\mathcal{W}_{\mu\nu\alpha\beta}(u,\ell,r)&\ni iL_{\mu\nu}L_{\alpha}R_{\beta}+iR_{\mu\nu}L_{\alpha}R_{\beta}+\sigma_{\mu}L_{\nu\alpha}L_{\beta}+\sigma_{\mu}R_{\nu\alpha}L_{\beta}+\sigma_{\mu}L_{\nu}L_{\alpha\beta}\\
&-(L\leftrightarrow R\,,\sigma\leftrightarrow\sigma^{\dagger})\,,
\end{align*}
with
\begin{equation*}
L_{\mu}=u\,\ell_{\mu}u^{\dagger}\,,\qquad L_{\mu\nu}=u(\partial_{\mu}\ell_{\nu})u^{\dagger}\,,\qquad R_{\mu}=u^{\dagger}r_{\mu}u\,,\qquad R_{\mu\nu}=u^{\dagger}(\partial_{\mu}r_{\nu})u
\end{equation*}
and
\begin{equation*}
\sigma_{\mu}=\{u^{\dagger},\partial_{\mu}u\}\,.
\end{equation*}

\item[c)] A classification of a minimal set of independent terms of the $\mathcal{L}_{\mathrm{WZW}}^{(6)}$ Lagrangian was initiated in \cite{Fearing:1994ga,Akhoury:1990px} and, according to \cite{Bijnens:2001bb,Ebertshauser:2001nj}, there are 23 terms in SU(3) in total, i.e.
\begin{equation}
\mathcal{L}_{\mathrm{WZW}}^{(6)}=\sum_{i=1}^{23}C_{i}^{W}\mathcal{O}_{i}^{W}=\varepsilon^{\mu\nu\alpha\beta}\sum_{i=1}^{23}C_{i}^{W}\big(\widehat{\mathcal{O}}_{i}^{W}\big)_{\mu\nu\alpha\beta}\,,\label{eq:ChPT_p6_Lagrangian}
\end{equation}
where $C_{i}^{W}$ are so-called low-energy constants (LECs, for short). In our case, only five terms are relevant and the corresponding operators read
\begin{align*}
\big(\widehat{\mathcal{O}}_{7}^{W}\big)_{\mu\nu\alpha\beta}&=i\big\langle\chi_{-}f_{+\mu\nu}f_{+\alpha\beta}\big\rangle\,,\\
\big(\widehat{\mathcal{O}}_{9}^{W}\big)_{\mu\nu\alpha\beta}&=i\big\langle\chi_{-}f_{-\mu\nu}f_{-\alpha\beta}\big\rangle\,,\\
\big(\widehat{\mathcal{O}}_{11}^{W}\big)_{\mu\nu\alpha\beta}&=i\big\langle\chi_{+}[f_{+\mu\nu},f_{-\alpha\beta}]\big\rangle\,,\\
\big(\widehat{\mathcal{O}}_{22}^{W}\big)_{\mu\nu\alpha\beta}&=\big\langle u_{\mu}\lbrace\nabla^{\gamma}f_{+\gamma\nu},f_{+\alpha\beta}\rbrace\big\rangle\,,\\
\big(\widehat{\mathcal{O}}_{23}^{W}\big)_{\mu\nu\alpha\beta}&=\big\langle u_{\mu}\lbrace\nabla^{\gamma}f_{-\gamma\nu},f_{-\alpha\beta}\rbrace\big\rangle\,.
\end{align*}
\end{itemize}

\subsection{Resonance chiral theory}\label{ssec:RChT}
Taking the large-$N_{c}$ limit we can construct the effective theory of QCD for an intermediate energy region that also satisfies all symmetries of the underlying theory \cite{Ecker:1988te}. This effective theory is called Resonance chiral theory (RChT) and is relevant for energies $M_{\rho}\leq E\leq 2\,\mathrm{GeV}$. 
RChT increases the number of degrees of freedom of ChPT by including massive $\mathrm{U}(3)$ multiplets of vector $V(1^{--})$, axial-vector $A(1^{++})$, scalar $S(0^{++})$ and pseudoscalar $P(0^{-+})$ resonances, denoted generically as a nonet field $R$. This field can be decomposed into singlet $R^{0}$ and octet $R^{a}$ such as \cite{Ecker:1988te,Cirigliano:2006hb}
\begin{equation}
R=\frac{R_{0}}{\sqrt{3}}+\sqrt{2}R_{a}T^{a}\,.\label{eq:nonetr}
\end{equation}
The vector and axial-vector resonances are spin-one particles and in order to describe them we will employ the so-called antisymmetric tensor formalism in the following.\footnote{Other formalisms used in this context are the Proca (vector) formalism \cite{Prades:1993ys} or the first order formalism \cite{Kampf:2006yf,Kampf:2007jf,Kampf:2009jh}.}

Using the large-$N_{c}$ approach, there is no limit to the number of resonances that can be included in the effective Lagrangians. Hence, we can construct the RChT Lagrangian as an expansion in the number of resonance fields, i.e.~\cite{Ecker:1988te}
\begin{equation}
\mathcal{L}_{\mathrm{RChT}}=\mathcal{L}_{\mathrm{ChPT}}+\mathcal{L}_{R_{1}R_{1}}^{(\mathrm{kin})}+\sum_{R_{1}}\mathcal{L}_{R_{1}}+\sum_{R_{1},\,R_{2}}\mathcal{L}_{R_{1}R_{2}}+\sum_{R_{1},\,R_{2},\,R_{3}}\mathcal{L}_{R_{1}R_{2}R_{3}}+\ldots\,,\label{eq:RchTlagrangian}
\end{equation}
where $R_{i}$ stands for the resonance fields $V_{\mu\nu}$, $A_{\mu\nu}$, $S$, $P$ and the dots denote terms with four or more resonances. The individual terms can also be classified by the chiral order, for which they contribute after integrating the resonances out. Then, the resonance Lagrangian schematically reads
\begin{equation}
\mathcal{L}_{\mathrm{RChT}}=\mathcal{L}_{\mathrm{RChT}}^{(4)}+\mathcal{L}_{\mathrm{RChT}}^{(6)}+\ldots\,,\label{eq:RchTlagrangian_v2}
\end{equation}
with the relevant contributions concisely presented below.

\begin{itemize}
\item[a)] The couplings of the lowest massive $\mathrm{U}(3)$ multiplets with the pseudoscalar fields and external sources in the leading order of $1/N_{c}$ are given by the linear part of the interaction resonance Lagrangian \eqref{eq:RchTlagrangian_v2}, i.e.~\cite{Ecker:1988te}
\begin{align}
\mathcal{L}_{\mathrm{RChT}}^{(4)}\ni&\frac{F_{V}}{2\sqrt{2}}\langle V_{\mu\nu}f^{\mu\nu}_{+}\rangle+\frac{F_{A}}{2\sqrt{2}}\langle A_{\mu\nu}f^{\mu\nu}_{-}\rangle+c_{m}\langle S\chi_{+}\rangle+id_{m}\langle P\chi_{-}\rangle\,.\label{eq:lagrangianr4}
\end{align}
\item[b)] The $\mathcal{O}(p^{6})$ Lagrangian, relevant in the odd-intrinsic parity sector, was classified for the first time in ref.~\cite{Kampf:2011ty}. Its form consists of 67 terms in total, parameterized as\footnote{Since ref.~\cite{Kampf:2011ty} uses an opposite convention for the Levi-Civita tensor, we have artificially modified the respective Lagrangian \eqref{eq:odd6-lagrangian} by the overall minus sign so as the sense of the corresponding coupling constants remains the same.}
\begin{equation}
\mathcal{L}_{\mathrm{RChT}}^{\mathrm{(6,\,odd)}}=-\sum_{(i,\,X)}\kappa_{i}^{X}\mathcal{O}_{i}^{X}=-\varepsilon^{\mu\nu\alpha\beta}\sum_{(i,\,X)}\kappa_{i}^{X}\big(\widehat{\mathcal{O}}_{i}^{X}\big)_{\mu\nu\alpha\beta}\,,\label{eq:odd6-lagrangian}
\end{equation}
where $\kappa_{i}^{X}$ are the coupling constants, with $i$ being a serial number of such an operator according to tables 1-7 in \cite{Kampf:2011ty} and $X$ denoting the relevant combination of resonance fields in the corresponding interaction vertex. In detail, the Lagrangian \eqref{eq:odd6-lagrangian} takes into account operators with the single ($V$, $A$, $S$, $P$), double ($VV$, $AA$, $SA$, $SV$, $VA$, $PA$, $PV$) and triple ($VVP$, $VAS$, $AAP$) resonance fields. The individual operators $\big(\widehat{\mathcal{O}}_{i}^{X}\big)_{\mu\nu\alpha\beta}$ can be found in ref.~\cite{Kampf:2011ty} (see tables 1-7 at pages 10-11 therein); nevertheless, we present the relevant ones in the sections \ref{sec:VVP}, \ref{sec:VAS} and \ref{sec:AAP} below as well.
\end{itemize}


\section{Green functions of chiral currents}\label{sec:GF}
The Green functions are formally defined as the vacuum expectation values of the time-ordered product of the composite local operators. In the context of this work, the three-point Green function of the chiral currents can be written down in the momentum representation as the Fourier transform of its form in the coordinate representation,
\begin{equation}
\Pi_{\mathcal{O}_{1}\mathcal{O}_{2}\mathcal{O}_{3}}(p,q;r)=\int\mathrm{d}^{4}x\,\mathrm{d}^{4}y\,e^{-i(p\cdot x+q\cdot y)}\,\big\langle 0\big\vert\mathrm{T}\,\mathcal{O}_{1}(x)\mathcal{O}_{2}(y)\mathcal{O}_{3}(0)\big\vert 0\big\rangle\,,\label{eq:gf-definition}
\end{equation}
where the operators $\mathcal{O}_{i}(x_{i})$ stand either for the chiral vector $V_{\mu}^{a}(x)=\overline{q}(x)\gamma^{\mu}T^{a}q(x)$ and axial-vector $A_{\mu}^{a}(x)=\overline{q}(x)\gamma^{\mu}\gamma_{5}T^{a}q(x)$ currents or the scalar $S^{a}(x)=\overline{q}(x)T^{a}q(x)$ and pseudoscalar $P^{a}(x)=i\overline{q}(x)\gamma_{5}T^{a}q(x)$ densities, with $q(x)$ being the triplet of the lightest quarks.
The symbol $\vert 0\rangle$ stands for the nonperturbative QCD vacuum, however, we will omit it using simply $\langle\mathcal{O}_{1}\mathcal{O}_{2}\mathcal{O}_{3}\rangle$ from now on.

\subsection{Odd-intrinsic parity sector of QCD}\label{ssec:odd_sector}
There exist fifteen different three-point Green functions made of the chiral currents. However, only five of them belong to the odd-intrinsic parity sector of QCD --- for clarity, these can be further divided into two sets. The first one consists of the anomalous $\langle VVA\rangle$ and $\langle AAA\rangle$ correlators, that obtain the perturbative contribution in the chiral limit. On the other hand, the second set is made of the order parameters of the chiral symmetry breaking in the chiral limit, in which the Green functions $\langle VVP\rangle$, $\langle VAS\rangle$ and $\langle AAP\rangle$ belong to.

As we have stated in the introductory paragraphs of this paper, we will be interested only in the Green functions of the second set from now on. In contrast with the anomalous correlators, these have remarkably simple structure due to their transversality. Together with the Lorentz and parity invariance, we can write down the decompositions of these correlators as follows:
\begin{align}
\big[\Pi_{VVP}(p,q;r)\big]_{\mu\nu}^{abc}&=\mathcal{F}_{VVP}(p^{2},q^{2},r^{2})d^{abc}\varepsilon^{\mu\nu(p)(q)}\,,\label{eq:VVP}\\
\big[\Pi_{VAS}(p,q;r)\big]_{\mu\nu}^{abc}&=\mathcal{F}_{VAS}(p^{2},q^{2},r^{2})f^{abc}\varepsilon^{\mu\nu(p)(q)}\,,\label{eq:VAS}\\
\big[\Pi_{AAP}(p,q;r)\big]_{\mu\nu}^{abc}&=\mathcal{F}_{AAP}(p^{2},q^{2},r^{2})d^{abc}\varepsilon^{\mu\nu(p)(q)}\,,\label{eq:AAP}
\end{align}
with the invariant functions $\mathcal{F}_{VVP}$ and $\mathcal{F}_{AAP}$ being symmetrical in the first two arguments, due to the Bose symmetry. In \eqref{eq:VVP}-\eqref{eq:AAP}, $d^{abc}$ is the totally symmetric $\mathrm{SU}(3)$ group invariant and $f^{abc}$ is the totally antisymmetric $\mathrm{SU}(3)$ structure constant, defined as
\begin{align*}
d^{abc}=2\,\mathrm{Tr}\big(\lbrace T^{a},T^{b}\rbrace T^{c}\big)\,,\qquad
f^{abc}=-2i\,\mathrm{Tr}\big([T^{a},T^{b}]T^{c}\big)\,.
\end{align*}

\subsection{Operator product expansion}\label{ssec:OPE}
At high energies, or at short distances in the case of QCD, the asymptotic freedom allows us to use the perturbative expansion in terms of the strong coupling constant $\alpha_{s}$. Then, in the Euclidean region, the asymptotics of the current correlators for large momenta is given by the operator product expansion (OPE) \cite{Wilson:1970ag}.

Let us consider a product of three gauge-invariant composite operators, as on the right-hand side of \eqref{eq:gf-definition}. Assuming the coordinates of these operators to be close to each other, the OPE allows us to rewrite such a product as a linear combination of gauge-invariant local operators, made of the quark and gluon fields, with $c$-number Wilson coefficients that contain all the information about short-distance physics, i.e.~the dynamics above some scale. These are calculable in perturbative QCD by means of the technique of Feynman diagrams. The vacuum averages of these local operators are known as the QCD condensates.

In our previous work \cite{Kadavy:2020hox}, we have studied the contributions of the QCD condensates to the OPE of the three-point Green functions in the chiral limit with all the momenta being simultaneously large. Also, we have restricted ourselves to taking into account only the QCD condensates with canonical dimension $D\leq 6$ (for details, see subsection 3.2 in ref.~\cite{Kadavy:2020hox} and discussion therein). In such a case, we have investigated the perturbative contribution followed by the contributions of the quark $\langle\overline{q}q\rangle$, gluon $\langle G^{2}\rangle$, quark-gluon $\langle\overline{q}\sigma\hspace{-1pt}\cdot\hspace{-1pt}Gq\rangle$ and four-quark $\langle\overline{q}Xq\,\overline{q}Xq\rangle$ condensates.\footnote{Here, we have employed a short-hand notation: $\langle G^{2}\rangle\equiv\langle G_{\mu\nu}^{a}G^{\mu\nu,a}\rangle$ and $\langle\overline{q}\sigma\hspace{-1pt}\cdot\hspace{-1pt}Gq\rangle\equiv\langle\overline{q}\sigma_{\mu\nu}G^{\mu\nu}q\rangle$, with $\sigma^{\mu\nu}=\frac{i}{2}[\gamma^{\mu},\gamma^{\nu}]$ and $G_{\mu\nu}=G_{\mu\nu}^{a}T^{a}$ being the gluon field strength tensor, $G_{\mu\nu}=\partial_{\mu}\mathcal{A}_{\nu}-\partial_{\nu}\mathcal{A}_{\mu}+i g_{s}[\mathcal{A}_{\mu},\mathcal{A}_{\nu}]$ and $G_{\mu\nu}^{a}=\partial_{\mu}\mathcal{A}_{\nu}^{a}-\partial_{\nu}\mathcal{A}_{\mu}^{a}-g_{s}f^{abc}\mathcal{A}_{\mu}^{b}\mathcal{A}_{\nu}^{c}$. } Usually, the validity of the vacuum saturation is supposed for the four-quark condensate and, in the literature, it is systematically written in a form proportional to $\langle\overline{q}q\rangle^{2}$. The accuracy of such an approximation is considered to be $1/N_{c}^{2}$.

As we have mentioned, the $\langle VVP\rangle$, $\langle VAS\rangle$ and $\langle AAP\rangle$ Green functions are the order parameters of the chiral symmetry breaking in the chiral limit. As such, their OPE's start with the contribution of the quark condensate, followed by the one of the quark-gluon condensate.\footnote{Equivalently, the OPE's of the $\langle VVA\rangle$ and $\langle AAA\rangle$ correlators start, in the chiral limit, with the perturbative contribution, succeeded by the contributions of the gluon and four-quark condensates, respectively.} Using the results from subsections 5.2 and 7.2 in ref.~\cite{Kadavy:2020hox}, they read\footnote{The parts proportional to the quark condensate have been known for a long time already --- see, for example, refs.~\cite{Moussallam:1994xp,Knecht:2001xc,Jamin:2008rm,Kampf:2011ty}. On the other hand, to the best of our knowledge, the parts proportional to the quark-gluon condensate were published for the first time in our previous article \cite{Kadavy:2020hox}. In this regard, let us also comment on the decision to restrict ourselves to a study of OPE in the chiral limit only. Having to consider the nonzero quark mass, even in the isospin approximation, would lead to a significant increase of the intricacy of the respective calculations. The reader should be aware of the fact that it would not be sufficient to simply add the mass into the propagator --- the mass would have needed to be accounted for also in the individual propagation formulae. We have, however, tried to study the OPE beyond the chiral limit as well --- at least for some QCD condensates. As an example, we present the result for the subleading contribution of the quark condensate to \eqref{eq:vvp_ope} beyond the chiral limit, which reads
\begin{equation*}
\frac{m^{2}\langle\overline{q}q\rangle}{12\lambda^{6}}\frac{p^{4}(q^{2}+r^{2})+p^{2}(q^{4}+r^{4})+q^{2}r^{2}(q^{2}+r^{2})}{p^{4}q^{4}r^{4}}+\mathcal{O}(m^{4})\,,
\end{equation*}
from which one sees that it starts at $\mathcal{O}(m^{2})$ and is, therefore, suppressed. Finally, we point out that a study of such an OPE with not all momenta large is quite complicated, and we have thus omitted this issue in our analysis.}
\begin{align}
\mathcal{F}_{VVP}^{\mathrm{OPE}}\Big((\lambda p)^{2},(\lambda q)^{2},(\lambda r)^{2}\Big)&=\frac{\langle\overline{q}q\rangle}{6\lambda^{4}}\frac{p^{2}+q^{2}+r^{2}}{p^{2}q^{2}r^{2}}\nonumber\\
&\hspace{-75pt}-\frac{g_{s}\langle\overline{q}\sigma\hspace{-1pt}\cdot\hspace{-1pt}Gq\rangle}{72\lambda^{6}}\frac{r^{2}(p^{4}+q^{4})+3(p^{2}-q^{2})^{2}(p^{2}+q^{2})+4r^{6}}{p^{4}q^{4}r^{4}}+\mathcal{O}\bigg(\frac{1}{\lambda^{8}}\bigg)\,,\label{eq:vvp_ope}\\
\mathcal{F}_{VAS}^{\mathrm{OPE}}\Big((\lambda p)^{2},(\lambda q)^{2},(\lambda r)^{2}\Big)&=\frac{\langle\overline{q}q\rangle}{6\lambda^{4}}\frac{p^{2}-q^{2}-r^{2}}{p^{2}q^{2}r^{2}}\nonumber\\
&\hspace{-75pt}-\frac{g_{s}\langle\overline{q}\sigma\hspace{-1pt}\cdot\hspace{-1pt}Gq\rangle}{72\lambda^{6}}\frac{r^{2}(p^{4}-q^{4})+3(p^{2}-q^{2})(p^{4}+q^{4})-4r^{6}}{p^{4}q^{4}r^{4}}+\mathcal{O}\bigg(\frac{1}{\lambda^{8}}\bigg)\,,\label{eq:vas_ope}\\
\mathcal{F}_{AAP}^{\mathrm{OPE}}\Big((\lambda p)^{2},(\lambda q)^{2},(\lambda r)^{2}\Big)&=\frac{\langle\overline{q}q\rangle}{6\lambda^{4}}\frac{p^{2}+q^{2}-r^{2}}{p^{2}q^{2}r^{2}}\nonumber\\
&\hspace{-75pt}-\frac{g_{s}\langle\overline{q}\sigma\hspace{-1pt}\cdot\hspace{-1pt}Gq\rangle}{72\lambda^{6}}\frac{r^{2}(p^{4}+q^{4})+3(p^{2}-q^{2})^{2}(p^{2}+q^{2})-4r^{6}}{p^{4}q^{4}r^{4}}+\mathcal{O}\bigg(\frac{1}{\lambda^{8}}\bigg)\label{eq:aap_ope}
\end{align}
for $\lambda\to\infty$.

The numerical values of the QCD condensates cannot be calculated perturbatively. They need to be obtained by other means, such as using calculations in the lattice QCD or extracting them from experimental measurements. The values of the quark and quark-gluon condensates depend on the renormalization scale. Usually, the values are given at $1\,\mathrm{GeV}$, for which we have
\begin{align}
\langle\overline{q}q\rangle=&-(240\pm 10)^{3}\,\mathrm{MeV}^{3}\,,\label{eq:quark_condensate_v1}\\
g_{s}\langle\overline{q}\sigma\hspace{-1pt}\cdot\hspace{-1pt}Gq\rangle=&\quad\,m_{0}^{2}\langle\overline{q}q\rangle\,,
\end{align}
with
$
m_{0}^{2}=(0.8\pm 0.2)\,\mathrm{GeV}^{2}\,.
$
The quark condensate is, within the framework of RChT, usually parameterized as
\begin{equation}
\langle\overline{q}q\rangle=-3B_{0}F^{2}\,,\label{eq:quark_condensate_v2}
\end{equation}
where the factor of three comes from the fact that we consider the quark field to be the flavour triplet.


\subsection{General form of resonance contributions}\label{ssec:general_structure}
The mass dimension of the Lorentz-invariant functions of the $\langle VVP\rangle$, $\langle VAS\rangle$ and $\langle AAP\rangle$ correlators is $-1$. Keeping that in mind, one may thus try to construct such Lorentz-invariant functions $\mathcal{F}_{\mathcal{O}_{1}\mathcal{O}_{2}\mathcal{O}_{3}}$ of the respective resonance contributions and find out, whether the expected properties of the proposed ans\"{a}tze are fulfilled.

The resonance contribution in the large $N_{c}$ limit is a rational function with the denominator reflecting the propagators of the individual fields, be it Goldstone bosons or resonances, and with the polynomial of the corresponding dimension in the numerator.

Such a rational ansatz for the Lorentz-invariant function $\mathcal{F}_{\mathcal{O}_{1}\mathcal{O}_{2}\mathcal{O}_{3}}$ is completely general and any Lagrangian, consistent with the chiral symmetry and including the corresponding resonance multiplets, must recover its form. The free parameters of this ansatz are then expressed in terms of the coupling constants of the respective Lagrangian. In practice, the number of these couplings may exceed considerably the number of free parameters of the ansatz. This means that omitting some of the operators in the Lagrangian does not necessarily change the general form of the result. In such a case, only the ``saturation'' of the free parameters of the ansatz by the Lagrangian couplings is modified. In the ideal case, when we know all the couplings of the Lagrangian, we can predict the values of the free parameters of the ansatz. In practice, however, the situation is the opposite --- the proliferation of the Lagrangian couplings at higher orders prevents us from determining them reliably. Therefore, it is hard to estimate the effect of omitting some of the operators, provided this omission does not force the free parameters of the ansatz to acquire definite values (e.g.~zero). The only apparent convenience of the Lagrangian approach is the possible interrelation between the parameters of the ansatz expressed in terms of the coupling constants (typically as a consequence of the symmetries). But this is not the case for our correlators since different non-intersecting subsets of the operators contribute to different correlators. For phenomenological purposes, it is, therefore, more convenient to work with the general ansatz instead of the Lagrangian.

In what follows, we provide a shortened overview of the properties that the invariant functions $\mathcal{F}_{\mathcal{O}_{1}\mathcal{O}_{2}\mathcal{O}_{3}}$ of resonance contributions to the individual Green functions are required to fulfil. A detailed discussion of the properties of the respective structures is then presented in appendix \ref{sec:struktura}, where also the results of all the intermediate steps are provided.

\paragraph{$\bm{\langle VVP\rangle}$ Green function.}
To find a suitable form of the Lorentz-invariant function of the resonance contribution to the $\langle VVP\rangle$ correlator, we require the following conditions to be satisfied. For a clearer reference to these properties in the forthcoming text, we present them in a brief and itemized manner.

\begin{itemize}
\item[1)] We require the high-energy behaviour in terms of the OPE to be satisfied, given by the expression \eqref{eq:vvp_ope}, i.e.~when all the momenta are simultaneously large.
\end{itemize}

There is, however, no argument that would suggest the OPE \eqref{eq:vvp_ope} to be the most preferable one. In other words, although we distinctly require the OPE of the type \eqref{eq:vvp_ope} to be fulfilled, there might be also other kinds of OPE that might be simultaneously valid for our ansatz. Therefore, the second required property of such a structure is as follows.

\begin{itemize}
\item[2)] The $\langle VVP\rangle$ Green function must satisfy also the OPE with only two momenta large. In detail, we require such a function to fulfil the condition \cite{Knecht:2001xc}
\begin{equation}
\mathcal{F}_{VVP}^{\mathrm{OPE}}\Big((\lambda p)^{2},(q-\lambda p)^{2},q^{2}\Big)=\frac{\langle\overline{q}q\rangle}{3\lambda^{2}}\frac{1}{p^{2}q^{2}}+\mathcal{O}\bigg(\frac{1}{\lambda^{3}}\bigg)\qquad \text{for }\lambda\to\infty\,.
\label{eq:vvp_ope_2-op}
\end{equation}
\end{itemize}

There are some other properties of such a correlator that can be of further help. As we have mentioned earlier, the $\langle VVP\rangle$ Green function is probably the most interesting correlator from the phenomenological point of view. It is directly connected to the transition form factor that describes the neutral pion decay. This object will be the center of our attention since there are additional requirements that the form factor is believed to satisfy. Since we work in the chiral limit, let us define the pion transition form factor for the off-shell photons and the on-shell pion as
\begin{equation}
\mathcal{F}_{\gamma^{\ast}\gamma^{\ast}\pi^{0}}(p^{2},q^{2})=\frac{2}{3B_{0}F}\lim_{r^{2}\,\to\,0}r^{2}\mathcal{F}_{VVP}(p^{2},q^{2},r^{2})\,,\label{eq:form_factor}
\end{equation}
where the factor in front of the limit sign is proportional to the overlap between the pion state and the pseudoscalar quark current \cite{Gasser:1983yg,Gasser:1984gg}. Then, the following constraints are expected to be satisfied.

\begin{itemize}
\item[3)] The pion transition form factor \eqref{eq:form_factor} is supposed to fulfill the Brodsky--Lepage behaviour \cite{Lepage:1980fj,Brodsky:1981rp} that we write down in the form as
\begin{equation}
\lim_{Q^{2}\,\to\,\infty}Q^{2}\mathcal{F}_{\gamma^{\ast}\gamma^{\ast}\pi^{0}}(0,-Q^{2})=2F\,.\label{eq:form_factor_BL}
\end{equation}
Alternatively, the Brodsky--Lepage behaviour \eqref{eq:form_factor_BL} can also be written down as
\begin{equation}
\frac{\mathcal{F}_{\gamma^{\ast}\gamma^{\ast}\pi^{0}}(0,-Q^{2})}{\mathcal{F}_{\gamma^{\ast}\gamma^{\ast}\pi^{0}}(0,0)}\rightarrow\frac{24\pi^{2}F^{2}}{N_{c}}\frac{1}{Q^{2}}\qquad\text{for}\,\,Q^{2}\rightarrow \infty\,,\label{eq:footnote_BL}
\end{equation}
which resembles the form presented in ref.~\cite{Husek:2015wta}, see eq.~(14) at page no.~5 therein.
\item[4)] Finally, the last condition demanded from the form factor is that it should recover the chiral anomaly at the photon point. Therefore, apart from the conventional factor of $2/3$, as introduced in \eqref{eq:form_factor}, the WZW anomaly term needs to be obtained:
\begin{equation}
\mathcal{F}_{\gamma^{\ast}\gamma^{\ast}\pi^{0}}(0,0)=\frac{N_{c}}{12\pi^{2}F}\,.\label{eq:form_factor_anomaly}
\end{equation}
\end{itemize}

There is no apparent way how to choose a rational ansatz that satisfies all the four requirements above --- one must therefore resort to the method of trial and error. To this end, we present here the table \ref{tab:VVP_ansatz_table}, in which the properties, with respect to the above-mentioned conditions, of variety ans\"{a}tze for the $\langle VVP\rangle$ correlators are shown.

These ans\"{a}tze are as follows. The LMD+P ansatz takes into account the Goldstone bosons and the lowest vector and pseudoscalar multiplets of resonances \cite{Moussallam:1994xp}. Its extension for a second multiplet of vector resonances then corresponds to the two-hadron saturation, or THS for short \cite{Husek:2015wta}. Finally, the THS+P stands, obviously, for an addition of another pseudoscalar multiplet to the THS parametrization.

As the table \ref{tab:VVP_ansatz_table} suggests, it is necessary to take into account two resonance multiplets of each kind in order to fulfill the demanded properties --- i.e.~the THS+P parametrization of the $\langle VVP\rangle$ correlator is the minimal suitable one.

\begin{table}[t!]
\centering
\begin{tabular}{|C{1.7cm}|C{1.7cm}|C{1.7cm}|C{1.7cm}|C{1.7cm}|C{1.7cm}|C{1.7cm}|C{1.7cm}|}
\hline
Ansatz & Degree of polynomial & No.~of parameters & 3-OPE \eqref{eq:vvp_ope} & 2-OPE \eqref{eq:vvp_ope_2-op} & B--L \eqref{eq:form_factor_BL} & No.~of free parameters \\ \hline\hline
LMD+P  & $4$           & $7$               & $\bm{\times}$ & $\bm{\times}$ & $\bm{\times}$ & 7 \\ \hline
THS    & $8$           & $22$              & \checkmark & \checkmark & $\bm{\times}$ & 7 \\ \hline
THS+P  & $10$          & $34$              & \checkmark & \checkmark & \checkmark & 8 \\ \hline
\end{tabular}
\caption{An overview of the properties of various ans\"{a}tze for the $\langle VVP\rangle$ correlator.}
\label{tab:VVP_ansatz_table}
\end{table}

\paragraph{$\bm{\langle VAS\rangle}$ Green function.}
Searching for a suitable parametrization of a general rational structure of the $\langle VAS\rangle$ correlator is affected by the following facts. Firstly, an absence of the exchange of the Goldstone boson lowers the mass dimension of the polynomial in the denominator by two.

Secondly, there is much less knowledge about the theoretical constraints that can be required from it --- apart from its OPE (see \cite{Jamin:2008rm,Kampf:2011ty,Kadavy:2020hox} for details), there is apparently no study regarding a respective form factor that could have been connected to some phenomenology. The reason can be only assumed to be the fact that $\langle VAS\rangle$ describes very rare processes that have not yet been measured at all or at least accurately enough. Also, the status of the particle content of some of these multiplets is still questionable, especially in the scalar sector.

Nevertheless, as can be seen in appendix \ref{sec:struktura} in detail, we found that in order to be able to satisfy the OPE \eqref{eq:vas_ope}, one is required to take into account two resonance multiplets of each kind in this case, too. Here, we introduce the table \ref{tab:VAS_ansatz_table} that briefly collects such a finding.

\begin{table}[t!]
\centering
\begin{tabular}{|C{3.00cm}|C{2.44cm}|C{2.44cm}|C{2.44cm}|C{2.44cm}|}
\hline
Ansatz                                 & Degree of polynomial     & No.~of parameters     & 3-OPE \eqref{eq:vas_ope} & No.~of free parameters \\ \hline\hline
$V_{1}\,\&\,A_{1}\,\&\,S_{1}$       & $2$                  & $4$                   & $\bm{\times}$ & $4$ \\ \hline
$V_{1,2}\,\&\,A_{1}\,\&\,S_{1}$     & \multirow{3}{*}{$4$} & \multirow{3}{*}{$10$} & \multirow{3}{*}{$\bm{\times}$} & \multirow{3}{*}{$10$} \\ \cline{0-0}
$V_{1}\,\&\,A_{1,2}\,\&\,S_{1}$     &                          &                       &  & \\ \cline{0-0}
$V_{1}\,\&\,A_{1}\,\&\,S_{1,2}$     &                          &                       &  & \\ \hline
$V_{1,2}\,\&\,A_{1,2}\,\&\,S_{1}$   & \multirow{3}{*}{$6$} & \multirow{3}{*}{$20$} & \multirow{3}{*}{$\bm{\times}$} & \multirow{3}{*}{$20$} \\ \cline{0-0}
$V_{1,2}\,\&\,A_{1}\,\&\,S_{1,2}$   &                          &                       &  & \\ \cline{0-0}
$V_{1}\,\&\,A_{1,2}\,\&\,S_{1,2}$   &                          &                       &  & \\ \hline
$V_{1,2}\,\&\,A_{1,2}\,\&\,S_{1,2}$ & $8$                  & $35$                  & \checkmark & $10$ \\ \hline
\end{tabular}
\caption{An overview of the properties of various ans\"{a}tze for the $\langle VAS\rangle$ correlator. Unlike in the previous case, there is no nomenclature used for various $\langle VAS\rangle$ parametrizations, at least to the best of our knowledge. Therefore, we have schematically denoted which resonance multiplets are taken into account in each case.}
\label{tab:VAS_ansatz_table}
\end{table}

\paragraph{$\bm{\langle AAP\rangle}$ Green function.}
The situation here, from the phenomenological point of view, is similar to the case of the $\langle VAS\rangle$. There is, however, one theoretical advantage, and that is the knowledge of OPE in two kinematical regimes, in equivalence to the $\langle VVP\rangle$ correlator. We thus require the structure of the $\langle AAP\rangle$ Green function to satisfy the two following conditions.

\begin{itemize}
\item[1)] The general structure of the resonance contribution to the $\langle AAP\rangle$ correlator needs to satisfy the OPE \eqref{eq:aap_ope}.
\item[2)] The $\langle AAP\rangle$ Green function must also fulfil the OPE with only two momenta large. Its explicit form is identical to \eqref{eq:vvp_ope_2-op} and reads \cite{Knecht:2001xc}
\begin{equation}
\mathcal{F}_{AAP}^{\mathrm{OPE}}\Big((\lambda p)^{2},(q-\lambda p)^{2},q^{2}\Big)=\frac{\langle\overline{q}q\rangle}{3\lambda^{2}}\frac{1}{p^{2}q^{2}}+\mathcal{O}\bigg(\frac{1}{\lambda^{3}}\bigg)\,.\label{eq:aap_ope_2-op}
\end{equation}
\end{itemize}

A detailed overview of all the relevant ans\"{a}tze for the $\langle AAP\rangle$ Green function is presented in appendix \ref{sec:struktura} --- here, we show a table \ref{tab:AAP_ansatz_table}, which summarizes the results obtained therein. Interestingly enough, it is important to point out that the above-mentioned conditions can be satisfied even if only the axial-vector resonances are duplicated. However, since the previous analysis of the $\langle VVP\rangle$ and $\langle VAS\rangle$ correlators suggest the necessity of taking into account two resonance multiplets of each kind, we duplicate the pseudoscalar multiplet in this case, too --- at least to have the complete resonance Lagrangian up to $\mathcal{O}(p^{6})$.

\begin{table}[t!]
\centering
\begin{tabular}{|C{2.08cm}|C{2.08cm}|C{2.08cm}|C{2.08cm}|C{2.08cm}|C{2.08cm}|}
\hline
Ansatz                                 & Degree of polynomial & No.~of parameters & 3-OPE \eqref{eq:aap_ope} & 2-OPE \eqref{eq:aap_ope_2-op} & No.~of free parameters \\ \hline\hline
$A_{1}\,\&\,P_{1}$       & $4$              & $7$               & $\bm{\times}$            & $\bm{\times}$                 & 7 \\ \hline
$A_{1,2}\,\&\,P_{1}$   & $8$              & $22$              & \checkmark               & \checkmark                    & 7 \\ \hline
$A_{1,2}\,\&\,P_{1,2}$ & $10$             & $34$              & \checkmark               & \checkmark                    & 12 \\ \hline
\end{tabular}
\caption{An overview of the properties of various ans\"{a}tze for the $\langle AAP\rangle$ correlator.}
\label{tab:AAP_ansatz_table}
\end{table}
%


\subsection{Duplication of resonance Lagrangians}\label{ssec:duplication}
The summary of the preceding subsection is thus that it is necessary to enlarge the Lagrangians \eqref{eq:lagrangianr4} and \eqref{eq:odd6-lagrangian} to two resonance multiplets in each channel in order to be able to construct the resonance contributions to the $\langle VVP\rangle$, $\langle VAS\rangle$ and $\langle AAP\rangle$ Green functions such that the respective OPE can be fulfilled. In what follows, we perform such a procedure.

Since the Lagrangian \eqref{eq:lagrangianr4} is linear in resonance fields, its relevant part is easily modified as $\mathcal{L}_{\mathrm{RChT}}^{(4)}\rightarrow\widetilde{\mathcal{L}}_{\mathrm{RChT}}^{(4)}$, with
\begin{align}
\widetilde{\mathcal{L}}_{\mathrm{RChT}}^{(4)}\ni\sum_{i=1}^{2}\bigg(\frac{F_{V_{i}}}{2\sqrt{2}}\langle V_{i\,\mu\nu}f^{\mu\nu}_{+}\rangle&+\frac{F_{A_{i}}}{2\sqrt{2}}\langle A_{i\,\mu\nu}f^{\mu\nu}_{-}\rangle+c_{m}^{(i)}\langle S_{i}\,\chi_{+}\rangle+id_{m}^{(i)}\langle P_{i}\,\chi_{-}\rangle\bigg)\,.\label{eq:lagrangianr4_v2}
\end{align}

The situation for the Lagrangian \eqref{eq:odd6-lagrangian} is a bit complicated. Let us remind the reader that the original 67 operators are accompanied by the respective 67 couplings, generally denoted as $\kappa_{i}^{X}$. The duplication of operators linear in resonance fields is done trivially as in the case above. However, similar manipulations with operators quadratic or cubic in resonance fields require one to proceed cautiously --- not only that one needs to consider all the relevant combinations, but it is necessary to make sure that only the linearly independent terms are kept. It can be expected that number of the newly-introduced operators exceeds the original amount significantly --- as we shall see in the next section, instead of $10$, $9$ and $13$ original operators for the $\langle VVP\rangle$, $\langle VAS\rangle$ and $\langle AAP\rangle$ correlators, one then has $39$, $44$ and $46$ operators in total, respectively. Nevertheless, not every operator necessarily contributes at the tree level after all.

In the tables \ref{tab:VVP_monomials_1}-\ref{tab:AAP_monomials_3}, we present all the relevant operators of the Lagrangian \eqref{eq:odd6-lagrangian} that generate nontrivial vertices contributing to the $\langle VVP\rangle$, $\langle VAS\rangle$ and $\langle AAP\rangle$ Green functions, accompanied by their additional variants with higher-mass resonances taken into account.
\begin{table}[t!]
\centering
\begin{tabular}{|C{4cm}|C{2cm}||C{4cm}|C{2cm}|}
\hline
Operator $\big(\widehat{\mathcal{O}}_{i}^{R}\big)_{\mu\nu\alpha\beta}$                          & Coupling constant & Operator $\big(\widehat{\mathcal{O}}_{i}^{R}\big)_{\mu\nu\alpha\beta}$                          & Coupling constant  \\ \hline\hline
$\langle V_{1}^{\mu\nu}\lbrace f_{+}^{\alpha\rho},h^{\beta\sigma}\rbrace\rangle g_{\rho\sigma}$ & $\kappa_{12}^{V}$ & $\langle V_{2}^{\mu\nu}\lbrace f_{+}^{\alpha\rho},h^{\beta\sigma}\rbrace\rangle g_{\rho\sigma}$ & $\lambda_{12}^{V}$ \\ \hline
$i\langle V_{1}^{\mu\nu}\lbrace f_{+}^{\alpha\beta},\chi_{-}\rbrace\rangle$                     & $\kappa_{14}^{V}$ & $i\langle V_{2}^{\mu\nu}\lbrace f_{+}^{\alpha\beta},\chi_{-}\rbrace\rangle$                     & $\lambda_{14}^{V}$ \\ \hline
$\langle V_{1}^{\mu\nu}\lbrace\nabla^{\alpha}f_{+}^{\beta\sigma},u_{\sigma}\rbrace\rangle$      & $\kappa_{16}^{V}$ & $\langle V_{2}^{\mu\nu}\lbrace\nabla^{\alpha}f_{+}^{\beta\sigma},u_{\sigma}\rbrace\rangle$      & $\lambda_{16}^{V}$ \\ \hline
$\langle V_{1}^{\mu\nu}\lbrace\nabla_{\sigma}f_{+}^{\alpha\sigma},u^{\beta}\rbrace\rangle$      & $\kappa_{17}^{V}$ & $\langle V_{2}^{\mu\nu}\lbrace\nabla_{\sigma}f_{+}^{\alpha\sigma},u^{\beta}\rbrace\rangle$      & $\lambda_{17}^{V}$ \\ \hline
$\langle P_{1}\lbrace f_{+}^{\mu\nu},f_{+}^{\alpha\beta}\rbrace\rangle$                         & $\kappa_{5}^{P}$  & $\langle P_{2}\lbrace f_{+}^{\mu\nu},f_{+}^{\alpha\beta}\rbrace\rangle$                         & $\lambda_{5}^{P}$ \\ \hline
\end{tabular}
\caption{Monomials of the extended RChT Lagrangian at $\mathcal{O}(p^{6})$ with one resonance field, relevant for the $\langle VVP\rangle$ Green function.}
\label{tab:VVP_monomials_1}
\end{table}
\begin{table}[t!]
\centering
\begin{tabular}{|C{4cm}|C{2cm}||C{4cm}|C{2cm}|}
\hline
Operator $\big(\widehat{\mathcal{O}}_{i}^{RR}\big)_{\mu\nu\alpha\beta}$                     & Coupling constant & Operator $\big(\widehat{\mathcal{O}}_{i}^{RR}\big)_{\mu\nu\alpha\beta}$                                 & Coupling constant   \\ \hline\hline
$i\langle\lbrace V_{1}^{\mu\nu},V_{1}^{\alpha\beta}\rbrace\chi_{-}\rangle$                  & $\kappa_{2}^{VV}$ & $i\langle\lbrace V_{1}^{\mu\nu},V_{2}^{\alpha\beta}\rbrace\chi_{-}\rangle$                              & $\lambda_{21}^{VV}$ \\
                                                                                            &                   & $i\langle\lbrace V_{2}^{\mu\nu},V_{2}^{\alpha\beta}\rbrace\chi_{-}\rangle$                              & $\lambda_{22}^{VV}$ \\  \hline
$\langle\lbrace \nabla_{\sigma}V_{1}^{\mu\nu},V_{1}^{\alpha\sigma}\rbrace u^{\beta}\rangle$ & $\kappa_{3}^{VV}$ & $\langle\lbrace\nabla_{\sigma}V_{1}^{\mu\nu},V_{2}^{\alpha\sigma}\rbrace u^{\beta}\rangle$              & $\lambda_{31}^{VV}$ \\
                                                                                            &                   & $\langle\lbrace\nabla_{\sigma}V_{1}^{\mu\nu},V_{2}^{\alpha\beta}\rbrace u^{\sigma}\rangle$              & $\lambda_{32}^{VV}$ \\
                                                                                            &                   & $\langle\lbrace\nabla_{\sigma}V_{2}^{\mu\nu},V_{1}^{\alpha\sigma}\rbrace u^{\beta}\rangle$              & $\lambda_{33}^{VV}$ \\
                                                                                            &                   & $\langle\lbrace\nabla_{\sigma}V_{2}^{\mu\nu},V_{2}^{\alpha\sigma}\rbrace u^{\beta}\rangle$              & $\lambda_{34}^{VV}$ \\ \hline
$\langle\lbrace \nabla^{\beta}V_{1}^{\mu\nu},V_{1}^{\alpha\sigma}\rbrace u_{\sigma}\rangle$ & $\kappa_{4}^{VV}$ & $\langle\lbrace\nabla^{\beta}V_{1}^{\mu\nu},V_{2}^{\alpha\sigma}\rbrace u_{\sigma}\rangle$              & $\lambda_{41}^{VV}$ \\
                                                                                            &                   & $\langle\lbrace\nabla^{\beta}V_{2}^{\mu\nu},V_{1}^{\alpha\sigma}\rbrace u_{\sigma}\rangle$              & $\lambda_{42}^{VV}$ \\
                                                                                            &                   & $\langle\lbrace\nabla^{\beta}V_{2}^{\mu\nu},V_{2}^{\alpha\sigma}\rbrace u_{\sigma}\rangle$              & $\lambda_{43}^{VV}$ \\ \hline
$\langle\lbrace V_{1}^{\mu\nu},P_{1}\rbrace f_{+}^{\alpha\beta}\rangle$                     & $\kappa_{3}^{PV}$ & $\langle\lbrace V_{2}^{\mu\nu},P_{1}\rbrace f_{+}^{\alpha\beta}\rangle$                                 & $\lambda_{31}^{PV}$ \\
                                                                                            &                   & $\langle\lbrace V_{1}^{\mu\nu},P_{2}\rbrace f_{+}^{\alpha\beta}\rangle$                                 & $\lambda_{32}^{PV}$ \\
                                                                                            &                   & $\langle\lbrace V_{2}^{\mu\nu},P_{2}\rbrace f_{+}^{\alpha\beta}\rangle$                                 & $\lambda_{33}^{PV}$ \\ \hline
\end{tabular}
\caption{Monomials of the extended RChT Lagrangian at $\mathcal{O}(p^{6})$ with two resonance fields, relevant for the $\langle VVP\rangle$ Green function. Note that another operator with both $V_{1}$ and $V_{2}$ resonances formally exists, namely $\langle\lbrace\nabla_{\sigma}V_{2}^{\mu\nu},V_{1}^{\alpha\beta}\rbrace u^{\sigma}\rangle$. It can be, however, rewritten as a combination of the remaining ones and, therefore, omitted.}
\label{tab:VVP_monomials_2}
\end{table}
\begin{table}[t!]
\centering
\begin{tabular}{|C{4cm}|C{2cm}||C{4cm}|C{2cm}|}
\hline
Operator $\big(\widehat{\mathcal{O}}_{i}^{RRR}\big)_{\mu\nu\alpha\beta}$ & Coupling constant & Operator $\big(\widehat{\mathcal{O}}_{i}^{RRR}\big)_{\mu\nu\alpha\beta}$ & Coupling constant   \\ \hline\hline
$\langle V_{1}^{\mu\nu}V_{1}^{\alpha\beta}P_{1}\rangle$                  & $\kappa^{VVP}$    & $\langle\lbrace V_{1}^{\mu\nu},V_{2}^{\alpha\beta}\rbrace P_{1}\rangle$  & $\lambda_{1}^{VVP}$ \\
                                                                         &                   & $\langle V_{1}^{\mu\nu}V_{1}^{\alpha\beta}P_{2}\rangle$                  & $\lambda_{2}^{VVP}$ \\
                                                                         &                   & $\langle V_{2}^{\mu\nu}V_{2}^{\alpha\beta}P_{1}\rangle$                  & $\lambda_{3}^{VVP}$ \\
                                                                         &                   & $\langle\lbrace V_{1}^{\mu\nu},V_{2}^{\alpha\beta}\rbrace P_{2}\rangle$  & $\lambda_{4}^{VVP}$ \\
                                                                         &                   & $\langle V_{2}^{\mu\nu}V_{2}^{\alpha\beta}P_{2}\rangle$                  & $\lambda_{5}^{VVP}$ \\ \hline
\end{tabular}
\caption{Monomials of the extended RChT Lagrangian at $\mathcal{O}(p^{6})$ with three resonance fields, relevant for the $\langle VVP\rangle$ Green function.}
\label{tab:VVP_monomials_3}
\end{table}
\begin{table}[t!]
\centering
\begin{tabular}{|C{4cm}|C{2cm}||C{4cm}|C{2cm}|}
\hline
Operator $\big(\widehat{\mathcal{O}}_{i}^{R}\big)_{\mu\nu\alpha\beta}$ & Coupling constant & Operator $\big(\widehat{\mathcal{O}}_{i}^{R}\big)_{\mu\nu\alpha\beta}$ & Coupling constant  \\ \hline\hline
$i\langle[V_{1}^{\mu\nu},\nabla^{\alpha}\chi_{+}]u^{\beta}\rangle$     & $\kappa_{4}^{V}$  & $i\langle[V_{2}^{\mu\nu},\nabla^{\alpha}\chi_{+}]u^{\beta}\rangle$     & $\lambda_{4}^{V}$ \\ \hline
$i\langle V_{1}^{\mu\nu}[f_{-}^{\alpha\beta},\chi_{+}]\rangle$         & $\kappa_{15}^{V}$ & $i\langle V_{2}^{\mu\nu}[f_{-}^{\alpha\beta},\chi_{+}]\rangle$         & $\lambda_{15}^{V}$ \\ \hline
$i\langle A_{1}^{\mu\nu}[f_{+}^{\alpha\beta},\chi_{+}]\rangle$         & $\kappa_{14}^{A}$ & $i\langle A_{2}^{\mu\nu}[f_{+}^{\alpha\beta},\chi_{+}]\rangle$         & $\lambda_{14}^{A}$ \\ \hline
$i\langle S_{1}[f_{+}^{\mu\nu},f_{-}^{\alpha\beta}]\rangle$            & $\kappa_{2}^{S}$  & $i\langle S_{2}[f_{+}^{\mu\nu},f_{-}^{\alpha\beta}]\rangle$            & $\lambda_{2}^{S}$ \\ \hline
\end{tabular}
\caption{Monomials of the extended RChT Lagrangian at $\mathcal{O}(p^{6})$ with one resonance field, relevant for the $\langle VAS\rangle$ Green function.}
\label{tab:VAS_monomials_1}
\end{table}
\begin{table}[t!]
\centering
\begin{tabular}{|C{4cm}|C{2cm}||C{4cm}|C{2cm}|}
\hline
Operator $\big(\widehat{\mathcal{O}}_{i}^{RR}\big)_{\mu\nu\alpha\beta}$ & Coupling constant & Operator $\big(\widehat{\mathcal{O}}_{i}^{RR}\big)_{\mu\nu\alpha\beta}$ & Coupling constant  \\ \hline\hline
$i\langle[A_{1}^{\mu\nu},S_{1}]f_{+}^{\alpha\beta}\rangle$              & $\kappa_{1}^{SA}$ & $i\langle[A_{2}^{\mu\nu},S_{1}]f_{+}^{\alpha\beta}\rangle$              & $\lambda_{11}^{SA}$ \\
                                                                        &                   & $i\langle[A_{1}^{\mu\nu},S_{2}]f_{+}^{\alpha\beta}\rangle$              & $\lambda_{12}^{SA}$ \\
                                                                        &                   & $i\langle[A_{2}^{\mu\nu},S_{2}]f_{+}^{\alpha\beta}\rangle$              & $\lambda_{13}^{SA}$ \\ \hline
$i\langle[V_{1}^{\mu\nu},S_{1}]f_{-}^{\alpha\beta}\rangle$              & $\kappa_{1}^{SV}$ & $i\langle[V_{2}^{\mu\nu},S_{1}]f_{-}^{\alpha\beta}\rangle$              & $\lambda_{11}^{SV}$ \\
                                                                        &                   & $i\langle[V_{1}^{\mu\nu},S_{2}]f_{-}^{\alpha\beta}\rangle$              & $\lambda_{12}^{SV}$ \\
                                                                        &                   & $i\langle[V_{2}^{\mu\nu},S_{2}]f_{-}^{\alpha\beta}\rangle$              & $\lambda_{13}^{SV}$ \\ \hline
$i\langle[V_{1}^{\mu\nu},\nabla^{\alpha}S_{1}]u^{\beta}\rangle$         & $\kappa_{2}^{SV}$ & $i\langle[V_{2}^{\mu\nu},\nabla^{\alpha}S_{1}]u^{\beta}\rangle$         & $\lambda_{21}^{SV}$ \\
                                                                        &                   & $i\langle[V_{1}^{\mu\nu},\nabla^{\alpha}S_{2}]u^{\beta}\rangle$         & $\lambda_{22}^{SV}$ \\
                                                                        &                   & $i\langle[V_{2}^{\mu\nu},\nabla^{\alpha}S_{2}]u^{\beta}\rangle$         & $\lambda_{23}^{SV}$ \\ \hline
$i\langle [V_{1}^{\mu\nu},A_{1}^{\alpha\beta}]\chi_{+}\rangle$          & $\kappa_{6}^{VA}$ & $i\langle [V_{2}^{\mu\nu},A_{1}^{\alpha\beta}]\chi_{+}\rangle$          & $\lambda_{61}^{VA}$ \\
                                                                        &                   & $i\langle [V_{1}^{\mu\nu},A_{2}^{\alpha\beta}]\chi_{+}\rangle$          & $\lambda_{62}^{VA}$ \\
                                                                        &                   & $i\langle [V_{2}^{\mu\nu},A_{2}^{\alpha\beta}]\chi_{+}\rangle$          & $\lambda_{63}^{VA}$ \\ \hline
\end{tabular}
\caption{Monomials of the extended RChT Lagrangian at $\mathcal{O}(p^{6})$ with two resonance fields, relevant for the $\langle VAS\rangle$ Green function.}
\label{tab:VAS_monomials_2}
\end{table}
\begin{table}[t!]
\centering
\begin{tabular}{|C{4cm}|C{2cm}||C{4cm}|C{2cm}|}
\hline
Operator $\big(\widehat{\mathcal{O}}_{i}^{RRR}\big)_{\mu\nu\alpha\beta}$ & Coupling constant & Operator $\big(\widehat{\mathcal{O}}_{i}^{RRR}\big)_{\mu\nu\alpha\beta}$ & Coupling constant   \\ \hline\hline
$i\langle[V_{1}^{\mu\nu},A_{1}^{\alpha\beta}]S_{1}\rangle$               & $\kappa^{VAS}$    & $i\langle[V_{2}^{\mu\nu},A_{1}^{\alpha\beta}]S_{1}\rangle$ & $\lambda_{1}^{VAS}$ \\
                                                                         &                   & $i\langle[V_{1}^{\mu\nu},A_{2}^{\alpha\beta}]S_{1}\rangle$ & $\lambda_{2}^{VAS}$ \\
                                                                         &                   & $i\langle[V_{1}^{\mu\nu},A_{1}^{\alpha\beta}]S_{2}\rangle$ & $\lambda_{3}^{VAS}$ \\
                                                                         &                   & $i\langle[V_{2}^{\mu\nu},A_{2}^{\alpha\beta}]S_{1}\rangle$ & $\lambda_{4}^{VAS}$ \\
                                                                         &                   & $i\langle[V_{2}^{\mu\nu},A_{1}^{\alpha\beta}]S_{2}\rangle$ & $\lambda_{5}^{VAS}$ \\
                                                                         &                   & $i\langle[V_{1}^{\mu\nu},A_{2}^{\alpha\beta}]S_{2}\rangle$ & $\lambda_{6}^{VAS}$ \\
                                                                         &                   & $i\langle[V_{2}^{\mu\nu},A_{2}^{\alpha\beta}]S_{2}\rangle$ & $\lambda_{7}^{VAS}$ \\ \hline
\end{tabular}
\caption{Monomials of the extended RChT Lagrangian at $\mathcal{O}(p^{6})$ with three resonance fields, relevant for the $\langle VAS\rangle$ Green function.}
\label{tab:VAS_monomials_3}
\end{table}
\begin{table}[t!]
\centering
\begin{tabular}{|C{4cm}|C{2cm}||C{4cm}|C{2cm}|}
\hline
Operator $\big(\widehat{\mathcal{O}}_{i}^{R}\big)_{\mu\nu\alpha\beta}$                     & Coupling constant & Operator $\big(\widehat{\mathcal{O}}_{i}^{R}\big)_{\mu\nu\alpha\beta}$                     & Coupling constant  \\ \hline\hline
$\langle A_{1}^{\mu\nu}\lbrace\nabla^{\alpha}h^{\beta\sigma},u_{\sigma}\rbrace\rangle$     & $\kappa_{3}^{A}$  & $\langle A_{2}^{\mu\nu}\lbrace\nabla^{\alpha}h^{\beta\sigma},u_{\sigma}\rbrace\rangle$     & $\lambda_{3}^{A}$  \\ \hline
$\langle A_{1}^{\mu\nu}\lbrace f_{-}^{\alpha\sigma},h^{\beta\sigma}\rbrace\rangle$         & $\kappa_{8}^{A}$  & $\langle A_{2}^{\mu\nu}\lbrace f_{-}^{\alpha\sigma},h^{\beta\sigma}\rbrace\rangle$         & $\lambda_{8}^{A}$  \\ \hline
$i\langle A_{1}^{\mu\nu}\lbrace f_{-}^{\alpha\beta},\chi_{-}\rbrace\rangle$                & $\kappa_{11}^{A}$ & $i\langle A_{2}^{\mu\nu}\lbrace f_{-}^{\alpha\beta},\chi_{-}\rbrace\rangle$                & $\lambda_{11}^{A}$ \\ \hline
$i\langle A_{1}^{\mu\nu}\lbrace\nabla^{\alpha}\chi_{-},u^{\beta}\rbrace\rangle$            & $\kappa_{12}^{A}$ & $i\langle A_{2}^{\mu\nu}\lbrace\nabla^{\alpha}\chi_{-},u^{\beta}\rbrace\rangle$            & $\lambda_{12}^{A}$ \\ \hline
$\langle A_{1}^{\mu\nu}\lbrace\nabla^{\alpha}f_{-}^{\beta\sigma},u_{\sigma}\rbrace\rangle$ & $\kappa_{15}^{A}$ & $\langle A_{2}^{\mu\nu}\lbrace\nabla^{\alpha}f_{-}^{\beta\sigma},u_{\sigma}\rbrace\rangle$ & $\lambda_{15}^{A}$ \\ \hline
$\langle A_{1}^{\mu\nu}\lbrace\nabla_{\sigma}f_{-}^{\alpha\sigma},u^{\beta}\rbrace\rangle$ & $\kappa_{16}^{A}$ & $\langle A_{2}^{\mu\nu}\lbrace\nabla_{\sigma}f_{-}^{\alpha\sigma},u^{\beta}\rbrace\rangle$ & $\lambda_{16}^{A}$ \\ \hline
$\langle P_{1}\lbrace f_{-}^{\mu\nu},f_{-}^{\alpha\beta}\rbrace\rangle$                    & $\kappa_{1}^{P}$  & $\langle P_{2}\lbrace f_{-}^{\mu\nu},f_{-}^{\alpha\beta}\rbrace\rangle$                    & $\lambda_{1}^{P}$  \\ \hline
\end{tabular}
\caption{Monomials of the extended RChT Lagrangian at $\mathcal{O}(p^{6})$ with one resonance field, relevant for the $\langle AAP\rangle$ Green function.}
\label{tab:AAP_monomials_1}
\end{table}
\begin{table}[t!]
\centering
\begin{tabular}{|C{4cm}|C{2cm}||C{4cm}|C{2cm}|}
\hline
Operator $\big(\widehat{\mathcal{O}}_{i}^{RR}\big)_{\mu\nu\alpha\beta}$                     & Coupling constant & Operator $\big(\widehat{\mathcal{O}}_{i}^{RR}\big)_{\mu\nu\alpha\beta}$                                 & Coupling constant   \\ \hline\hline
$i\langle\lbrace A_{1}^{\mu\nu},A_{1}^{\alpha\beta}\rbrace\chi_{-}\rangle$                  & $\kappa_{2}^{AA}$ & $i\langle\lbrace A_{1}^{\mu\nu},A_{2}^{\alpha\beta}\rbrace\chi_{-}\rangle$                              & $\lambda_{21}^{AA}$ \\
                                                                                            &                   & $i\langle\lbrace A_{2}^{\mu\nu},A_{2}^{\alpha\beta}\rbrace\chi_{-}\rangle$                              & $\lambda_{22}^{AA}$ \\ \hline
$\langle\lbrace \nabla_{\sigma}A_{1}^{\mu\nu},A_{1}^{\alpha\sigma}\rbrace u^{\beta}\rangle$ & $\kappa_{3}^{AA}$ & $\langle\lbrace\nabla_{\sigma}A_{1}^{\mu\nu},A_{2}^{\alpha\sigma}\rbrace u^{\beta}\rangle$              & $\lambda_{31}^{AA}$ \\
                                                                                            &                   & $\langle\lbrace\nabla_{\sigma}A_{1}^{\mu\nu},A_{2}^{\alpha\beta}\rbrace u^{\sigma}\rangle$              & $\lambda_{32}^{AA}$ \\
                                                                                            &                   & $\langle\lbrace\nabla_{\sigma}A_{2}^{\mu\nu},A_{1}^{\alpha\sigma}\rbrace u^{\beta}\rangle$              & $\lambda_{33}^{AA}$ \\
                                                                                            &                   & $\langle\lbrace\nabla_{\sigma}A_{2}^{\mu\nu},A_{2}^{\alpha\sigma}\rbrace u^{\beta}\rangle$              & $\lambda_{34}^{AA}$ \\ \hline
$\langle\lbrace \nabla^{\beta}A_{1}^{\mu\nu},A_{1}^{\alpha\sigma}\rbrace u_{\sigma}\rangle$ & $\kappa_{4}^{AA}$ & $\langle\lbrace\nabla^{\beta}A_{1}^{\mu\nu},A_{2}^{\alpha\sigma}\rbrace u_{\sigma}\rangle$              & $\lambda_{41}^{AA}$ \\ 
                                                                                            &                   & $\langle\lbrace\nabla^{\beta}A_{2}^{\mu\nu},A_{1}^{\alpha\sigma}\rbrace u_{\sigma}\rangle$              & $\lambda_{42}^{AA}$ \\
                                                                                            &                   & $\langle\lbrace\nabla^{\beta}A_{2}^{\mu\nu},A_{2}^{\alpha\sigma}\rbrace u_{\sigma}\rangle$              & $\lambda_{43}^{AA}$ \\ \hline
$\langle\lbrace A_{1}^{\mu\nu},P_{1}\rbrace f_{-}^{\alpha\beta}\rangle$                     & $\kappa_{1}^{PA}$ & $\langle\lbrace A_{2}^{\mu\nu},P_{1}\rbrace f_{-}^{\alpha\beta}\rangle$                                 & $\lambda_{11}^{PA}$ \\
                                                                                            &                   & $\langle\lbrace A_{1}^{\mu\nu},P_{2}\rbrace f_{-}^{\alpha\beta}\rangle$                                 & $\lambda_{12}^{PA}$ \\
                                                                                            &                   & $\langle\lbrace A_{2}^{\mu\nu},P_{2}\rbrace f_{-}^{\alpha\beta}\rangle$                                 & $\lambda_{13}^{PA}$ \\ \hline
$\langle\lbrace A_{1}^{\mu\nu},\nabla^{\alpha}P_{1}\rbrace u^{\beta}\rangle$                & $\kappa_{2}^{PA}$ & $\langle\lbrace A_{2}^{\mu\nu},\nabla^{\alpha}P_{1}\rbrace u^{\beta}\rangle$                            & $\lambda_{21}^{PA}$ \\
                                                                                            &                   & $\langle\lbrace A_{1}^{\mu\nu},\nabla^{\alpha}P_{2}\rbrace u^{\beta}\rangle$                            & $\lambda_{22}^{PA}$ \\
                                                                                            &                   & $\langle\lbrace A_{2}^{\mu\nu},\nabla^{\alpha}P_{2}\rbrace u^{\beta}\rangle$                            & $\lambda_{23}^{PA}$ \\ \hline
\end{tabular}
\caption{Monomials of the extended RChT Lagrangian at $\mathcal{O}(p^{6})$ with two resonance fields, relevant for the $\langle AAP\rangle$ Green function. Note that another operator with both $V_{1}$ and $V_{2}$ resonances formally exists, namely $\langle\lbrace\nabla_{\sigma}A_{2}^{\mu\nu},A_{1}^{\alpha\beta}\rbrace u^{\sigma}\rangle$. It can be, however, rewritten as a combination of the remaining ones and, therefore, omitted.}
\label{tab:AAP_monomials_2}
\end{table}
\begin{table}[t!]
\centering
\begin{tabular}{|C{4cm}|C{2cm}||C{4cm}|C{2cm}|}
\hline
Operator $\big(\widehat{\mathcal{O}}_{i}^{RRR}\big)_{\mu\nu\alpha\beta}$ & Coupling constant & Operator $\big(\widehat{\mathcal{O}}_{i}^{RRR}\big)_{\mu\nu\alpha\beta}$ & Coupling constant   \\ \hline\hline
$\langle A_{1}^{\mu\nu}A_{1}^{\alpha\beta}P_{1}\rangle$                  & $\kappa^{AAP}$    & $\langle\lbrace A_{1}^{\mu\nu},A_{2}^{\alpha\beta}\rbrace P_{1}\rangle$  & $\lambda_{1}^{AAP}$ \\
                                                                         &                   & $\langle A_{1}^{\mu\nu}A_{1}^{\alpha\beta}P_{2}\rangle$                  & $\lambda_{2}^{AAP}$ \\
                                                                         &                   & $\langle A_{2}^{\mu\nu}A_{2}^{\alpha\beta}P_{1}\rangle$                  & $\lambda_{3}^{AAP}$ \\
                                                                         &                   & $\langle\lbrace A_{1}^{\mu\nu},A_{2}^{\alpha\beta}\rbrace P_{2}\rangle$  & $\lambda_{4}^{AAP}$ \\
                                                                         &                   & $\langle A_{2}^{\mu\nu}A_{2}^{\alpha\beta}P_{2}\rangle$                  & $\lambda_{5}^{AAP}$ \\ \hline
\end{tabular}
\caption{Monomials of the extended RChT Lagrangian at $\mathcal{O}(p^{6})$ with three resonance fields, relevant for the $\langle AAP\rangle$ Green function.}
\label{tab:AAP_monomials_3}
\end{table}
%


\section{Two-multiplet resonance contribution within RChT}\label{sec:two_multiplets}

\subsection{\texorpdfstring{$\langle VVP\rangle$}{} Green function}\label{sec:VVP}
In this section, we first reintroduce the ChPT contribution to the $\langle VVP\rangle$ and then we revisit its RChT contribution. The obtained results are further subjected to the matching and the respective high-energy behaviour within the framework of OPE is discussed.

\begin{figure}[t!]
\centering
    \begin{subfigure}[t]{0.27\textwidth}
        \hspace{6.9pt}\includegraphics[width=0.85\textwidth]{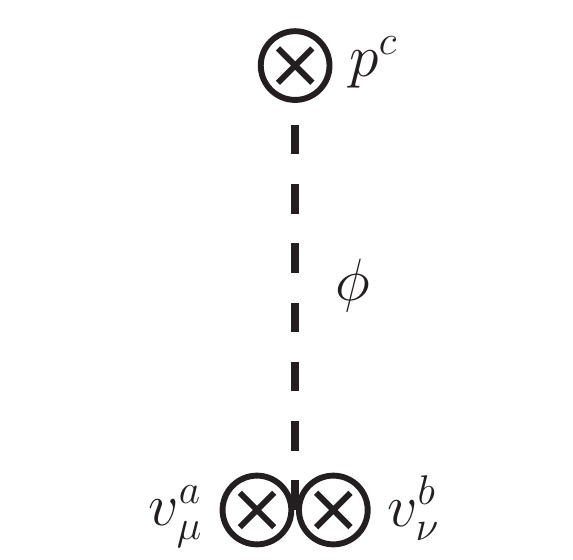}
        \caption{}
        \label{fig:VVP_ChPT_graph_1}
    \end{subfigure}
    \begin{subfigure}[t]{0.27\textwidth}
        \hspace{7.1pt}\includegraphics[width=0.85\textwidth]{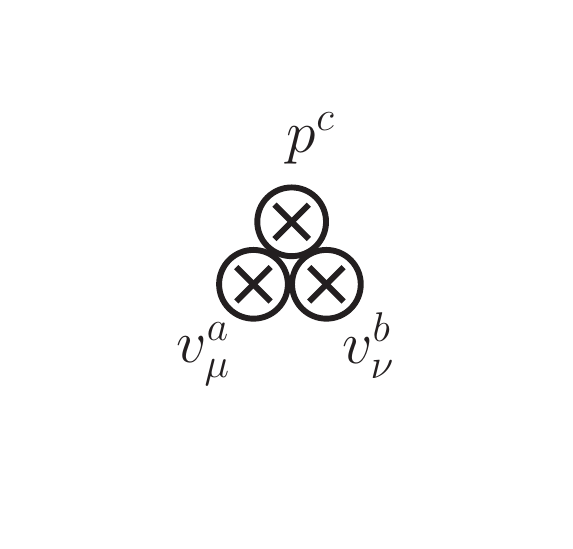}
        \caption{}
        \label{fig:VVP_ChPT_graph_2}
    \end{subfigure}
    \caption{Tree-level Feynman diagrams of the ChPT contribution to the $\langle VVP\rangle$ Green function up to $\mathcal{O}(p^{6})$.}
    \label{fig:VVP_ChPT_graphs}
\end{figure}

\paragraph{Contribution of ChPT.}
In the limit $N_{c}\to\infty$, the ChPT contribution up to $\mathcal{O}(p^{6})$ to the $\langle VVP\rangle$ Green function is given by two diagram topologies, see fig.~\ref{fig:VVP_ChPT_graphs}, and reads
\begin{align}
\mathcal{F}_{VVP}^{\mathrm{ChPT}}(p^{2},q^{2},r^{2})=\frac{B_{0}N_{c}}{8\pi^{2}r^{2}}-32B_{0}C_{7}^{W}+8B_{0}C_{22}^{W}\frac{p^{2}+q^{2}}{r^{2}}\,,\label{eq:VVP_ChPT}
\end{align}
i.e.~it is determined by two low-energy constants, $C_{7}^{W}$ and $C_{22}^{W}$.

\paragraph{Contribution of RChT.}
On the other hand, in RChT, the contribution of the LECs is accounted for in terms of the resonances. Therefore, only the anomaly $\mathcal{O}(p^{4})$ term, i.e.~the first term at the right-hand side of \eqref{eq:VVP_ChPT}, needs to be added into the resonance contribution as an intrinsic property of the $\langle VVP\rangle$ Green function.

\begin{figure}[t!]
\centering
    \begin{subfigure}[t]{0.23\textwidth}
        \hspace{-1.85pt}\includegraphics[width=1\textwidth]{figures/figure_vvp_graph_type_2-eps-converted-to.pdf}
        \caption{}
        \label{fig:VVP_RChT_tensor_graph_1}
    \end{subfigure}
    \begin{subfigure}[t]{0.23\textwidth}
        \hspace{-2pt}\includegraphics[width=1\textwidth]{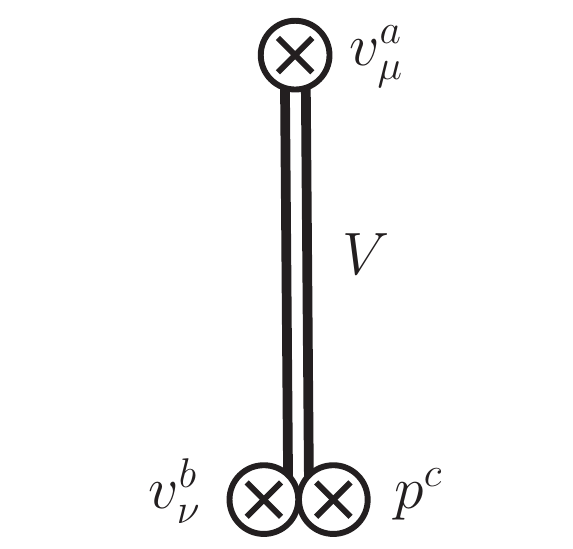}
        \caption{}
        \label{fig:VVP_RChT_tensor_graph_2}
    \end{subfigure}
    \begin{subfigure}[t]{0.23\textwidth}
        \hspace{-2pt}\includegraphics[width=1\textwidth]{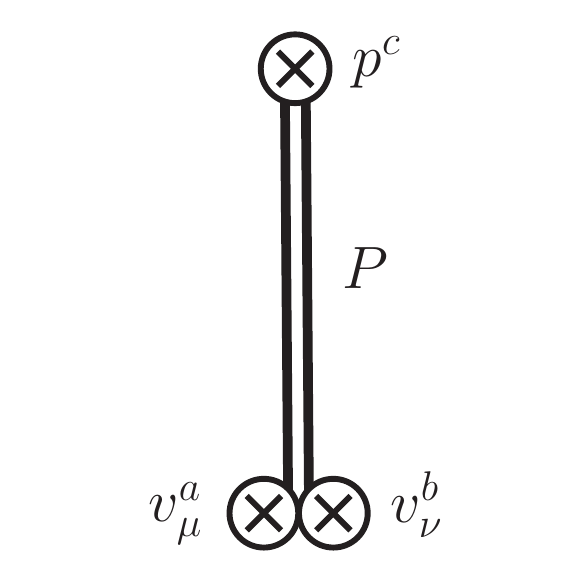}
        \caption{}
        \label{fig:VVP_RChT_tensor_graph_3}
    \end{subfigure}
    \begin{subfigure}[t]{0.23\textwidth}
        \hspace{-1pt}\includegraphics[width=1\textwidth]{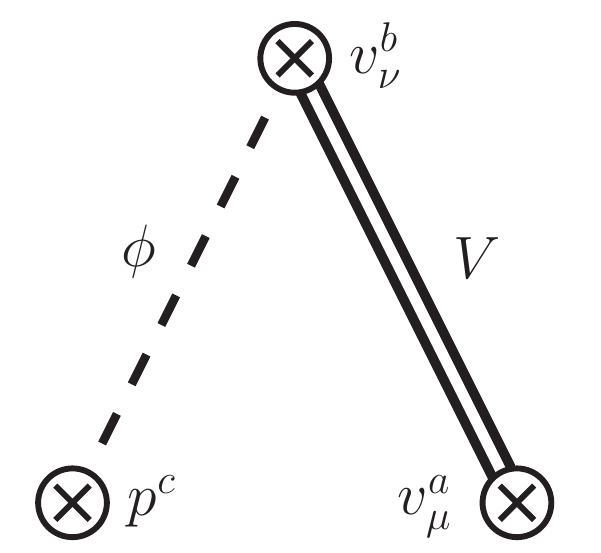}
        \caption{}
        \label{fig:VVP_RChT_tensor_graph_4}
    \end{subfigure}\vspace{20pt}
    
    \begin{subfigure}[t]{0.23\textwidth}
        \includegraphics[width=1\textwidth]{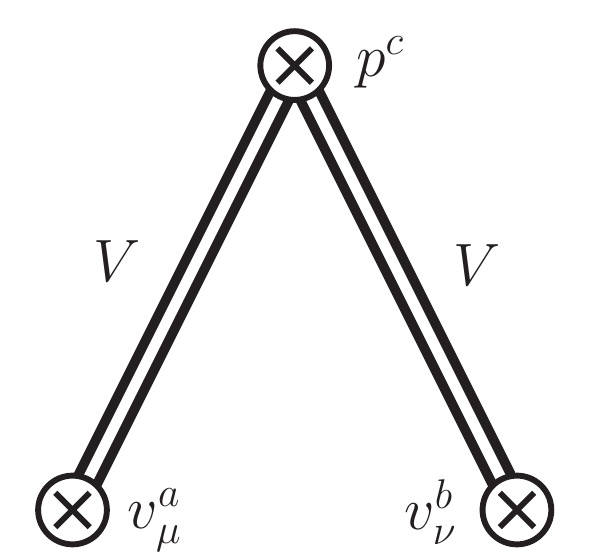}
        \caption{}
        \label{fig:VVP_RChT_tensor_graph_5}
    \end{subfigure}
    \begin{subfigure}[t]{0.23\textwidth}
        \includegraphics[width=1\textwidth]{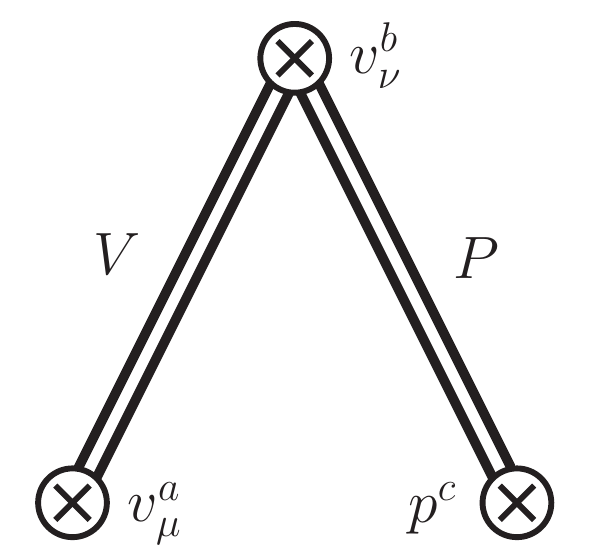}
        \caption{}
        \label{fig:VVP_RChT_tensor_graph_6}
    \end{subfigure}
    \begin{subfigure}[t]{0.23\textwidth}
        \hspace{-1.6pt}\includegraphics[width=1\textwidth]{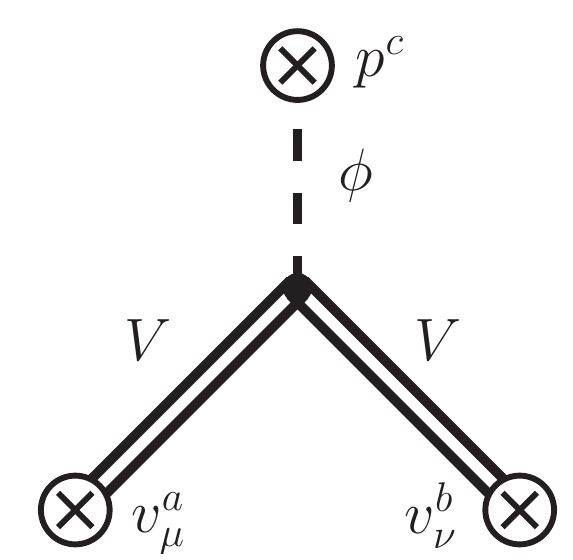}
        \caption{}
        \label{fig:VVP_RChT_tensor_graph_7}
    \end{subfigure}
    \begin{subfigure}[t]{0.23\textwidth}
        \hspace{-2.1pt}\includegraphics[width=1\textwidth]{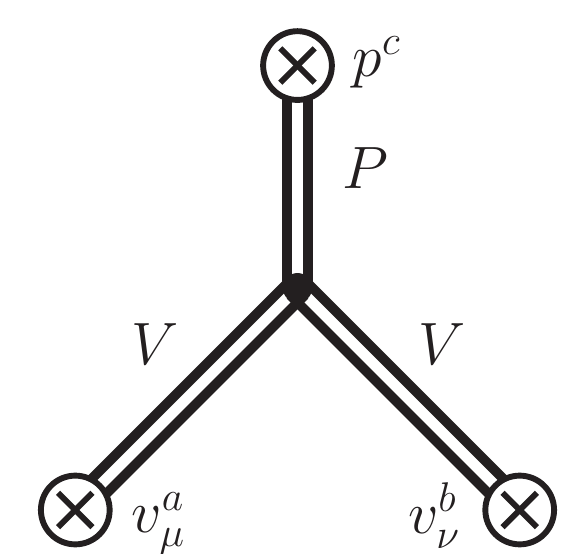}
        \caption{}
        \label{fig:VVP_RChT_tensor_graph_8}
    \end{subfigure}
    \caption{Tree-level Feynman diagrams of the nontrivial RChT contribution to the $\langle VVP\rangle$ Green function up to $\mathcal{O}(p^{6})$ in the tensor formalism.}
    \label{fig:VVP_RChT_tensor_graphs}
\end{figure}

The contribution of the vector and pseudoscalar resonances to the $\langle VVP\rangle$ correlator, within RChT up to $\mathcal{O}(p^{6})$, is given by the operators shown in the tables \ref{tab:VVP_monomials_1}-\ref{tab:VVP_monomials_3}. Let us firstly restrict ourselves only to the contributions of the operators on the left-hand sides of such tables, i.e.~those contributing to the Feynman graphs with just the first multiplets $V_{1}$ and $P_{1}$. Evaluating the contribution of these operators, that generate the Feynman diagrams shown in fig.~\ref{fig:VVP_RChT_tensor_graphs}, gives us (cf.~ref.~\cite{Kampf:2011ty})
\begin{align}
\left[\mathcal{F}_{VVP}^{\mathrm{RChT}}(p^{2},q^{2},r^{2})\right]_{(1)}&=\frac{B_{0}N_{c}}{8\pi^{2}r^{2}}\nonumber\\
&\hspace{-80pt}+\frac{2\sqrt{2}B_{0}F_{V_{1}}}{(p^{2}-M_{V_{1}}^{2})r^{2}}\big[(2\kappa_{12}^{V}+\kappa_{16}^{V})(p^{2}-q^{2}-r^{2})+2\kappa_{17}^{V}q^{2}-8\kappa_{14}^{V}r^{2}\big]\nonumber\\
&\hspace{-80pt}+\frac{2\sqrt{2}B_{0}F_{V_{1}}}{(q^{2}-M_{V_{1}}^{2})r^{2}}\big[(2\kappa_{12}^{V}+\kappa_{16}^{V})(q^{2}-p^{2}-r^{2})+2\kappa_{17}^{V}p^{2}-8\kappa_{14}^{V}r^{2}\big]\nonumber\\
&\hspace{-80pt}+\frac{64B_{0}d_{m}^{(1)}\kappa_{5}^{P}}{r^{2}-M_{P_{1}}^{2}}-\frac{4B_{0}F_{V_{1}}^{2}(8\kappa_{2}^{VV}-\kappa_{3}^{VV})}{(p^{2}-M_{V_{1}}^{2})(q^{2}-M_{V_{1}}^{2})}-\frac{4B_{0}F_{V_{1}}^{2}\kappa_{3}^{VV}(p^{2}+q^{2})}{(p^{2}-M_{V_{1}}^{2})(q^{2}-M_{V_{1}}^{2})r^{2}}\nonumber\\
&\hspace{-80pt}+\frac{16\sqrt{2}B_{0}F_{V_{1}}d_{m}^{(1)}\kappa_{3}^{PV}}{(p^{2}-M_{V_{1}}^{2})(r^{2}-M_{P_{1}}^{2})}+\frac{16\sqrt{2}B_{0}F_{V_{1}}d_{m}^{(1)}\kappa_{3}^{PV}}{(q^{2}-M_{V_{1}}^{2})(r^{2}-M_{P_{1}}^{2})}\nonumber\\
&\hspace{-80pt}+\frac{16B_{0}F_{V_{1}}^{2}d_{m}^{(1)}\kappa^{VVP}}{(p^{2}-M_{V_{1}}^{2})(q^{2}-M_{V_{1}}^{2})(r^{2}-M_{P_{1}}^{2})}\,.\label{eq:VVP_RChT_tensor}
\end{align}
It is interesting to point out that although there are 10 nontrivial operators at this level, only 9 of them actually contribute to the result \eqref{eq:VVP_RChT_tensor} after all since the relevant contribution of the operator proportional to the coupling $\kappa_{4}^{VV}$ vanishes here.

As we have suggested in subsection \ref{ssec:general_structure} and in detail in  appendix \ref{sec:struktura}, the resonance contribution \eqref{eq:VVP_RChT_tensor} fails the matching onto the respective high-energy behaviour \eqref{eq:vvp_ope} due to the missing polynomial structures and one is then required to generate these structures by other means. Therein, we have, however, found out that in order to be able to satisfy the requirements \eqref{eq:vvp_ope}, \eqref{eq:vvp_ope_2-op}, \eqref{eq:form_factor_BL} and \eqref{eq:form_factor_anomaly}, the THS+P ansatz must be acquired, i.e.~another multiplet of vector and pseudoscalar resonances needs to be added into consideration.

Now, we can carry on in an evaluation of the contributions of the additional operators with heavier resonance multiplets. These, of course, generate the Feynman graphs topologically equivalent to those at fig.~\ref{fig:VVP_RChT_tensor_graphs}. A straightforward calculation thus gives us their contribution in the form
\begin{align}
\left[\mathcal{F}_{VVP}^{\mathrm{RChT}}(p^{2},q^{2},r^{2})\right]_{(2)}&=\frac{2\sqrt{2}B_{0}F_{V_{2}}}{(p^{2}-M_{V_{2}}^{2})r^{2}}\big[(2\lambda_{12}^{V}+\lambda_{16}^{V})(p^{2}-q^{2}-r^{2})+2\lambda_{17}^{V}q^{2}-8\lambda_{14}^{V}r^{2}\big]\nonumber\\
&\hspace{-80pt}+\frac{2\sqrt{2}B_{0}F_{V_{2}}}{(q^{2}-M_{V_{2}}^{2})r^{2}}\big[(2\lambda_{12}^{V}+\lambda_{16}^{V})(q^{2}-p^{2}-r^{2})+2\lambda_{17}^{V}p^{2}-8\lambda_{14}^{V}r^{2}\big]+\frac{64B_{0}d_{m}^{(2)}\lambda_{5}^{P}}{r^{2}-M_{P_{2}}^{2}}\nonumber\\
&\hspace{-80pt}-\frac{16B_{0}F_{V_{1}}F_{V_{2}}\lambda_{21}^{VV}}{(p^{2}-M_{V_{1}}^{2})(q^{2}-M_{V_{2}}^{2})}-\frac{16B_{0}F_{V_{1}}F_{V_{2}}\lambda_{21}^{VV}}{(p^{2}-M_{V_{2}}^{2})(q^{2}-M_{V_{1}}^{2})}\nonumber\\
&\hspace{-80pt}-\frac{4B_{0}F_{V_{2}}^{2}(8\lambda_{22}^{VV}-\lambda_{34}^{VV})}{(p^{2}-M_{V_{2}}^{2})(q^{2}-M_{V_{2}}^{2})}-\frac{4B_{0}F_{V_{2}}^{2}\lambda_{34}^{VV}(p^{2}+q^{2})}{(p^{2}-M_{V_{2}}^{2})(q^{2}-M_{V_{2}}^{2})r^{2}}\nonumber\\
&\hspace{-80pt}-\frac{2B_{0}F_{V_{1}}F_{V_{2}}}{(p^{2}-M_{V_{1}}^{2})(q^{2}-M_{V_{2}}^{2})r^{2}}\big[(p^{2}+q^{2}-r^{2})(\lambda_{31}^{VV}+2\lambda_{32}^{VV}+\lambda_{33}^{VV})-4\lambda_{32}^{VV}p^{2}\big]\nonumber\\
&\hspace{-80pt}-\frac{2B_{0}F_{V_{1}}F_{V_{2}}}{(p^{2}-M_{V_{2}}^{2})(q^{2}-M_{V_{1}}^{2})r^{2}}\big[(p^{2}+q^{2}-r^{2})(\lambda_{31}^{VV}+2\lambda_{32}^{VV}+\lambda_{33}^{VV})-4\lambda_{32}^{VV}q^{2}\big]\nonumber\\
&\hspace{-80pt}+\frac{16\sqrt{2}B_{0}F_{V_{2}}d_{m}^{(1)}\lambda_{31}^{PV}}{(q^{2}-M_{V_{2}}^{2})(r^{2}-M_{P_{1}}^{2})}+\frac{16\sqrt{2}B_{0}F_{V_{2}}d_{m}^{(1)}\lambda_{31}^{PV}}{(p^{2}-M_{V_{2}}^{2})(r^{2}-M_{P_{1}}^{2})}\nonumber\\
&\hspace{-80pt}+\frac{16\sqrt{2}B_{0}F_{V_{1}}d_{m}^{(2)}\lambda_{32}^{PV}}{(q^{2}-M_{V_{1}}^{2})(r^{2}-M_{P_{2}}^{2})}+\frac{16\sqrt{2}B_{0}F_{V_{1}}d_{m}^{(2)}\lambda_{32}^{PV}}{(p^{2}-M_{V_{1}}^{2})(r^{2}-M_{P_{2}}^{2})}\nonumber\\
&\hspace{-80pt}+\frac{16\sqrt{2}B_{0}F_{V_{2}}d_{m}^{(2)}\lambda_{33}^{PV}}{(q^{2}-M_{V_{2}}^{2})(r^{2}-M_{P_{2}}^{2})}+\frac{16\sqrt{2}B_{0}F_{V_{2}}d_{m}^{(2)}\lambda_{33}^{PV}}{(p^{2}-M_{V_{2}}^{2})(r^{2}-M_{P_{2}}^{2})}\nonumber\\
&\hspace{-80pt}+\frac{16B_{0}F_{V_{1}}F_{V_{2}}d_{m}^{(1)}\lambda_{1}^{VVP}}{(p^{2}-M_{V_{1}}^{2})(q^{2}-M_{V_{2}}^{2})(r^{2}-M_{P_{1}}^{2})}+\frac{16B_{0}F_{V_{1}}F_{V_{2}}d_{m}^{(1)}\lambda_{1}^{VVP}}{(p^{2}-M_{V_{2}}^{2})(q^{2}-M_{V_{1}}^{2})(r^{2}-M_{P_{1}}^{2})}\nonumber\\
&\hspace{-80pt}+\frac{16B_{0}F_{V_{1}}^{2}d_{m}^{(2)}\lambda_{2}^{VVP}}{(p^{2}-M_{V_{1}}^{2})(q^{2}-M_{V_{1}}^{2})(r^{2}-M_{P_{2}}^{2})}+\frac{16B_{0}F_{V_{2}}^{2}d_{m}^{(1)}\lambda_{3}^{VVP}}{(p^{2}-M_{V_{2}}^{2})(q^{2}-M_{V_{2}}^{2})(r^{2}-M_{P_{1}}^{2})}\nonumber\\
&\hspace{-80pt}+\frac{16B_{0}F_{V_{1}}F_{V_{2}}d_{m}^{(2)}\lambda_{4}^{VVP}}{(p^{2}-M_{V_{1}}^{2})(q^{2}-M_{V_{2}}^{2})(r^{2}-M_{P_{2}}^{2})}+\frac{16B_{0}F_{V_{1}}F_{V_{2}}d_{m}^{(2)}\lambda_{4}^{VVP}}{(p^{2}-M_{V_{2}}^{2})(q^{2}-M_{V_{1}}^{2})(r^{2}-M_{P_{2}}^{2})}\nonumber\\
&\hspace{-80pt}+\frac{16B_{0}F_{V_{2}}^{2}d_{m}^{(2)}\lambda_{5}^{VVP}}{(p^{2}-M_{V_{2}}^{2})(q^{2}-M_{V_{2}}^{2})(r^{2}-M_{P_{2}}^{2})}\,,\label{eq:VVP_RChT_tensor_v2}
\end{align}
from which we see that out of 22 relevant operators, only 19 actually contribute here.

As it turns out, the summed-up contribution of \eqref{eq:VVP_RChT_tensor} and \eqref{eq:VVP_RChT_tensor_v2} does not represent the final resonance contribution that could be matched onto \eqref{eq:vvp_ope} successfully. The reason is purely algebraic --- the present result is missing suitable rational terms that would improve its UV behaviour. In order to compensate for such terms, an additional set of operators needs to be added, and these operators must be of a higher dimension than those already considered above.\footnote{Obviously, taking into account only the first two resonance multiplets of each kind, one cannot introduce any other $\mathcal{O}(p^{6})$ operators that would not be linear combinations of those already considered.} Nevertheless, having to extend the already duplicated $\mathcal{O}(p^{6})$ Lagrangian \eqref{eq:lagrangianr4_v2} for the complete set of $\mathcal{O}(p^{8})$ or even $\mathcal{O}(p^{10})$ operators would undoubtedly lead to a significant increase of the number of unknown couplings of such Lagrangian and predictability of this model would be extremely limited if any at all. For this reason, we refrain from trying to completely classify the $\mathcal{O}(p^{8})$ and $\mathcal{O}(p^{10})$ operators and, instead, rather give an example of a minimal set which enables us to satisfy the matching.

It appears that only a small set of those additional operators, shown in table~\ref{tab:VVP_monomials_4}, is sufficient to fulfil the requirements \eqref{eq:vvp_ope}, \eqref{eq:vvp_ope_2-op}, \eqref{eq:form_factor_BL} and \eqref{eq:form_factor_anomaly}. The choice of these particular operators is of course ad hoc --- we have introduced explicitly only those that contribute in such a way that their behaviour within the matching procedure complements the missing polynomial terms. As one can see, they are given by only merely modifying some of the operators presented in tables \ref{tab:VVP_monomials_2} and \ref{tab:VVP_monomials_3}, and omitting any of them leads to the violation of the fulfillment of the above-mentioned conditions.\footnote{The similar argumentation is valid, of course, also for the tables \ref{tab:VAS_monomials_4} and \ref{tab:AAP_monomials_4} for the $\langle VAS\rangle$ and $\langle AAP\rangle$ correlators, respectively. For clarity, we will thus refrain from mentioning such a strategy in these cases once again and show the tables straight away.} In this sense, these operators indeed form a minimal set, as we have claimed.

\begin{table}[t!]
\centering
\begin{tabular}{|C{4cm}|C{2cm}||C{4cm}|C{2cm}|}
\hline
Operator $\big(\widehat{\mathcal{O}}_{i}^{RR}\big)_{\mu\nu\alpha\beta}$           & Coupling constant & Operator $\big(\widehat{\mathcal{O}}_{i}^{RRR}\big)_{\mu\nu\alpha\beta}$ & Coupling constant   \\ \hline\hline
$i\langle\lbrace V_{1}^{\mu\nu},V_{1}^{\alpha\beta}\rbrace\Box\chi_{-}\rangle$ & $\mu_{2}^{VV}$    & $\langle\lbrace\Box^{2}V_{1}^{\mu\nu},V_{1}^{\alpha\beta}\rbrace P_{1}\rangle$ & $\mu_{1}^{VVP}$ \\
$i\langle\lbrace V_{1}^{\mu\nu},V_{2}^{\alpha\beta}\rbrace\Box\chi_{-}\rangle$ & $\mu_{21}^{VV}$   & $\langle\lbrace\Box^{2}V_{1}^{\mu\nu},V_{1}^{\alpha\beta}\rbrace P_{2}\rangle$ & $\mu_{2}^{VVP}$ \\
$i\langle\lbrace V_{2}^{\mu\nu},V_{2}^{\alpha\beta}\rbrace\Box\chi_{-}\rangle$ & $\mu_{22}^{VV}$   & $\langle\lbrace\Box^{2}V_{2}^{\mu\nu},V_{2}^{\alpha\beta}\rbrace P_{1}\rangle$ & $\mu_{3}^{VVP}$ \\
                                                                               &                   & $\langle\lbrace\Box^{2}V_{2}^{\mu\nu},V_{2}^{\alpha\beta}\rbrace P_{2}\rangle$ & $\mu_{4}^{VVP}$ \\ \hline
\end{tabular}
\caption{Additional higher-derivative monomials of the extended RChT Lagrangian at $\mathcal{O}(p^{6})$ with three resonance fields, relevant for the $\langle VVP\rangle$ Green function.}
\label{tab:VVP_monomials_4}
\end{table}

The contribution of the derivative terms in the table~\ref{tab:VVP_monomials_4} is easily obtained as
\begin{align}
\left[\mathcal{F}_{VVP}^{\mathrm{RChT}}(p^{2},q^{2},r^{2})\right]_{(3)}&=\frac{32B_{0}F_{V_{1}}^{2}\mu_{2}^{VV}r^{2}}{(p^{2}-M_{V_{1}}^{2})(q^{2}-M_{V_{1}}^{2})}+\frac{16B_{0}F_{V_{1}}F_{V_{2}}\mu_{21}^{VV}r^{2}}{(p^{2}-M_{V_{2}}^{2})(q^{2}-M_{V_{1}}^{2})}\nonumber\\
&\hspace{-80pt}+\frac{16B_{0}F_{V_{1}}F_{V_{2}}\mu_{21}^{VV}r^{2}}{(p^{2}-M_{V_{1}}^{2})(q^{2}-M_{V_{2}}^{2})}+\frac{32B_{0}F_{V_{2}}^{2}\mu_{22}^{VV}r^{2}}{(p^{2}-M_{V_{2}}^{2})(q^{2}-M_{V_{2}}^{2})}\nonumber\\
&\hspace{-80pt}+\frac{16B_{0}F_{V_{1}}^{2}d_{m}^{(1)}\mu_{1}^{VVP}(p^{4}+q^{4})}{(p^{2}-M_{V_{1}}^{2})(q^{2}-M_{V_{1}}^{2})(r^{2}-M_{P_{1}}^{2})}+\frac{16B_{0}F_{V_{1}}^{2}d_{m}^{(2)}\mu_{2}^{VVP}(p^{4}+q^{4})}{(p^{2}-M_{V_{1}}^{2})(q^{2}-M_{V_{1}}^{2})(r^{2}-M_{P_{2}}^{2})}\nonumber\\
&\hspace{-80pt}+\frac{16B_{0}F_{V_{2}}^{2}d_{m}^{(1)}\mu_{3}^{VVP}(p^{4}+q^{4})}{(p^{2}-M_{V_{2}}^{2})(q^{2}-M_{V_{2}}^{2})(r^{2}-M_{P_{1}}^{2})}+\frac{16B_{0}F_{V_{2}}^{2}d_{m}^{(2)}\mu_{4}^{VVP}(p^{4}+q^{4})}{(p^{2}-M_{V_{2}}^{2})(q^{2}-M_{V_{2}}^{2})(r^{2}-M_{P_{2}}^{2})}\,,\label{eq:VVP_RChT_tensor_v3}
\end{align}
with all of them actually contributing.

Finally, we thus have the minimal RChT contribution to the $\langle VVP\rangle$ Green function that can be indeed matched onto \eqref{eq:vvp_ope}. The desired result reads, schematically,
\begin{equation}
\mathcal{F}_{VVP}^{\mathrm{RChT}}(p^{2},q^{2},r^{2})=\sum_{i=1}^{3}\left[\mathcal{F}_{VVP}^{\mathrm{RChT}}(p^{2},q^{2},r^{2})\right]_{(i)}\,,\label{eq:VVP_RChT_tensor_final}
\end{equation}
and depends on 35 coupling constants in total.

\paragraph{Matching RChT with OPE.}
Naturally, the matching between the total resonance contribution \eqref{eq:VVP_RChT_tensor_final} and the OPE \eqref{eq:vvp_ope} can now be performed. The result can be written down in form of ``sum rules'' --- the coefficients $c_{i}$'s, as introduced in appendix \ref{sec:struktura}, on the left-hand side of the equation and the respective coefficients in terms of the parameters of the resonance Lagrangian on the right-hand side. Needless to say, the result of such a matching is somewhat lengthy, as can be expected from the length of the individual resonance contributions. In order to present it as economically as possible, we refer the reader to ref.~\cite{KadavyThesis:2022}, where we have provided an auxiliary file in the form of the \mathematica notebook, in which the results can be found.

\paragraph{Resonance saturation.}
Similarly, as above, the next step is to perform the matching between the ChPT contribution \eqref{eq:VVP_ChPT} and the total resonance contribution \eqref{eq:VVP_RChT_tensor_final}. We then obtain the expressions for the two low-energy constants $C_{7}^{W}$ and $C_{22}^{W}$ in terms of the couplings of the resonance Lagrangian. The results are significantly shorter than the ones in the case above and, for such a reason, we present them explicitly in appendix \ref{sec:resonance_saturation}.


\subsection{\texorpdfstring{$\langle VAS\rangle$}{} Green function}\label{sec:VAS}
Similarly as in the previous section, both the ChPT and the RChT contributions to the $\langle VAS\rangle$ Green functions will be reintroduced and the respective matchings will be performed.

\begin{figure}[t!]
\centering
\includegraphics[scale=0.6]{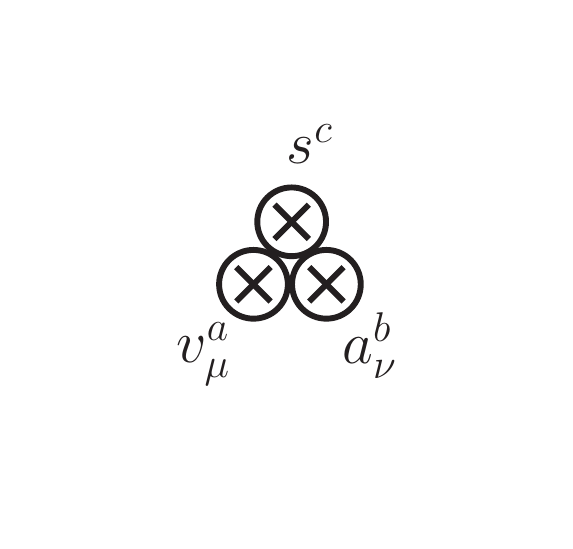}
\vspace{-30pt}\caption{Tree-level Feynman diagram of the ChPT contribution to the $\langle VAS\rangle$ Green function up to $\mathcal{O}(p^{6})$.}
\label{fig:VAS_ChPT_graph}
\end{figure}

\paragraph{Contribution of ChPT.}
The ChPT contribution to the $\langle VAS\rangle$ Green function up to $\mathcal{O}(p^{6})$ consists of one diagram topology, see fig.~\ref{fig:VAS_ChPT_graph}. The contribution of such a contact term reads simply
\begin{align}
\mathcal{F}_{VAS}^{\mathrm{ChPT}}(p^{2},q^{2},r^{2})=32B_{0}C_{11}^{W}\,.\label{eq:VAS_ChPT}
\end{align}

\begin{figure}[t!]
\centering
    \begin{subfigure}[t]{0.23\textwidth}
        \hspace{-2pt}\includegraphics[width=1\textwidth]{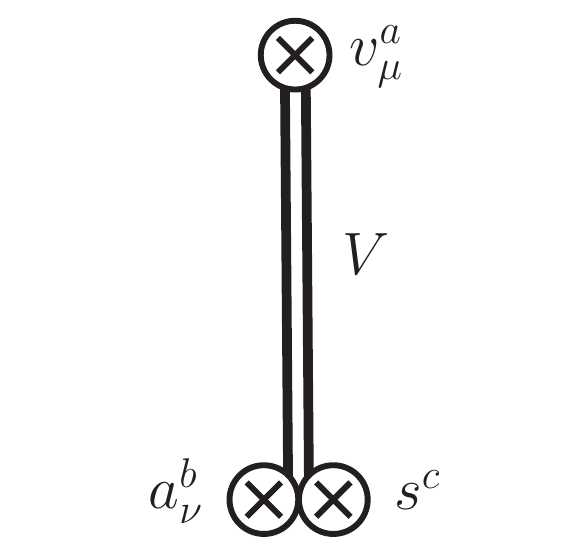}
        \caption{}
        \label{fig:VAS_RChT_tensor_graph_1}
    \end{subfigure}
    \begin{subfigure}[t]{0.23\textwidth}
        \hspace{-2pt}\includegraphics[width=1\textwidth]{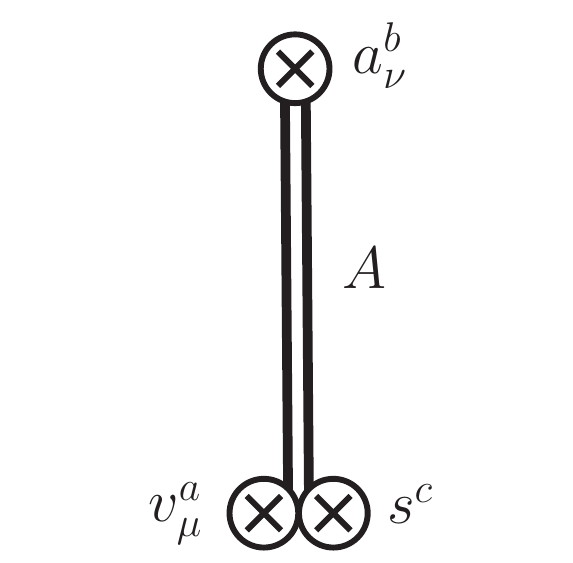}
        \caption{}
        \label{fig:VAS_RChT_tensor_graph_2}
    \end{subfigure}
    \begin{subfigure}[t]{0.23\textwidth}
        \hspace{-2pt}\includegraphics[width=1\textwidth]{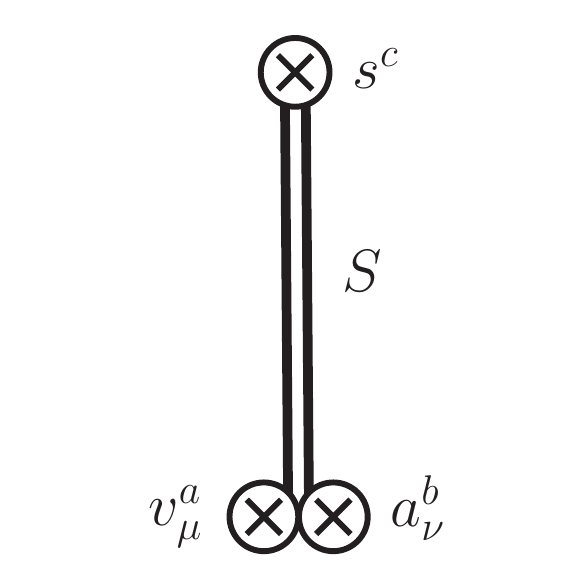}
        \caption{}
        \label{fig:VAS_RChT_tensor_graph_3}
    \end{subfigure}
    \begin{subfigure}[t]{0.23\textwidth}
        \includegraphics[width=1\textwidth]{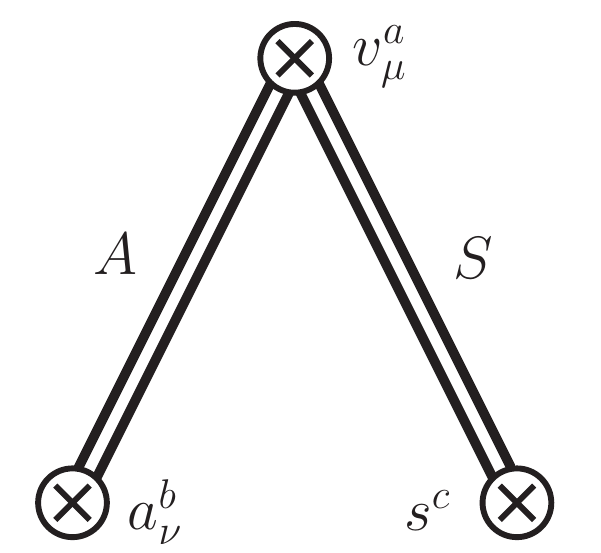}
        \caption{}
        \label{fig:VAS_RChT_tensor_graph_4}
    \end{subfigure}\vspace{20pt}
    
    \begin{subfigure}[t]{0.23\textwidth}
        \includegraphics[width=1\textwidth]{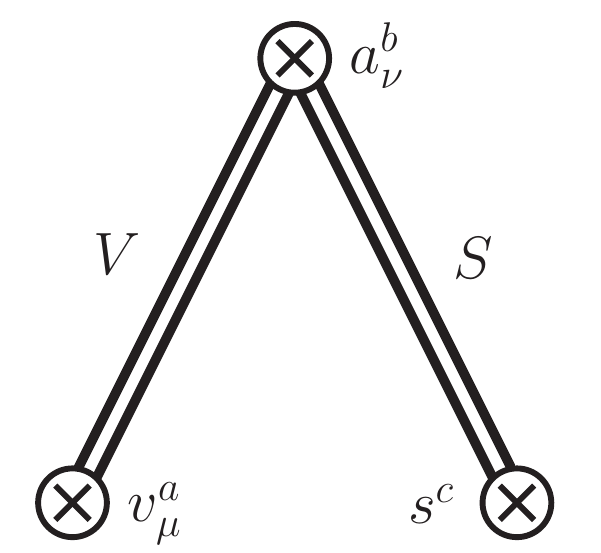}
        \caption{}
        \label{fig:VAS_RChT_tensor_graph_5}
    \end{subfigure}
    \begin{subfigure}[t]{0.23\textwidth}
        \includegraphics[width=1\textwidth]{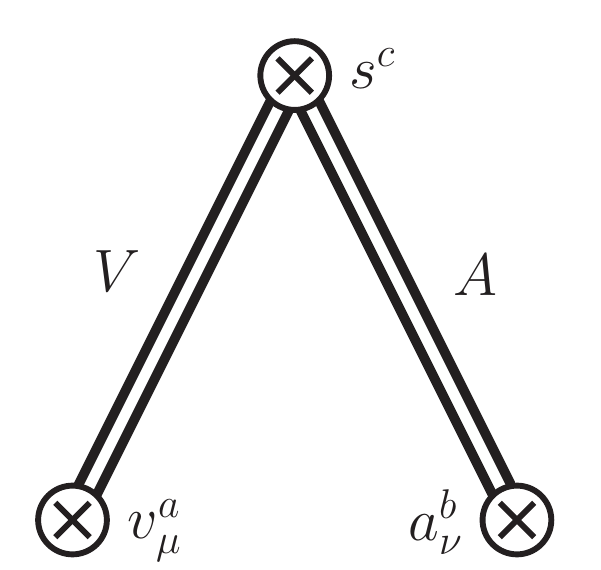}
        \caption{}
        \label{fig:VAS_RChT_tensor_graph_6}
    \end{subfigure}
    \begin{subfigure}[t]{0.23\textwidth}
        \hspace{-2pt}\includegraphics[width=1\textwidth]{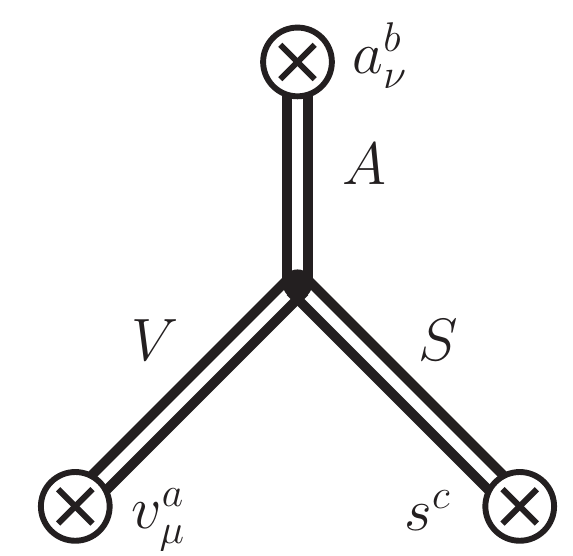}
        \caption{}
        \label{fig:VAS_RChT_tensor_graph_7}
    \end{subfigure}
    \caption{Tree-level Feynman diagrams of the nontrivial RChT contribution to the $\langle VAS\rangle$ Green function up to $\mathcal{O}(p^{6})$ in the tensor formalism.}
    \label{fig:VAS_RChT_tensor_graphs}
\end{figure}

\paragraph{Contribution of RChT.}
The resonance contribution to the $\langle VAS\rangle$ Green function is a bit easier than in the previous case. Let us start with the contribution of the lowest resonance multiplets, i.e.~with reproducing the result that has been already presented in ref.~\cite{Kampf:2011ty}. The corresponding operators are listed on the left-hand side of tables~\ref{tab:VAS_monomials_1}-\ref{tab:VAS_monomials_3} and generate the Feynman diagrams shown in fig.~\ref{fig:VAS_RChT_tensor_graphs}. Eventually, one easily arrives at the contribution of the lowest resonances in the form
\begin{align}
\left[\mathcal{F}_{VAS}^{\mathrm{RChT}}(p^{2},q^{2},r^{2})\right]_{(1)}&=-\frac{8\sqrt{2}B_{0}F_{V_{1}}(\kappa_{4}^{V}-2\kappa_{15}^{V})}{p^{2}-M_{V_{1}}^{2}}-\frac{16\sqrt{2}B_{0}F_{A_{1}}\kappa_{14}^{A}}{q^{2}-M_{A_{1}}^{2}}-\frac{32B_{0}c_{m}^{(1)}\kappa_{2}^{S}}{r^{2}-M_{S_{1}}^{2}}\nonumber\\
&\hspace{-80pt}-\frac{16\sqrt{2}B_{0}F_{A_{1}}c_{m}^{(1)}\kappa_{1}^{SA}}{(q^{2}-M_{A_{1}}^{2})(r^{2}-M_{S_{1}}^{2})}+\frac{8\sqrt{2}B_{0}F_{V_{1}}c_{m}^{(1)}(2\kappa_{1}^{SV}+\kappa_{2}^{SV})}{(p^{2}-M_{V_{1}}^{2})(r^{2}-M_{S_{1}}^{2})}+\frac{16B_{0}F_{V_{1}}F_{A_{1}}\kappa_{6}^{VA}}{(p^{2}-M_{V_{1}}^{2})(q^{2}-M_{A_{1}}^{2})}\nonumber\\
&\hspace{-80pt}-\frac{16B_{0}F_{V_{1}}F_{A_{1}}c_{m}^{(1)}\kappa^{VAS}}{(p^{2}-M_{V_{1}}^{2})(q^{2}-M_{A_{1}}^{2})(r^{2}-M_{S_{1}}^{2})}\,,\label{eq:VAS_RChT_tensor}
\end{align}
which indeed coincides with ref.~\cite{Kampf:2011ty} (see eq.~(81) at page no.~21 therein). We point out that from nine relevant operators, all of them actually contribute to \eqref{eq:VAS_RChT_tensor}.

\begin{figure}[t!]
\centering
    \begin{subfigure}[t]{0.27\textwidth}
        \hspace{10pt}\includegraphics[width=0.85\textwidth]{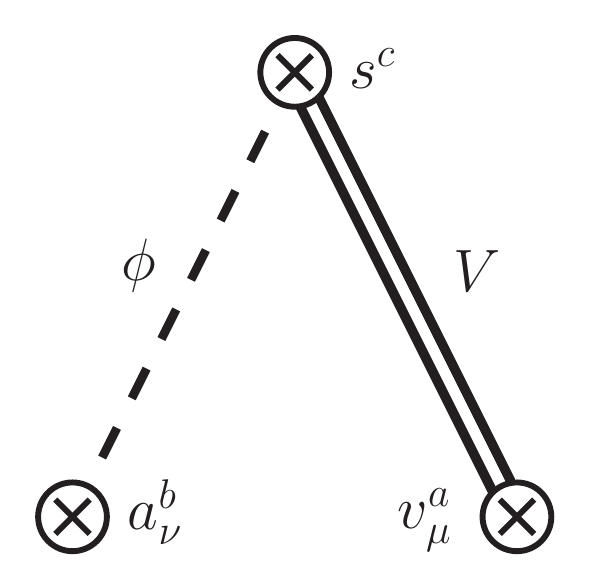}
        \caption{}
        \label{fig:VAS_RChT_tensor_graph_8}
    \end{subfigure}
    \begin{subfigure}[t]{0.27\textwidth}
        \hspace{6pt}\includegraphics[width=0.85\textwidth]{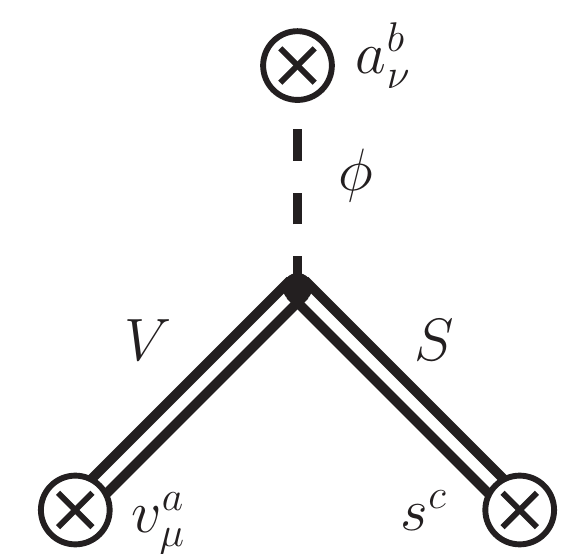}
        \caption{}
        \label{fig:VAS_RChT_tensor_graph_9}
    \end{subfigure}
    \caption{Tree-level Feynman diagrams of the vanishing RChT contribution to the $\langle VAS\rangle$ Green function up to $\mathcal{O}(p^{6})$ in the tensor formalism.}
    \label{fig:VAS_RChT_tensor_graphs_vanishing}
\end{figure}

We note that the $\langle VAS\rangle$ Green function is the first example where the resonance Lagrangian \eqref{eq:odd6-lagrangian} suggests a presence of the Feynman diagrams with the axial-vector source coupled to the pseudoscalar field, see fig.~\ref{fig:VAS_RChT_tensor_graphs_vanishing}. Although the Feynman rules are nontrivial, the contributions of these diagrams vanish. We emphasize that the same situation occurs also in the case of the $\langle AAP\rangle$ Green function, see fig.~\ref{fig:AAP_RChT_tensor_graphs_vanishing}.\footnote{Nevertheless, it is important to point out that although such types of diagrams vanish in the above-mentioned cases, the similar graphs represent an important and nonvanishing contribution to the anomalous $\langle VVA\rangle$ and $\langle AAA\rangle$ Green functions, without which the result would be incomplete and the respective Ward identities would be violated.}

Let us now advance to an evaluation of the contribution of the operators on the right-hand side of the tables~\ref{tab:VAS_monomials_1}-\ref{tab:VAS_monomials_3} that gives us the Feynman diagrams equivalent to those at fig.~\ref{fig:VAS_RChT_tensor_graphs}. Their contribution reads
\begin{align}
\left[\mathcal{F}_{VAS}^{\mathrm{RChT}}(p^{2},q^{2},r^{2})\right]_{(2)}&=-\frac{8\sqrt{2}B_{0}F_{V_{2}}(\lambda_{4}^{V}-2\lambda_{15}^{V})}{p^{2}-M_{V_{2}}^{2}}-\frac{16\sqrt{2}B_{0}F_{A_{2}}\lambda_{14}^{A}}{q^{2}-M_{A_{2}}^{2}}-\frac{32B_{0}c_{m}^{(2)}\lambda_{2}^{S}}{r^{2}-M_{S_{2}}^{2}}\nonumber\\
&\hspace{-80pt}-\frac{16\sqrt{2}B_{0}F_{A_{2}}c_{m}^{(1)}\lambda_{11}^{SA}}{(q^{2}-M_{A_{2}}^{2})(r^{2}-M_{S_{1}}^{2})}-\frac{16\sqrt{2}B_{0}F_{A_{1}}c_{m}^{(2)}\lambda_{12}^{SA}}{(q^{2}-M_{A_{1}}^{2})(r^{2}-M_{S_{2}}^{2})}-\frac{16\sqrt{2}B_{0}F_{A_{2}}c_{m}^{(2)}\lambda_{13}^{SA}}{(q^{2}-M_{A_{2}}^{2})(r^{2}-M_{S_{2}}^{2})}\nonumber\\
&\hspace{-80pt}+\frac{8\sqrt{2}B_{0}F_{V_{2}}c_{m}^{(1)}(2\lambda_{11}^{SV}+\lambda_{21}^{SV})}{(p^{2}-M_{V_{2}}^{2})(r^{2}-M_{S_{1}}^{2})}+\frac{8\sqrt{2}B_{0}F_{V_{1}}c_{m}^{(2)}(2\lambda_{12}^{SV}+\lambda_{22}^{SV})}{(p^{2}-M_{V_{1}}^{2})(r^{2}-M_{S_{2}}^{2})}\nonumber\\
&\hspace{-80pt}+\frac{8\sqrt{2}B_{0}F_{V_{2}}c_{m}^{(2)}(2\lambda_{13}^{SV}+\lambda_{23}^{SV})}{(p^{2}-M_{V_{2}}^{2})(r^{2}-M_{S_{2}}^{2})}+\frac{16B_{0}F_{V_{2}}F_{A_{1}}\lambda_{61}^{VA}}{(p^{2}-M_{V_{2}}^{2})(q^{2}-M_{A_{1}}^{2})}+\frac{16B_{0}F_{V_{1}}F_{A_{2}}\lambda_{62}^{VA}}{(p^{2}-M_{V_{1}}^{2})(q^{2}-M_{A_{2}}^{2})}\nonumber\\
&\hspace{-80pt}+\frac{16B_{0}F_{V_{2}}F_{A_{2}}\lambda_{63}^{VA}}{(p^{2}-M_{V_{2}}^{2})(q^{2}-M_{A_{2}}^{2})}-\frac{16B_{0}F_{V_{2}}F_{A_{1}}c_{m}^{(1)}\lambda_{1}^{VAS}}{(p^{2}-M_{V_{2}}^{2})(q^{2}-M_{A_{1}}^{2})(r^{2}-M_{S_{1}}^{2})}\nonumber\\
&\hspace{-80pt}-\frac{16B_{0}F_{V_{1}}F_{A_{2}}c_{m}^{(1)}\lambda_{2}^{VAS}}{(p^{2}-M_{V_{1}}^{2})(q^{2}-M_{A_{2}}^{2})(r^{2}-M_{S_{1}}^{2})}-\frac{16B_{0}F_{V_{1}}F_{A_{1}}c_{m}^{(2)}\lambda_{3}^{VAS}}{(p^{2}-M_{V_{1}}^{2})(q^{2}-M_{A_{1}}^{2})(r^{2}-M_{S_{2}}^{2})}\nonumber\\
&\hspace{-80pt}-\frac{16B_{0}F_{V_{2}}F_{A_{2}}c_{m}^{(1)}\lambda_{4}^{VAS}}{(p^{2}-M_{V_{2}}^{2})(q^{2}-M_{A_{2}}^{2})(r^{2}-M_{S_{1}}^{2})}-\frac{16B_{0}F_{V_{2}}F_{A_{1}}c_{m}^{(2)}\lambda_{5}^{VAS}}{(p^{2}-M_{V_{2}}^{2})(q^{2}-M_{A_{1}}^{2})(r^{2}-M_{S_{2}}^{2})}\nonumber\\
&\hspace{-80pt}-\frac{16B_{0}F_{V_{1}}F_{A_{2}}c_{m}^{(2)}\lambda_{6}^{VAS}}{(p^{2}-M_{V_{1}}^{2})(q^{2}-M_{A_{2}}^{2})(r^{2}-M_{S_{2}}^{2})}-\frac{16B_{0}F_{V_{2}}F_{A_{2}}c_{m}^{(2)}\lambda_{7}^{VAS}}{(p^{2}-M_{V_{2}}^{2})(q^{2}-M_{A_{2}}^{2})(r^{2}-M_{S_{2}}^{2})}\,,\label{eq:VAS_RChT_tensor_v2}
\end{align}
from which we see that from 23 relevant operators, every one of them actually contributes.

Finally, similarly as in the case of the $\langle VVP\rangle$ correlator, we are required to take into account the derivative operators, since the summed up contribution of \eqref{eq:VAS_RChT_tensor} and \eqref{eq:VAS_RChT_tensor_v2} can not be matched onto \eqref{eq:vas_ope} successfully. Once again, as it turns out, the operators shown in the table~\ref{tab:VAS_monomials_4} are needed. Their contribution reads
\begin{align}
\left[\mathcal{F}_{VAS}^{\mathrm{RChT}}(p^{2},q^{2},r^{2})\right]_{(3)}&=\nonumber\\
&\hspace{-80pt}-\frac{16B_{0}c_{m}^{(1)}F_{V_{2}}F_{A_{1}}\mu_{11}^{VAS}p^{4}}{(p^{2}-M_{V_{2}}^{2})(q^{2}-M_{A_{1}}^{2})(r^{2}-M_{S_{1}}^{2})}-\frac{16B_{0}c_{m}^{(1)}F_{V_{2}}F_{A_{2}}\mu_{12}^{VAS}p^{4}}{(p^{2}-M_{V_{2}}^{2})(q^{2}-M_{A_{2}}^{2})(r^{2}-M_{S_{1}}^{2})}\nonumber\\
&\hspace{-80pt}-\frac{16B_{0}c_{m}^{(2)}F_{V_{2}}F_{A_{1}}\mu_{13}^{VAS}p^{4}}{(p^{2}-M_{V_{2}}^{2})(q^{2}-M_{A_{1}}^{2})(r^{2}-M_{S_{2}}^{2})}-\frac{16B_{0}c_{m}^{(2)}F_{V_{2}}F_{A_{2}}\mu_{14}^{VAS}p^{4}}{(p^{2}-M_{V_{2}}^{2})(q^{2}-M_{A_{2}}^{2})(r^{2}-M_{S_{2}}^{2})}\nonumber\\
&\hspace{-80pt}-\frac{16B_{0}c_{m}^{(1)}F_{V_{1}}F_{A_{2}}\mu_{21}^{VAS}q^{4}}{(p^{2}-M_{V_{1}}^{2})(q^{2}-M_{A_{2}}^{2})(r^{2}-M_{S_{1}}^{2})}-\frac{16B_{0}c_{m}^{(1)}F_{V_{2}}F_{A_{2}}\mu_{22}^{VAS}q^{4}}{(p^{2}-M_{V_{2}}^{2})(q^{2}-M_{A_{2}}^{2})(r^{2}-M_{S_{1}}^{2})}\nonumber\\
&\hspace{-80pt}-\frac{16B_{0}c_{m}^{(2)}F_{V_{1}}F_{A_{2}}\mu_{23}^{VAS}q^{4}}{(p^{2}-M_{V_{1}}^{2})(q^{2}-M_{A_{2}}^{2})(r^{2}-M_{S_{2}}^{2})}-\frac{16B_{0}c_{m}^{(2)}F_{V_{2}}F_{A_{2}}\mu_{24}^{VAS}q^{4}}{(p^{2}-M_{V_{2}}^{2})(q^{2}-M_{A_{2}}^{2})(r^{2}-M_{S_{2}}^{2})}\nonumber\\
&\hspace{-80pt}-\frac{16B_{0}c_{m}^{(2)}F_{V_{1}}F_{A_{1}}\mu_{31}^{VAS}r^{4}}{(p^{2}-M_{V_{1}}^{2})(q^{2}-M_{A_{1}}^{2})(r^{2}-M_{S_{2}}^{2})}-\frac{16B_{0}c_{m}^{(2)}F_{V_{2}}F_{A_{1}}\mu_{32}^{VAS}r^{4}}{(p^{2}-M_{V_{2}}^{2})(q^{2}-M_{A_{1}}^{2})(r^{2}-M_{S_{2}}^{2})}\nonumber\\
&\hspace{-80pt}-\frac{16B_{0}c_{m}^{(2)}F_{V_{1}}F_{A_{2}}\mu_{33}^{VAS}r^{4}}{(p^{2}-M_{V_{1}}^{2})(q^{2}-M_{A_{2}}^{2})(r^{2}-M_{S_{2}}^{2})}-\frac{16B_{0}c_{m}^{(2)}F_{V_{2}}F_{A_{2}}\mu_{34}^{VAS}r^{4}}{(p^{2}-M_{V_{2}}^{2})(q^{2}-M_{A_{2}}^{2})(r^{2}-M_{S_{2}}^{2})}\,.\label{eq:VAS_RChT_tensor_v3}
\end{align}
\begin{table}[t!]
\centering
\begin{tabular}{|C{4cm}|C{2cm}||C{4cm}|C{2cm}|}
\hline
Operator $\big(\widehat{\mathcal{O}}_{i}^{RR}\big)_{\mu\nu\alpha\beta}$           & Coupling constant & Operator $\big(\widehat{\mathcal{O}}_{i}^{RRR}\big)_{\mu\nu\alpha\beta}$ & Coupling constant   \\ \hline\hline
$i\langle[\Box^{2}V_{2}^{\mu\nu},A_{1}^{\alpha\beta}]S_{1}\rangle$ & $\mu_{11}^{VAS}$ & $i\langle[V_{1}^{\mu\nu},\Box^{2}A_{2}^{\alpha\beta}]S_{2}\rangle$ & $\mu_{23}^{VAS}$ \\
$i\langle[\Box^{2}V_{2}^{\mu\nu},A_{2}^{\alpha\beta}]S_{1}\rangle$ & $\mu_{12}^{VAS}$ & $i\langle[V_{2}^{\mu\nu},\Box^{2}A_{2}^{\alpha\beta}]S_{2}\rangle$ & $\mu_{24}^{VAS}$ \\
$i\langle[\Box^{2}V_{2}^{\mu\nu},A_{1}^{\alpha\beta}]S_{2}\rangle$ & $\mu_{13}^{VAS}$ & $i\langle[V_{1}^{\mu\nu},A_{1}^{\alpha\beta}]\Box^{2}S_{2}\rangle$ & $\mu_{31}^{VAS}$ \\
$i\langle[\Box^{2}V_{2}^{\mu\nu},A_{2}^{\alpha\beta}]S_{2}\rangle$ & $\mu_{14}^{VAS}$ & $i\langle[V_{2}^{\mu\nu},A_{1}^{\alpha\beta}]\Box^{2}S_{2}\rangle$ & $\mu_{32}^{VAS}$ \\
$i\langle[V_{1}^{\mu\nu},\Box^{2}A_{2}^{\alpha\beta}]S_{1}\rangle$ & $\mu_{21}^{VAS}$ & $i\langle[V_{1}^{\mu\nu},A_{2}^{\alpha\beta}]\Box^{2}S_{2}\rangle$ & $\mu_{33}^{VAS}$ \\
$i\langle[V_{2}^{\mu\nu},\Box^{2}A_{2}^{\alpha\beta}]S_{1}\rangle$ & $\mu_{22}^{VAS}$ & $i\langle[V_{2}^{\mu\nu},A_{2}^{\alpha\beta}]\Box^{2}S_{2}\rangle$ & $\mu_{34}^{VAS}$ \\ \hline
\end{tabular}
\caption{Additional derivative monomials of the extended RChT Lagrangian at $\mathcal{O}(p^{6})$ with three resonance fields, relevant for the $\langle VAS\rangle$ Green function.}
\label{tab:VAS_monomials_4}
\end{table}

Taking all the individual resonance contributions \eqref{eq:VAS_RChT_tensor}-\eqref{eq:VAS_RChT_tensor_v3} together, we have the final result, written down in a compact form as
\begin{equation}
\mathcal{F}_{VAS}^{\mathrm{RChT}}(p^{2},q^{2},r^{2})=\sum_{i=1}^{3}\left[\mathcal{F}_{VAS}^{\mathrm{RChT}}(p^{2},q^{2},r^{2})\right]_{(i)}\,,\label{eq:VAS_RChT_tensor_final}
\end{equation}
that is capable of satisfying the required OPE behaviour \eqref{eq:vas_ope}.

\paragraph{Matching RChT with OPE.}
The matching between the total resonance contribution \eqref{eq:VAS_RChT_tensor_final} and the OPE \eqref{eq:vas_ope} can now be performed. Once again, we refer the reader to \cite{KadavyThesis:2022} for the explicit results.

\paragraph{Resonance saturation.}
The matching between the ChPT contribution \eqref{eq:VAS_ChPT} and the total resonance contribution \eqref{eq:VAS_RChT_tensor_final} gives us the expression for the low-energy constant $C_{11}^{W}$ in terms of the couplings of the resonance Lagrangian. Such a result can be found in appendix \ref{sec:resonance_saturation}.


\subsection{\texorpdfstring{$\langle AAP\rangle$}{} Green function}\label{sec:AAP}

\begin{figure}[t!]
\centering
    \begin{subfigure}[t]{0.27\textwidth}
        \hspace{6.2pt}\includegraphics[width=0.85\textwidth]{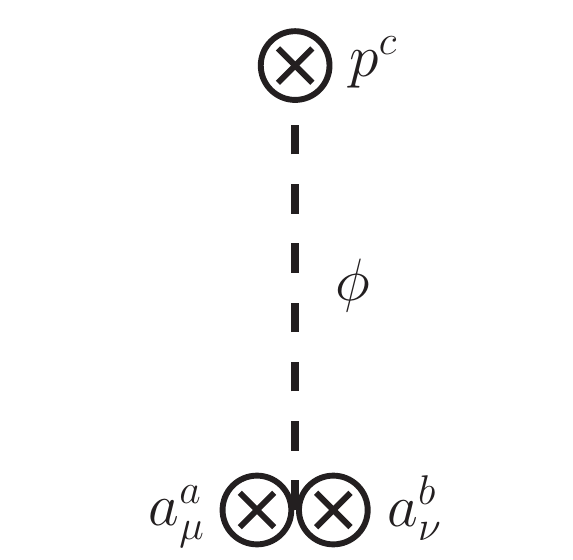}
        \caption{}
        \label{fig:AAP_ChPT_graph_1}
    \end{subfigure}
    \begin{subfigure}[t]{0.27\textwidth}
        \hspace{6.8pt}\includegraphics[width=0.85\textwidth]{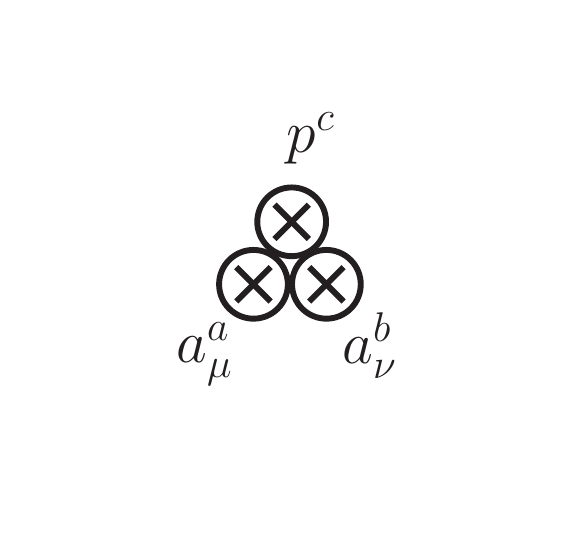}
        \caption{}
        \label{fig:AAP_ChPT_graph_2}
    \end{subfigure}
    \caption{Tree-level Feynman diagrams of the ChPT contribution to the $\langle AAP\rangle$ Green function up to $\mathcal{O}(p^{6})$.}
    \label{fig:AAP_ChPT_graphs}
\end{figure}

\paragraph{Contribution of ChPT.}
The ChPT contribution to the $\langle AAP\rangle$ correlator up to $\mathcal{O}(p^{6})$ is given by three individual contributions, represented by two diagram topologies, see fig.~\ref{fig:AAP_ChPT_graphs}. One thus finds the ChPT contribution in the form
\begin{equation}
\mathcal{F}_{AAP}^{\mathrm{ChPT}}(p^{2},q^{2},r^{2})=\frac{B_{0}N_{c}}{24\pi^{2}r^{2}}-32B_{0}C_{9}^{W}+8B_{0}C_{23}^{W}\frac{p^{2}+q^{2}}{r^{2}}\,,\label{eq:AAP_ChPT}
\end{equation}
which resembles the result \eqref{eq:VVP_ChPT} for the $\langle VVP\rangle$ correlator.

\begin{figure}[t!]
\centering
    \begin{subfigure}[t]{0.23\textwidth}
        \hspace{-1.5pt}\includegraphics[width=1\textwidth]{figures/figure_aap_graph_type_2-eps-converted-to.pdf}
        \caption{}
        \label{fig:AAP_RChT_tensor_graph_1}
    \end{subfigure}
    \begin{subfigure}[t]{0.23\textwidth}
        \hspace{-2pt}\includegraphics[width=1\textwidth]{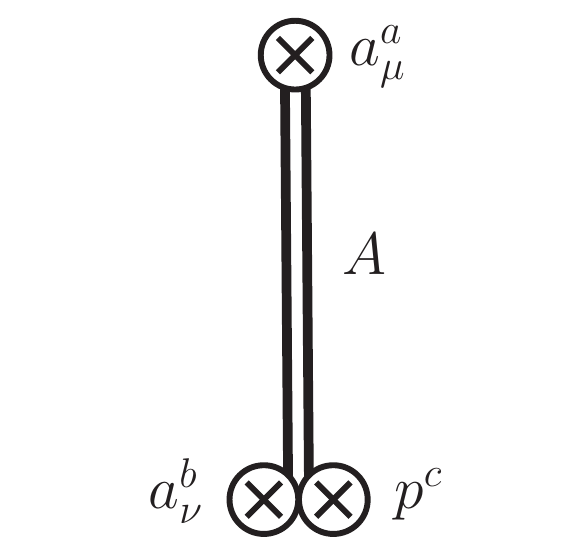}
        \caption{}
        \label{fig:AAP_RChT_tensor_graph_2}
    \end{subfigure}
    \begin{subfigure}[t]{0.23\textwidth}
        \hspace{-2pt}\includegraphics[width=1\textwidth]{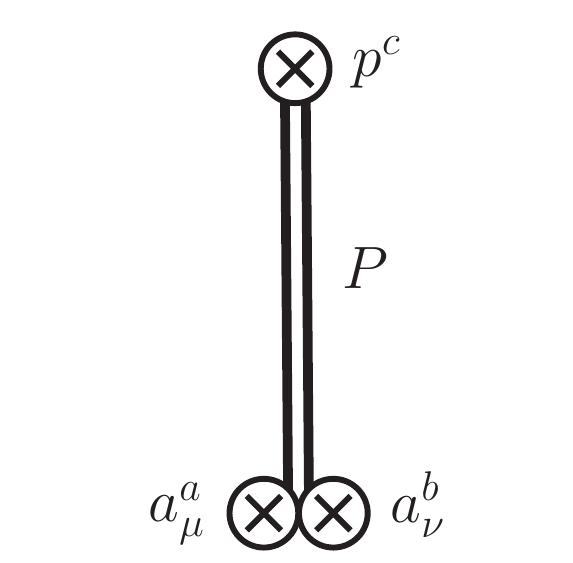}
        \caption{}
        \label{fig:AAP_RChT_tensor_graph_3}
    \end{subfigure}
    \begin{subfigure}[t]{0.23\textwidth}
        \includegraphics[width=1\textwidth]{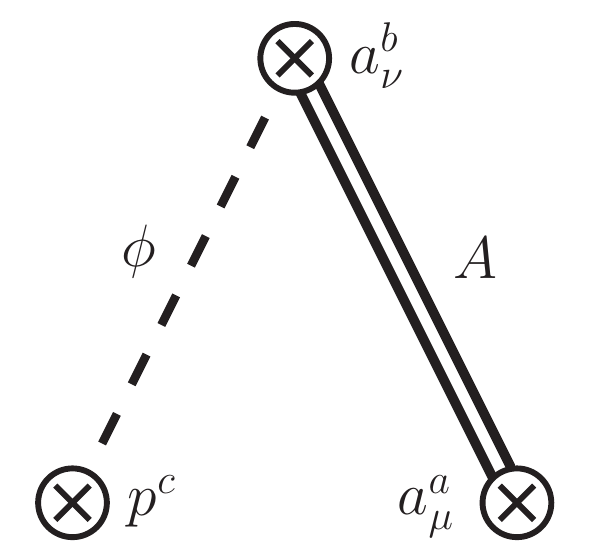}
        \caption{}
        \label{fig:AAP_RChT_tensor_graph_4}
    \end{subfigure}\vspace{20pt}
    
    \begin{subfigure}[t]{0.23\textwidth}
        \includegraphics[width=1\textwidth]{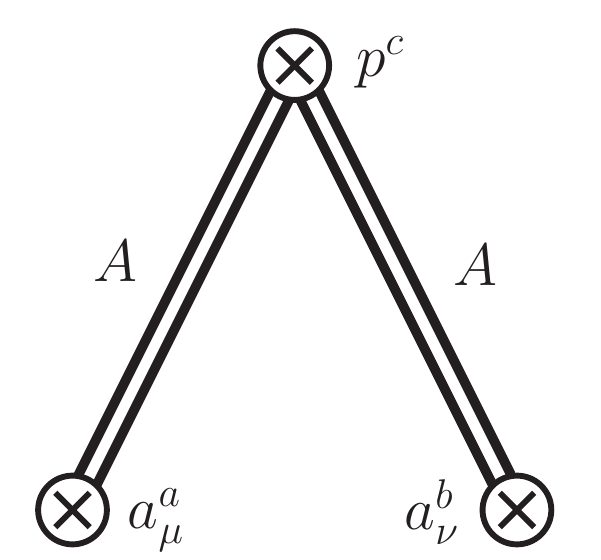}
        \caption{}
        \label{fig:AAP_RChT_tensor_graph_5}
    \end{subfigure}
    \begin{subfigure}[t]{0.23\textwidth}
        \includegraphics[width=1\textwidth]{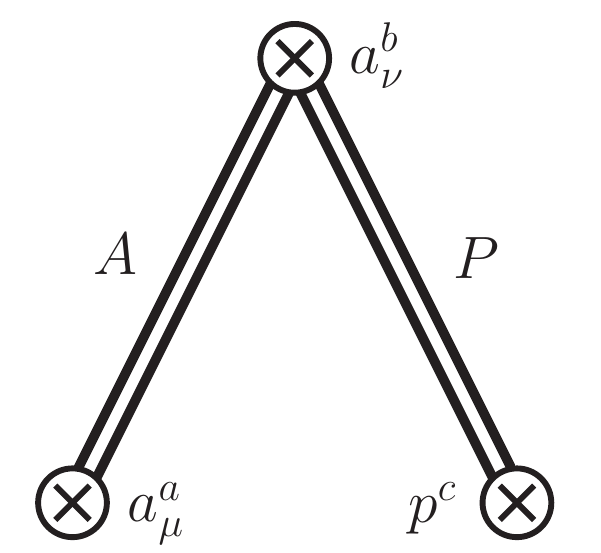}
        \caption{}
        \label{fig:AAP_RChT_tensor_graph_6}
    \end{subfigure}
    \begin{subfigure}[t]{0.23\textwidth}
        \hspace{-2pt}\includegraphics[width=1\textwidth]{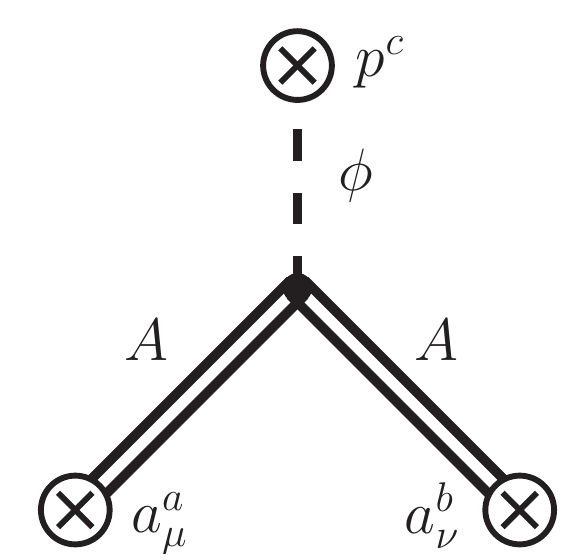}
        \caption{}
        \label{fig:AAP_RChT_tensor_graph_7}
    \end{subfigure}
    \begin{subfigure}[t]{0.23\textwidth}
        \hspace{-2pt}\includegraphics[width=1\textwidth]{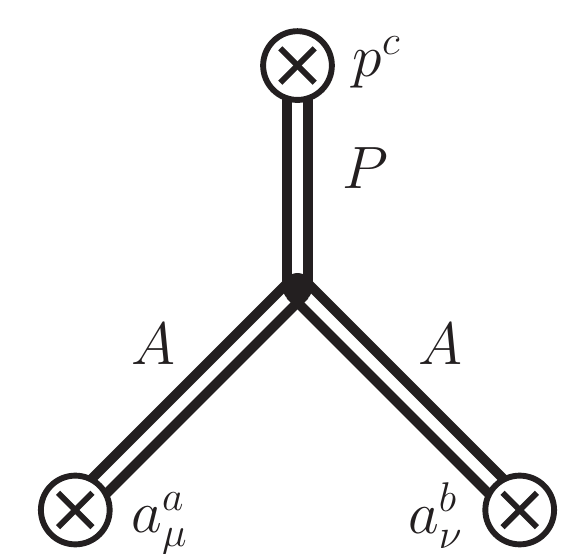}
        \caption{}
        \label{fig:AAP_RChT_tensor_graph_8}
    \end{subfigure}
    \caption{Tree-level Feynman diagrams of the nontrivial RChT contribution to the $\langle AAP\rangle$ Green function up to $\mathcal{O}(p^{6})$ in the tensor formalism.}
    \label{fig:AAP_RChT_tensor_graphs}
\end{figure}

\paragraph{Contribution of RChT.}
Let us now evaluate the resonance contribution of the lowest axial-vector and pseudoscalar resonances. The contributing operators are shown on the left-hand side of the tables~\ref{tab:AAP_monomials_1}-\ref{tab:AAP_monomials_3}. Except for the same anomalous ChPT vertex, these operators generate seven different contributing diagram topologies, see fig.~\ref{fig:AAP_RChT_tensor_graphs}. A straightforward evaluation of the contributions of these operators leads to the result
\begin{align}
\left[\mathcal{F}_{AAP}^{\mathrm{RChT}}(p^{2},q^{2},r^{2})\right]_{(1)}&=\frac{B_{0}N_{c}}{24\pi^{2}r^{2}}\nonumber\\
&\hspace{-50pt}+\frac{2\sqrt{2}B_{0}F_{A_{1}}}{(q^{2}-M_{A_{1}}^{2})r^{2}}\big[\big(q^{2}-p^{2}-r^{2}\big)\big(\kappa_{3}^{A}+2\kappa_{8}^{A}+\kappa_{15}^{A}\big)+2p^{2}\kappa_{16}^{A}\big]\nonumber\\
&\hspace{-50pt}+\frac{2\sqrt{2}B_{0}F_{A_{1}}}{(p^{2}-M_{A_{1}}^{2})r^{2}}\big[\big(p^{2}-q^{2}-r^{2}\big)\big(\kappa_{3}^{A}+2\kappa_{8}^{A}+\kappa_{15}^{A}\big)+2q^{2}\kappa_{16}^{A}\big]\nonumber\\
&\hspace{-50pt}-\frac{8\sqrt{2}B_{0}F_{A_{1}}(2\kappa_{11}^{A}+\kappa_{12}^{A})}{q^{2}-M_{A_{1}}^{2}}-\frac{8\sqrt{2}B_{0}F_{A_{1}}(2\kappa_{11}^{A}+\kappa_{12}^{A})}{p^{2}-M_{A_{1}}^{2}}+\frac{64B_{0}d_{m}^{(1)}\kappa_{1}^{P}}{r^{2}-M_{P_{1}}^{2}}\nonumber\\
&\hspace{-50pt}-\frac{32B_{0}F_{A_{1}}^{2}\kappa_{2}^{AA}}{(p^{2}-M_{A_{1}}^{2})(q^{2}-M_{A_{1}}^{2})}-\frac{4B_{0}F_{A_{1}}^{2}\kappa_{3}^{AA}(p^{2}+q^{2}-r^{2})}{(p^{2}-M_{A_{1}}^{2})(q^{2}-M_{A_{1}}^{2})r^{2}}\nonumber\\
&\hspace{-50pt}+\frac{8\sqrt{2}B_{0}F_{A_{1}}d_{m}^{(1)}(2\kappa_{1}^{PA}+\kappa_{2}^{PA})}{(p^{2}-M_{A_{1}}^{2})(r^{2}-M_{P_{1}}^{2})}+\frac{8\sqrt{2}B_{0}F_{A_{1}}d_{m}^{(1)}(2\kappa_{1}^{PA}+\kappa_{2}^{PA})}{(q^{2}-M_{A_{1}}^{2})(r^{2}-M_{P_{1}}^{2})}\nonumber\\
&\hspace{-50pt}+\frac{16B_{0}F_{A_{1}}^{2}d_{m}^{(1)}\kappa^{AAP}}{(p^{2}-M_{A_{1}}^{2})(q^{2}-M_{A_{1}}^{2})(r^{2}-M_{P_{1}}^{2})}\,.\label{eq:AAP_RChT_tensor}
\end{align}

We note that there are thirteen nontrivial operators at this stage, however, only twelve of them contribute to the result \eqref{eq:AAP_RChT_tensor} after all since the contribution of the operator proportional to the coupling constant $\kappa_{4}^{AA}$ vanishes. Once again, such a behaviour of the $\langle AAP\rangle$ correlator resembles the one of the $\langle VVP\rangle$.

\begin{figure}[t!]
\centering
    \begin{subfigure}[t]{0.27\textwidth}
        \hspace{12pt}\includegraphics[width=0.8\textwidth]{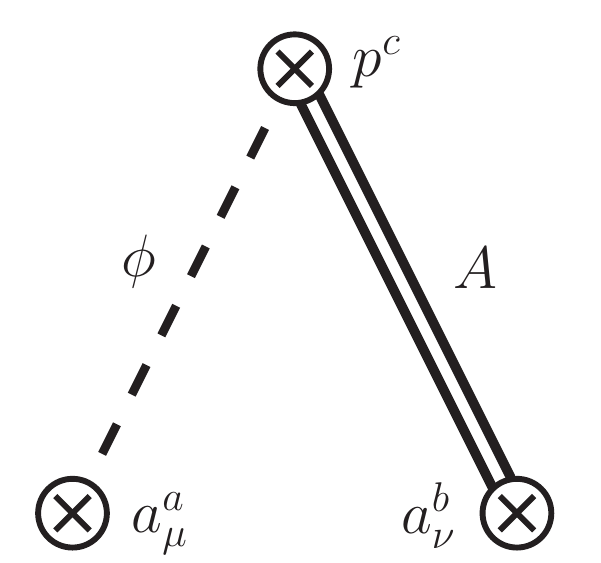}
        \caption{}
        \label{fig:AAP_RChT_tensor_graph_9}
    \end{subfigure}
    \begin{subfigure}[t]{0.27\textwidth}
        \hspace{11pt}\includegraphics[width=0.8\textwidth]{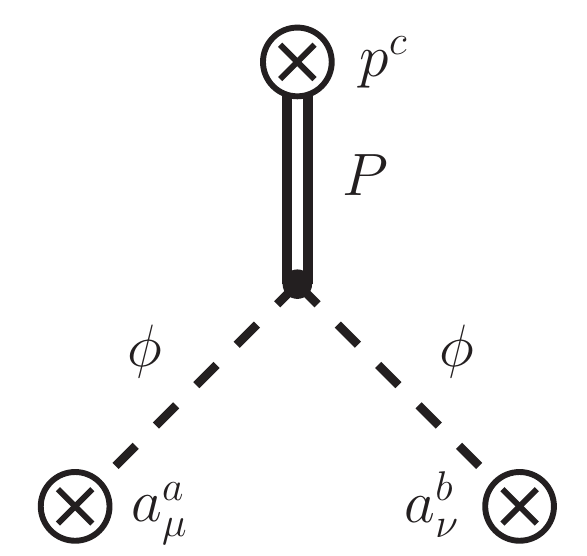}
        \caption{}
        \label{fig:AAP_RChT_tensor_graph_10}
    \end{subfigure}
    \begin{subfigure}[t]{0.27\textwidth}
        \hspace{5pt}\includegraphics[width=0.85\textwidth]{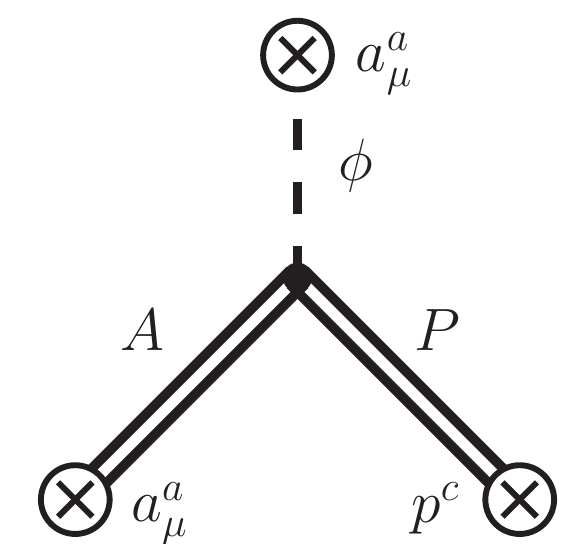}
        \caption{}
        \label{fig:AAP_RChT_tensor_graph_11}
    \end{subfigure}
    \caption{Tree-level Feynman diagrams of the vanishing RChT contribution to the $\langle AAP\rangle$ Green function up to $\mathcal{O}(p^{6})$ in the tensor formalism.}
    \label{fig:AAP_RChT_tensor_graphs_vanishing}
\end{figure}

Also, we point out that the relevant operators give arise to the Feynman diagrams at fig.~\ref{fig:AAP_RChT_tensor_graphs_vanishing} that actually vanish. In this regard, such a behaviour resembles the one of the $\langle VAS\rangle$ correlator.


An evaluation of the operators with higher-mass resonance multiplets, listed on the right-hand side of the tables~\ref{tab:AAP_monomials_1}-\ref{tab:AAP_monomials_3}, is also an easy task to perform. Similarly, as in the previous case, we obtain their contribution in the form
\begin{align}
\left[\mathcal{F}_{AAP}^{\mathrm{RChT}}(p^{2},q^{2},r^{2})\right]_{(2)}&=\frac{2\sqrt{2}B_{0}F_{A_{2}}}{(q^{2}-M_{A_{2}}^{2})r^{2}}\big[\big(q^{2}-p^{2}-r^{2}\big)\big(\lambda_{3}^{A}+2\lambda_{8}^{A}+\lambda_{15}^{A}\big)+2p^{2}\lambda_{16}^{A}\big]\nonumber\\
&\hspace{-80pt}+\frac{2\sqrt{2}B_{0}F_{A_{2}}}{(p^{2}-M_{A_{2}}^{2})r^{2}}\big[\big(p^{2}-q^{2}-r^{2}\big)\big(\lambda_{3}^{A}+2\lambda_{8}^{A}+\lambda_{15}^{A}\big)+2q^{2}\lambda_{16}^{A}\big]\nonumber\\
&\hspace{-80pt}-\frac{8\sqrt{2}B_{0}F_{A_{2}}(2\lambda_{11}^{A}+\lambda_{12}^{A})}{q^{2}-M_{A_{2}}^{2}}-\frac{8\sqrt{2}B_{0}F_{A_{2}}(2\lambda_{11}^{A}+\lambda_{12}^{A})}{p^{2}-M_{A_{2}}^{2}}+\frac{64B_{0}d_{m}^{(2)}\lambda_{1}^{P}}{r^{2}-M_{P_{2}}^{2}}\nonumber\\
&\hspace{-80pt}-\frac{16B_{0}F_{A_{1}}F_{A_{2}}\lambda_{21}^{AA}}{(p^{2}-M_{A_{1}}^{2})(q^{2}-M_{A_{2}}^{2})}-\frac{16B_{0}F_{A_{1}}F_{A_{2}}\lambda_{21}^{AA}}{(p^{2}-M_{A_{2}}^{2})(q^{2}-M_{A_{1}}^{2})}\nonumber\\
&\hspace{-80pt}-\frac{4B_{0}F_{A_{2}}^{2}(8\lambda_{22}^{AA}-\lambda_{34}^{AA})}{(p^{2}-M_{A_{2}}^{2})(q^{2}-M_{A_{2}}^{2})}-\frac{4B_{0}F_{A_{2}}^{2}\lambda_{34}^{AA}(p^{2}+q^{2})}{(p^{2}-M_{A_{2}}^{2})(q^{2}-M_{A_{2}}^{2})r^{2}}\nonumber\\
&\hspace{-80pt}-\frac{2B_{0}F_{A_{1}}F_{A_{2}}}{(p^{2}-M_{A_{1}}^{2})(q^{2}-M_{A_{2}}^{2})r^{2}}\big[(p^{2}+q^{2}-r^{2})(\lambda_{31}^{AA}+2\lambda_{32}^{AA}+\lambda_{33}^{AA})-4\lambda_{32}^{AA}p^{2}\big]\nonumber\\
&\hspace{-80pt}-\frac{2B_{0}F_{A_{1}}F_{A_{2}}}{(p^{2}-M_{A_{2}}^{2})(q^{2}-M_{A_{1}}^{2})r^{2}}\big[(p^{2}+q^{2}-r^{2})(\lambda_{31}^{AA}+2\lambda_{32}^{AA}+\lambda_{33}^{AA})-4\lambda_{32}^{AA}q^{2}\big]\nonumber\\
&\hspace{-80pt}+\frac{8\sqrt{2}B_{0}F_{A_{2}}d_{m}^{(1)}(2\lambda_{11}^{PA}+\lambda_{21}^{PA})}{(p^{2}-M_{A_{2}}^{2})(r^{2}-M_{P_{1}}^{2})}+\frac{8\sqrt{2}B_{0}F_{A_{2}}d_{m}^{(1)}(2\lambda_{11}^{PA}+\lambda_{21}^{PA})}{(q^{2}-M_{A_{2}}^{2})(r^{2}-M_{P_{1}}^{2})}\nonumber\\
&\hspace{-80pt}+\frac{8\sqrt{2}B_{0}F_{A_{1}}d_{m}^{(2)}(2\lambda_{12}^{PA}+\lambda_{22}^{PA})}{(p^{2}-M_{A_{1}}^{2})(r^{2}-M_{P_{2}}^{2})}+\frac{8\sqrt{2}B_{0}F_{A_{1}}d_{m}^{(2)}(2\lambda_{12}^{PA}+\lambda_{22}^{PA})}{(q^{2}-M_{A_{1}}^{2})(r^{2}-M_{P_{2}}^{2})}\nonumber\\
&\hspace{-80pt}+\frac{8\sqrt{2}B_{0}F_{A_{2}}d_{m}^{(2)}(2\lambda_{13}^{PA}+\lambda_{23}^{PA})}{(p^{2}-M_{A_{2}}^{2})(r^{2}-M_{P_{2}}^{2})}+\frac{8\sqrt{2}B_{0}F_{A_{2}}d_{m}^{(2)}(2\lambda_{13}^{PA}+\lambda_{23}^{PA})}{(q^{2}-M_{A_{2}}^{2})(r^{2}-M_{P_{2}}^{2})}\nonumber\\
&\hspace{-80pt}+\frac{16B_{0}F_{A_{1}}F_{A_{2}}d_{m}^{(1)}\lambda_{1}^{AAP}}{(p^{2}-M_{A_{1}}^{2})(q^{2}-M_{A_{2}}^{2})(r^{2}-M_{P_{1}}^{2})}+\frac{16B_{0}F_{A_{1}}F_{A_{2}}d_{m}^{(1)}\lambda_{1}^{AAP}}{(p^{2}-M_{A_{2}}^{2})(q^{2}-M_{A_{1}}^{2})(r^{2}-M_{P_{1}}^{2})}\nonumber\\
&\hspace{-80pt}+\frac{16B_{0}F_{A_{1}}^{2}d_{m}^{(2)}\lambda_{2}^{AAP}}{(p^{2}-M_{A_{1}}^{2})(q^{2}-M_{A_{1}}^{2})(r^{2}-M_{P_{2}}^{2})}+\frac{16B_{0}F_{A_{2}}^{2}d_{m}^{(1)}\lambda_{3}^{AAP}}{(p^{2}-M_{A_{2}}^{2})(q^{2}-M_{A_{2}}^{2})(r^{2}-M_{P_{1}}^{2})}\nonumber\\
&\hspace{-80pt}+\frac{16B_{0}F_{A_{1}}F_{A_{2}}d_{m}^{(2)}\lambda_{4}^{AAP}}{(p^{2}-M_{A_{1}}^{2})(q^{2}-M_{A_{2}}^{2})(r^{2}-M_{P_{2}}^{2})}+\frac{16B_{0}F_{A_{1}}F_{A_{2}}d_{m}^{(2)}\lambda_{4}^{AAP}}{(p^{2}-M_{A_{2}}^{2})(q^{2}-M_{A_{1}}^{2})(r^{2}-M_{P_{2}}^{2})}\nonumber\\
&\hspace{-80pt}+\frac{16B_{0}F_{A_{2}}^{2}d_{m}^{(2)}\lambda_{5}^{AAP}}{(p^{2}-M_{A_{2}}^{2})(q^{2}-M_{A_{2}}^{2})(r^{2}-M_{P_{2}}^{2})}\,.\label{eq:AAP_RChT_tensor_v2}
\end{align}

Finally, the derivative operators must be included in an equivalent way as for the previous Green functions. The list of the sufficient operators, needed to obtain the full resonance contribution that could be matched onto OPE \eqref{eq:aap_ope}, is given in the table~\ref{tab:AAP_monomials_4}. The contribution of these operators can be written down in the form
\begin{align}
\left[\mathcal{F}_{AAP}^{\mathrm{RChT}}(p^{2},q^{2},r^{2})\right]_{(3)}&=\frac{32B_{0}F_{A_{1}}^{2}\mu_{2}^{AA}r^{2}}{(p^{2}-M_{A_{1}}^{2})(q^{2}-M_{A_{1}}^{2})}+\frac{16B_{0}F_{A_{1}}F_{A_{2}}\mu_{21}^{AA}r^{2}}{(p^{2}-M_{A_{2}}^{2})(q^{2}-M_{A_{1}}^{2})}\nonumber\\
&\hspace{-80pt}+\frac{16B_{0}F_{A_{1}}F_{A_{2}}\mu_{21}^{AA}r^{2}}{(p^{2}-M_{A_{1}}^{2})(q^{2}-M_{A_{2}}^{2})}+\frac{32B_{0}F_{A_{2}}^{2}\mu_{22}^{AA}r^{2}}{(p^{2}-M_{A_{2}}^{2})(q^{2}-M_{A_{2}}^{2})}\nonumber\\
&\hspace{-80pt}+\frac{16B_{0}F_{A_{1}}^{2}d_{m}^{(1)}\mu_{1}^{AAP}(p^{4}+q^{4})}{(p^{2}-M_{A_{1}}^{2})(q^{2}-M_{A_{1}}^{2})(r^{2}-M_{P_{1}}^{2})}+\frac{16B_{0}F_{A_{1}}^{2}d_{m}^{(2)}\mu_{2}^{AAP}(p^{4}+q^{4})}{(p^{2}-M_{A_{1}}^{2})(q^{2}-M_{A_{1}}^{2})(r^{2}-M_{P_{2}}^{2})}\nonumber\\
&\hspace{-80pt}+\frac{16B_{0}F_{A_{2}}^{2}d_{m}^{(1)}\mu_{3}^{AAP}(p^{4}+q^{4})}{(p^{2}-M_{A_{2}}^{2})(q^{2}-M_{A_{2}}^{2})(r^{2}-M_{P_{1}}^{2})}\,.\label{eq:AAP_RChT_tensor_v3}
\end{align}
\begin{table}[t!]
\centering
\begin{tabular}{|C{4cm}|C{2cm}||C{4cm}|C{2cm}|}
\hline
Operator $\big(\widehat{\mathcal{O}}_{i}^{RR}\big)_{\mu\nu\alpha\beta}$           & Coupling constant & Operator $\big(\widehat{\mathcal{O}}_{i}^{RRR}\big)_{\mu\nu\alpha\beta}$ & Coupling constant   \\ \hline\hline
$i\langle\lbrace A_{1}^{\mu\nu},A_{1}^{\alpha\beta}\rbrace\Box\chi_{-}\rangle$ & $\mu_{2}^{AA}$    & $\langle\lbrace\Box^{2}A_{1}^{\mu\nu},A_{1}^{\alpha\beta}\rbrace P_{1}\rangle$ & $\mu_{1}^{AAP}$ \\
$i\langle\lbrace A_{1}^{\mu\nu},A_{2}^{\alpha\beta}\rbrace\Box\chi_{-}\rangle$ & $\mu_{21}^{AA}$   & $\langle\lbrace\Box^{2}A_{1}^{\mu\nu},A_{1}^{\alpha\beta}\rbrace P_{2}\rangle$ & $\mu_{2}^{AAP}$ \\
$i\langle\lbrace A_{2}^{\mu\nu},A_{2}^{\alpha\beta}\rbrace\Box\chi_{-}\rangle$ & $\mu_{22}^{AA}$   & $\langle\lbrace\Box^{2}A_{2}^{\mu\nu},A_{2}^{\alpha\beta}\rbrace P_{1}\rangle$ & $\mu_{3}^{AAP}$ \\ \hline
\end{tabular}
\caption{Additional derivative monomials of the extended RChT Lagrangian at $\mathcal{O}(p^{6})$ with three resonance fields, relevant for the $\langle AAP\rangle$ Green function.}
\label{tab:AAP_monomials_4}
\end{table}

Taking into account all the respective contributions \eqref{eq:AAP_RChT_tensor}-\eqref{eq:AAP_RChT_tensor_v3}, we denote the desired final result as
\begin{equation}
\mathcal{F}_{AAP}^{\mathrm{RChT}}(p^{2},q^{2},r^{2})=\sum_{i=1}^{3}\left[\mathcal{F}_{AAP}^{\mathrm{RChT}}(p^{2},q^{2},r^{2})\right]_{(i)}\,.\label{eq:AAP_RChT_tensor_final}
\end{equation}

\paragraph{Matching RChT with OPE.}
The matching between the total resonance contribution \eqref{eq:AAP_RChT_tensor_final} and the OPE \eqref{eq:aap_ope} can now be performed. As in all the previous cases above, we refer the reader to ref.~\cite{KadavyThesis:2022}.

\paragraph{Resonance saturation.}
Finally, the matching between the ChPT contribution \eqref{eq:AAP_ChPT} and the total resonance contribution \eqref{eq:AAP_RChT_tensor_final} gives us the expressions for the two low-energy constants $C_{9}^{W}$ and $C_{23}^{W}$ in terms of the couplings of the resonance Lagrangian. These can be found explicitly in appendix \ref{sec:resonance_saturation}.


\section{Three-multiplet resonance contribution within RChT: \texorpdfstring{$\bm{\langle VVP\rangle}$}{}}\label{sec:three_multiplets}
Until now, we have been interested in the contributions of resonances to the Green functions, which included the first two lowest-lying multiplets of the vector, axial-vector, scalar and pseudoscalar kind. Especially in the case of the $\langle VVP\rangle$ correlator, we have found that the respective THS+P parametrization merely reproduces the pion transition form factor in the form given by the THS. Further, there exist other high-energy conditions that one could demand from the $\langle VVP\rangle$ Green function to fulfil and which, however, could not be satisfied with the THS+P parametrization.

Since such a correlator possesses many interesting phenomenological applications, let us now get back to this case once more and add another set of vector and pseudoscalar multiplets of resonances. In a gritty language of acronyms, we might speak of the (THS+P)+V+P parametrization, however, we restrain ourselves from using such a denomination in the following text for clarity.

Formally, one can write down the general parametrization of the Lorentz-invariant function of the $\langle VVP\rangle$ correlator, with three vector and pseudoscalar resonance multiplets taken into account, in the form
\begin{align}
\mathcal{F}_{VVP}(p^{2},q^{2},r^{2})&=\frac{\mathcal{S}_{VVP}(p^{2},q^{2},r^{2})}{(p^{2}-M_{V_{1}}^{2})(p^{2}-M_{V_{2}}^{2})(p^{2}-M_{V_{3}}^{2})(q^{2}-M_{V_{1}}^{2})(q^{2}-M_{V_{2}}^{2})(q^{2}-M_{V_{3}}^{2})}\nonumber\\
&\times\frac{B_{0}F^{2}}{r^{2}(r^{2}-M_{P_{1}}^{2})(r^{2}-M_{P_{2}}^{2})(r^{2}-M_{P_{3}}^{2})}\,,\label{eq:VVP_polynomial_dim16}
\end{align}
where $\mathcal{S}_{VVP}(p^{2},q^{2},r^{2})$ is a dimension-16 polynomial, consisting of $95$ Bose-symmetric terms. The requirements \eqref{eq:vvp_ope}, \eqref{eq:vvp_ope_2-op}, \eqref{eq:form_factor_BL} and \eqref{eq:form_factor_anomaly}, i.e.~the items 1) to 4) for $\langle VVP\rangle$ in subsection \ref{ssec:general_structure}, lower the number of the unknown parameters, say $\alpha_{i}$, of such a polynomial from $95$ to $40$.

\subsection{Pion transition form factor}
The number of 40 parameters in the above-obtained Lorentz-invariant structure \eqref{eq:VVP_polynomial_dim16} is reduced down for the pion transition form factor \eqref{eq:form_factor}, which reads
\begin{align}
\mathcal{F}_{\gamma^{\ast}\gamma^{\ast}\pi^{0}}(p^{2},q^{2})&=\frac{N_{c}}{12\pi^{2}F}\label{eq:form_factor_dim16}\\
&\hspace{-40pt}\times\frac{M_{V_{1}}^{4}M_{V_{2}}^{4}M_{V_{3}}^{4}}{(p^{2}-M_{V_{1}}^{2})(p^{2}-M_{V_{2}}^{2})(p^{2}-M_{V_{3}}^{2})(q^{2}-M_{V_{1}}^{2})(q^{2}-M_{V_{2}}^{2})(q^{2}-M_{V_{3}}^{2})}\nonumber\\
&\hspace{-40pt}\times\bigg[1+\frac{4\pi^{2}F^{2}}{N_{c}M_{V_{1}}^{4}M_{V_{2}}^{4}M_{V_{3}}^{4}}\Big(6M_{V_{1}}^{2}M_{V_{2}}^{2}M_{V_{3}}^{2}(p^{4}+q^{4})-p^{4}q^{4}(p^{2}+q^{2})\Big)\nonumber\\
&\hspace{-40pt}-\frac{8\pi^{2}F^{2}}{N_{c}M_{V_{1}}^{4}M_{V_{2}}^{4}M_{V_{3}}^{4}M_{P_{1}}^{2} M_{P_{2}}^{2}M_{P_{3}}^{2}}\Big(\alpha_{2}(p^{2}+q^{2})+\alpha_{6}p^{2}q^{2}+\alpha_{10}p^{2}q^{2}(p^{2}+q^{2})\nonumber\\
&\hspace{-40pt}+\alpha_{16}p^{2}q^{2}(p^{4}+q^{4})+\alpha_{20}p^{4}q^{4}+\alpha_{25}p^{2}q^{2}(p^{2}-q^{2})^{2}(p^{2}+q^{2})+\alpha_{37}p^{2}q^{2}(p^{2}-q^{2})^{4}\Big)\bigg].\nonumber
\end{align}

The presence of free parameters allows us to require additional constraints that the $\langle VVP\rangle$ Green function is supposed to satisfy. In detail, we are going to utilize the ``subleading'' Brodsky--Lepage behaviour of the pion transition form factor and the expected properties of some other form factors. Continuing on items 1) to 4) in section~\ref{ssec:general_structure}, these are as follows.

\begin{itemize}
\item[5)] The $\langle VVP\rangle$ correlator is required to fulfil further constraint based on the higher-twist terms in OPE \cite{Shuryak:1981kj,Novikov:1983jt}, which goes beyond the leading term of the Brodsky--Lepage behaviour for symmetric configuration of virtualities. In detail, the double off-shell pion transition form factor, normalized by the anomaly, is supposed to behave as
\begin{equation}
\frac{\mathcal{F}_{\gamma^{\ast}\gamma^{\ast}\pi^{0}}(-Q^{2},-Q^{2})}{\mathcal{F}_{\gamma^{\ast}\gamma^{\ast}\pi^{0}}(0,0)}=\frac{8\pi^{2}F^{2}}{3}\left[\frac{1}{Q^{2}}-\frac{8\delta^{2}}{9Q^{4}}+\mathcal{O}\left(\frac{1}{Q^{6}}\right)\right],\label{eq:delta2}
\end{equation}
with
\begin{equation}
\delta^{2}=(0.20\pm 0.02)\,\mathrm{GeV}^{2}\,.\label{eq:delta2_value}
\end{equation}
\item[6)] Secondly, let us define the form factor
\begin{align}
\langle V_{1}|V|\pi^{0}\rangle&=\lim_{\substack{p^{2}\,\rightarrow\,M_{V_{1}}^{2} \\ r^{2}\,\rightarrow\,0}}(p^{2}-M_{V_{1}}^{2})r^{2}\mathcal{F}_{VVP}(p^{2},q^{2},r^{2})\nonumber\\
&=\frac{3B_{0}F}{2}\lim_{p^{2}\,\rightarrow\,M_{V_{1}}^{2}}(p^{2}-M_{V_{1}}^{2})\mathcal{F}_{\gamma^{\ast}\gamma^{\ast}\pi^{0}}(p^{2},q^{2})\,.\label{eq:VVP_constraints_part2}
\end{align}
To ensure its assumed asymptotic behaviour, we require that
\begin{equation}
\langle V_{1}|V|\pi^{0}\rangle\sim\frac{1}{q^{2}}\qquad\text{for}\,\,q^{2}\rightarrow \infty\,.\label{eq:VVP_constraints_part2_behaviour}
\end{equation}
\item[7)] As the last constraint, we introduce the semi-on-shell Brodsky--Lepage behaviour for pseudoscalar resonances \cite{Hoferichter:2020lap}. In detail, we define a set of form factors
\begin{equation}
\mathcal{F}_{\gamma^{\ast}\gamma^{\ast}P_{i}}(Q^{2})=\lim_{r^{2}\,\rightarrow\,M_{P_{i}}^{2}}(r^{2}-M_{P_{i}}^{2})\mathcal{F}_{VVP}(0,-Q^{2},r^{2})\,,\label{eq:VVP_constraints_part3}
\end{equation}
with factorization considered in all pseudoscalar channels individually (i.e.~$i=1,2,3$), for which we require that it asymptotically behaves as
\begin{equation}
\mathcal{F}_{\gamma^{\ast}\gamma^{\ast}P_{i}}(Q^{2})\sim\frac{1}{Q^{2}}\qquad\text{for}\,\,Q^{2}\rightarrow \infty\,.
\end{equation}
\end{itemize}

Utilizing the above-mentioned constraints leads, after some simple algebraic manipulations, to the following solution:
\begin{align}
\alpha_{16}&=-\frac{1}{2}M_{V_{1}}^{2}M_{P_{1}}^{2}M_{P_{2}}^{2}M_{P_{3}}^{2}\,,\nonumber\\
\alpha_{20}&=-M_{P_{1}}^{2}M_{P_{2}}^{2}M_{P_{3}}^{2}\left(M_{V_{1}}^{2}+2M_{V_{2}}^{2}+2M_{V_{3}}^{2}-\frac{8}{9}\delta^{2}\right),\nonumber\\
\alpha_{25}=\alpha_{37}&=0\,.\label{eq:VVP_additional_constraints}
\end{align}

Having the constraints \eqref{eq:VVP_additional_constraints} at our disposal, one can substitute them back into the pion transition form factor \eqref{eq:form_factor_dim16}, which is the central quantity of our analysis. Its form reads
\begin{align}
\mathcal{F}_{\gamma^{\ast}\gamma^{\ast}\pi^{0}}(p^{2},q^{2})&=\frac{N_{c}}{12\pi^{2}F}\frac{M_{V_{1}}^{4}M_{V_{2}}^{4}}{(p^{2}-M_{V_{1}}^{2})(p^{2}-M_{V_{2}}^{2})(q^{2}-M_{V_{1}}^{2})(q^{2}-M_{V_{2}}^{2})}\nonumber\\
&\hspace{-40pt}\times\Bigg[1+\frac{1}{(p^{2}-M_{V_{3}}^{2})(q^{2}-M_{V_{3}}^{2})}\Bigg(M_{V_{3}}^{2}\beta_{2}(p^{2}+q^{2})-\beta_{6}p^{2}q^{2}-\frac{\beta_{10}}{M_{V_{3}}^{2}}(p^{2}+q^{2})p^{2}q^{2}\nonumber\\
&\hspace{-40pt}+\frac{8\pi^{2}F^{2}}{9N_{c}M_{V_{1}}^{4}M_{V_{2}}^{4}}(9M_{V_{1}}^{2}+18M_{V_{2}}^{2}+18M_{V_{3}}^{2}-8\delta^{2})p^{4}q^{4}+\frac{4\pi^{2}F^{2}}{N_{c}M_{V_{1}}^{2}M_{V_{2}}^{4}}(p^{4}+q^{4})p^{2}q^{2}\nonumber\\
&\hspace{-40pt}+\frac{24\pi^{2}F^{2}M_{V_{3}}^{2}}{N_{c}M_{V_{1}}^{2}M_{V_{2}}^{2}}(p^{4}+q^{4})-\frac{4\pi^{2}F^{2}}{N_{c}M_{V_{1}}^{4}M_{V_{2}}^{4}}(p^{2}+q^{2})p^{4}q^{4}\Bigg)\Bigg]\,,\label{eq:form_factor_dim16_v2}
\end{align}
where we have introduced the dimensionless parameters
\begin{align}
\beta_{2}&\equiv 1-\frac{8\pi^{2}F^{2}}{N_{c}}\frac{\alpha_{2}}{M_{V_{1}}^{4}M_{V_{2}}^{4}M_{V_{3}}^{2}M_{P_{1}}^{2}M_{P_{2}}^{2}M_{P_{3}}^{2}}\,,\nonumber\\
\beta_{6}&\equiv 1+\frac{8\pi^{2}F^{2}}{N_{c}}\frac{\alpha_{6}}{M_{V_{1}}^{4}M_{V_{2}}^{4}M_{P_{1}}^{2}M_{P_{2}}^{2}M_{P_{3}}^{2}}\,,\nonumber\\
\beta_{10}&\equiv\frac{8\pi^{2}F^{2}}{N_{c}}\frac{M_{V_{3}}^{2}\alpha_{10}}{M_{V_{1}}^{4}M_{V_{2}}^{4}M_{P_{1}}^{2}M_{P_{2}}^{2}M_{P_{3}}^{2}}\,.\label{eq:beta_parameters}
\end{align}

So far, we have been able to reduce the number of unknown parameters of the pion transition form factor to only three: $\beta_{2}$, $\beta_{6}$ and $\beta_{10}$. Obviously, once one of the momentum squared is set to zero, i.e.~when one of the photons is on-shell, the form factor \eqref{eq:form_factor_dim16_v2} is then a function of only $\beta_{2}$. The determination of $\beta_{6}$ and $\beta_{10}$ thus requires to use of a fully off-shell form factor, as presented above. A further ascertainment of the numerical values of these constants requires phenomenological inputs, such as those presented below.

To this end, we present the values of physical constants that are used in the following analysis: $\alpha=1/137$, $N_{c}=3$, $M_{e}=0.511\,\mathrm{MeV}$, $M_{\mu}=105.658\,\mathrm{MeV}$, $M_{\pi^{+}}=139.57\,\mathrm{MeV}$, $M_{\omega}=782.65\,\mathrm{MeV}$, $F=92.22\,\mathrm{MeV}$, $F_{\omega}=140\,\mathrm{MeV}$ and $F_{V_{1}}=146.3\,\mathrm{MeV}$. Regarding the masses of the individual multiplets, only the masses of the vector resonances are actually needed for numerical analysis in the forthcoming subsections, since the masses of the pseudoscalar ones are, at least in the context of the form factor \eqref{eq:form_factor_dim16_v2}, absorbed into the coefficients \eqref{eq:beta_parameters}. To this end, we take the masses of the first two vector resonance multiplets as the masses of $\rho(770)$ and $\rho(1450)$, i.e.~$M_{V_{1}}=775.26\,\mathrm{MeV}$ and $M_{V_{2}}=1.45\,\mathrm{GeV}$, respectively. The mass of the third vector multiplet is considered to be effective, i.e.~we take its central value as an average of the lightest ($\rho(1570)$) and the heaviest ($\phi(1680)$) resonance of such a multiplet, with a respective range taken into account: 
\begin{equation}
M_{V_{3}}=(1.63\pm 0.06)\,\mathrm{GeV}.
\label{eq:MV3}
\end{equation}.


\subsection{Determination of \texorpdfstring{$\beta_{2}$}{}, \texorpdfstring{$\beta_{6}$}{} and \texorpdfstring{$\beta_{10}$}{}}
In order to fix the unknown parameters $\beta_{2}$, $\beta_{6}$ and $\beta_{10}$, we will use the lattice calculation provided in ref.~\cite{Gerardin:2019vio}. 
Therein, the authors have fitted the lattice data to the ten parametric modified $z$-expansion
\begin{multline}
P(Q_1^2,Q_2^2) \ {\cal F}_{\gamma^*\gamma^*\pi^0}(-Q_1^2, -Q_2^2)  = \\  \sum_{n,\,m=0}^{N} c_{nm} \, \left( z_1^n - (-1)^{N+n+1} \frac{n}{N+1} \, z_1^{N+1} \right) \, \left( z_2^m -  (-1)^{N+m+1} \frac{m}{N+1} \, z_2^{N+1} \right),
\end{multline}
where
\begin{equation}
P(Q_1^2,Q_2^2) = 1 + \frac{Q_1^2 + Q_2^2}{M_V^2} \,,
\end{equation}
and
\begin{equation}
z_{1,\,2} = \frac{ \sqrt{4 M_\pi^2+Q_{1,\,2}^2} - \sqrt{4 M_\pi^2 - t_0} }{ \sqrt{4 M_\pi^2+Q_{1,\,2}^2} + \sqrt{4 M_\pi^2 - t_0} }\,, \qquad t_0 = 4 M_\pi^2 \left( 1 - \sqrt{1+ \frac{Q_{\max}^2}{4 M_\pi^2} } \right),
\end{equation}
with $Q_\text{max}^2= 4\,\text{GeV}^2$.
The result of their fit of $c_{nm}$ and the corresponding correlation matrix is summarized in Tab.~III and in eq.~(45) in ref.~\cite{Gerardin:2019vio}, respectively.

Since in our model, the $c_{nm}$ parameters are functions of  $\beta_2$, $\beta_6$ and $\beta_{10}$, we have used the lattice values of $c_{nm}$ to fit the free parameters $\beta_i$ with the result
\begin{equation}
\beta_2 = -1.31\pm 0.10\pm 0.10\,,\quad
\beta_6 = -8.27\pm 0.37\pm 0.95\,,\quad
\beta_{10}= 6.59\pm 0.34\pm 0.98\,.
\label{eq:fit_result}
\end{equation}
Here, the first uncertainty corresponds to the fit obtained by fixing $M_{V_3}$ and $\delta^2$ as their mean values \eqref{eq:MV3} and \eqref{eq:delta2_value}, respectively, and the second one represents the uncertainty induced by the variation of the fit with $M_{V_3}$ and $\delta^{2}$ varying within their error bars\footnote{To estimate this, we have generated $10^5$ random values of $M_{V_3}$ and $\delta^2$ normally distributed with the corresponding mean values and standard deviations according to \eqref{eq:MV3} and \eqref{eq:delta2_value}, respectively, and for each of them we fitted the parameters $\beta_i$. The error estimates then correspond to the standard deviations of such samples. A similar approach is used for the error estimates of other observables derived from the pion transition formfactor.}.
The correlation matrix for the fit with  $M_{V_3}$ and $\delta^2$ fixed at their mean values $M_{V_3}=1.63\,\mathrm{GeV}$ and $\delta^{2}=0.2\,\mathrm{GeV}^{2}$  reads
\begin{equation}
\text{cor}(\beta_i) = 
\begin{pmatrix}
\phantom{+}1.00 &-0.51&-0.83\\
-0.51&\phantom{+}1.00 &\phantom{+}0.47\\
-0.83&\phantom{+}0.47 &\phantom{+}1.00
\end{pmatrix}.
\end{equation}
The quality of the fit is given by $\chi^2/\text{d.o.f}= 1.7$.
It would be interesting to use the lattice data directly without the $z$ parametric expansion.

As a rough check of the consistency of our result, we have also fitted  the form-factor~\eqref{eq:form_factor_dim16_v2} directly to 20\,100 artificial data points as created from the fitted modified $z$-expansion in  \cite{Gerardin:2019vio}.
For simplicity, in this indicative fit, the weight of each of the data points was given by its individual inverse error squared.
However, these points are highly correlated, and without taking into account the correlation such a fit cannot be taken seriously. Nevertheless, we obtain  compatible values:
\begin{equation}
\beta_{2}=-1.41\pm 0.10\,,\qquad\beta_{6}=-7.84\pm 0.78\,,\qquad\beta_{10}=6.39\pm 0.90\,.
\end{equation}
%


In what follows, we consider the values (\ref{eq:fit_result}) above as the reference ones and only these will be used for the determination of the $g-2$ and the quantity $\chi^{(\mathrm{r})}$ later on. However, to verify the reliability of the values of the parameters in \eqref{eq:fit_result}, one may consider several consistency checks that we shall discuss below.

\subsubsection{Consistency check: \texorpdfstring{$\beta_{2}$}{}}

\paragraph{Pion transition form factor.}
To follow up on the previous analysis of fitting the respective data with the pion transition form factor, we start with its single off-shell variant in the space-like region. Therein, such a form factor can be compared to the data obtained by the BABAR \cite{Aubert:2009mc}, BELLE \cite{Uehara:2012ag} and CLEO \cite{Gronberg:1997fj} collaborations. To this end, using the expression \eqref{eq:form_factor_dim16_v2} and the relevant kinematics in the Euclidean region, the pion transition form factor gets simplified significantly and reads
\begin{align}
\mathcal{F}_{\gamma^{\ast}\gamma^{\ast}\pi^{0}}(0,-Q^{2})&=\frac{N_{c}}{12\pi^{2}F}\frac{M_{V_{1}}^{2}M_{V_{2}}^{2}M_{V_{3}}^{2}}{(Q^{2}+M_{V_{1}}^{2})(Q^{2}+M_{V_{2}}^{2})(Q^{2}+M_{V_{3}}^{2})}\nonumber\\
&\hspace{-35pt}\times\left(1+\frac{1-\beta_{2}}{M_{V_{3}}^{2}}Q^{2}+\frac{24\pi^{2}F^{2}}{N_{c}M_{V_{1}}^{2}M_{V_{2}}^{2}M_{V_{3}}^{2}}Q^{4}\right).\label{eq:form_factor_dim16_v3}
\end{align}
The situation is thus a bit easier than it was in the case above --- an expression \eqref{eq:form_factor_dim16_v3} does not contain the parameter $\delta^{2}$ anymore and one is, therefore, required to vary only $M_{V_{3}}$.

\begin{figure}[t!]
\small
\centering
\includegraphics[width=1\textwidth]{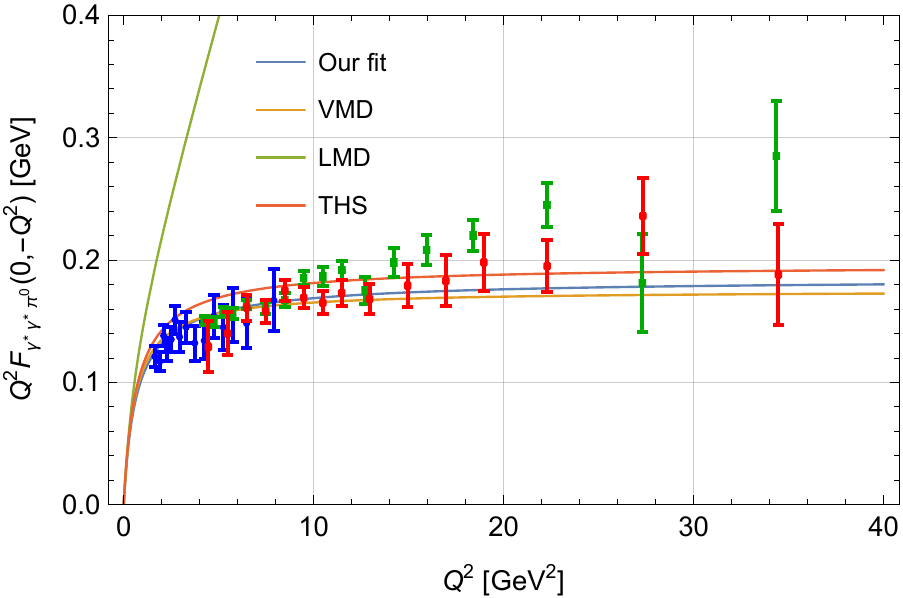}
\caption{A plot of BABAR, BELLE and CLEO (green, red and blue bars, respectively) data and the form factor $\mathcal{F}_{\gamma^{\ast}\gamma^{\ast}\pi^{0}}(0,-Q^{2})$ within the LMD, VMD, THS and the three-resonance parametrization (green, orange, red and blue lines, respectively).}
\label{fig:form_factor_plot}
\end{figure}

Taking into account the experimentally found values for the form factor (multiplied by the momentum transfer squared, to be precise) by the above-mentioned collaborations and its analytical form \eqref{eq:form_factor_dim16_v3} leads us to the prediction for $\beta_{2}$ as\footnote{$\chi^{2}/\text{d.o.f.}=1.97$.}
\begin{equation}
\beta_{2}=-1.31\pm 0.09\,.\label{eq:B2_value}
\end{equation}
Once again, we point out that the weight of each of the data points was determined by the individual inverse error squared and the method of minimizing the $\chi^{2}$ was employed as well. Also, for a comparison with other models (specifically with LMD, VMD and THS), we present the graph shown in fig.~\ref{fig:form_factor_plot}, where the experimental data sets fitted with our form factor \eqref{eq:form_factor_dim16_v3} for the value \eqref{eq:B2_value} can be seen.

A note is perhaps in order here. As one can see from fig.~\ref{fig:form_factor_plot}, the last five or six pairs of data are not quite in agreement and such a discrepancy has been a subject of discussion before. In our procedure of extracting the value of $\beta_{2}$, we have taken into account all the forty-seven data points --- fifteen from BELLE and CLEO each and seventeen from BABAR. One might argue, whether there would be any significant change in the value of $\beta_{2}$ when the last couple of disparate data points are ignored. To this end, we have performed the above-mentioned fit on the data in the region below $Q^{2}=15\,\mathrm{GeV}^{2}$, where the experimental values seem to be consistent with each other. Such a truncation leaves us with thirty-seven data points (twelve from BABAR, ten from BELLE and the CLEO data being untouched) and we get the value of $\beta_{2}$ as\footnote{$\chi^{2}/\text{d.o.f.}=1.25$.}
\begin{equation}
\beta_{2}=-1.28\pm 0.09\,.\label{eq:B2_value_v2}
\end{equation}

\paragraph{The decay $\bm{\rho^{+}\to\pi^{+}\gamma}$.}
Now, let us examine the decay of the charged rho meson, namely the channel $\rho^{+}\to\pi^{+}\gamma$, for which the decay width can be written down as
\begin{equation}
\Gamma_{\rho^{+}\,\to\,\pi^{+}\gamma}=\frac{M_{\rho^{+}}^{2}-M_{\pi^{+}}^{2}}{48\pi M_{\rho^{+}}^{3}}\sum_{\mathrm{pol.}}\left\vert\mathcal{A}_{\rho^{+}\to\,\pi^{+}\gamma}\,\varepsilon^{\mu\nu(p)(q)}\epsilon_{\mu}(p)\epsilon_{\nu}^{\ast}(q)\right\vert^{2},\label{eq:rho_decay_width}
\end{equation}
where, as denoted, we sum over the polarizations of the $\rho^{+}$ meson and the photon, given by the polarization tensors $\epsilon_{\mu}(p)$ and $\epsilon_{\nu}^{\ast}(k)$, respectively. Since the mass of the $\rho^{+}$ meson is assumed to correspond to the mass of the first vector resonance multiplet, we take $M_{\rho^{+}}=M_{V_{1}}$ in the forthcoming discussion so as to be consistent with the notation used in the above-mentioned form factors etc. The amplitude of the process in question then reads
\begin{equation}
\mathcal{A}_{\rho^{+}\,\to\,\pi^{+}\gamma}=\frac{e}{2F_{V_{1}}M_{V_{1}}}\lim_{q^{2}\,\rightarrow\,M_{V_{1}}^{2}}(q^{2}-M_{V_{1}}^{2})\mathcal{F}_{\gamma^{\ast}\gamma^{\ast}\pi^{0}}(0,q^{2})\,,\label{eq:rho_decay_amplitude}
\end{equation}
which, as one can suspect already, depends on $\beta_{2}$ and can be easily rewritten to
\begin{equation}
\mathcal{A}_{\rho^{+}\,\to\,\pi^{+}\gamma}=-\frac{e M_{V_{1}}}{24\pi^{2}F_{V_{1}}}\frac{M_{V_{1}}^{2} \left((\beta_{2}-1)M_{V_{2}}^{2} N_{c}+24\pi^{2}F^{2}\right)+M_{V_{2}}^{2}M_{V_{3}}^{2}N_{c}}{F(M_{V_{1}}^{2}-M_{V_{2}}^{2})(M_{V_{1}}^{2}-M_{V_{3}}^{2})}\,.\label{eq:rho_decay_amplitude_v2}
\end{equation}
Finally, employing the amplitude \eqref{eq:rho_decay_amplitude_v2} into \eqref{eq:rho_decay_width} and performing standard algebraic manipulations with polarization tensors, one readily obtains
\begin{equation}
\Gamma_{\rho^{+}\,\to\,\pi^{+}\gamma}^{\mathrm{theory}}=\frac{\alpha}{M_{V_{1}}}\left(\frac{M_{V_{1}}^{2}-M_{\pi^{+}}^{2}}{24}\right)^{3}\left(\frac{M_{V_{1}}^{2}\left((\beta_{2}-1)M_{V_{2}}^{2} N_{c}+24\pi^{2}F^{2}\right)+M_{V_{2}}^{2}M_{V_{3}}^{2}N_{c}}{\pi^{2}F_{V_{1}}F(M_{V_{1}}^{2}-M_{V_{2}}^{2})(M_{V_{1}}^{2}-M_{V_{3}}^{2})}\right)^{2}.
\end{equation}
Upon substituting for $\beta_{2}$ according to \eqref{eq:fit_result}, we get our prediction (as above, here and in what follows, the first uncertainty comes from the fit with $M_{V_3}$ and $\delta^2$ fixed, with the correlation of $\beta_i$ taking into account,  while the second one corresponds to the variation of these parameters)
\begin{equation} \Gamma_{\rho^{+}\,\to\,\pi^{+}\gamma}^{\mathrm{theory}}  =(6.73\pm 0.54\pm 0.03)\cdot 10^{-5}\,\mathrm{GeV}\,,
\end{equation}
which can be compared  with the experimental value presented in PDG \cite{ParticleDataGroup:2020ssz},
\begin{equation}
\Gamma_{\rho^{+}\,\to\,\pi^{+}\gamma}^{\mathrm{PDG}}=(6.71\pm 0.75)\cdot 10^{-5}\,\mathrm{GeV}\,.
\end{equation}

\paragraph{The decay $\bm{\omega\to\pi^{0}\gamma}$.}
Finally, we will examine the decay of the $\omega(782)$ meson. To make the analysis a bit easier, similarly, as the authors of ref.~\cite{Husek:2015wta} have proceeded, we will assume that its quark component carrying strangeness is negligible.\footnote{Under such an assumption, the $\omega(782)$ meson is then very nearly a pure symmetric state with the quark content given approximately as $\frac{1}{\sqrt{2}}(u\overline{u}+d\overline{d})$.} Then, let us focus on the decay channel $\omega\to\pi^{0}\gamma$, for which the relevant form factor reads \cite{Husek:2015wta}
\begin{equation}
\mathcal{F}_{V\omega\pi^{0}}(q^{2})=\frac{1}{B_{0}F}\frac{1}{M_{\omega}F_{\omega}}\lim_{\substack{p^{2}\,\rightarrow\,M_{V_{1}}^{2} \\ r^{2}\,\rightarrow\,0}}(p^{2}-M_{V_{1}}^{2})r^{2}\mathcal{F}_{VVP}(p^{2},q^{2},r^{2})\,,\label{eq:form_factor_Vwp}
\end{equation}
where $M_{\omega}$ and $F_{\omega}$ stand for the mass and the decay constant of the $\omega$ meson, respectively.

The decay width of such a process can be calculated easily and one obtains \cite{Husek:2015wta}
\begin{equation}
\Gamma_{\omega\,\to\,\pi^{0}\gamma}^{\mathrm{theory}}=\frac{\alpha M_{\omega}^{3}}{24}|\mathcal{F}_{V\omega\pi^{0}}(0)|^{2}\left(1-\frac{M_{\pi}^{2}}{M_{\omega}^{2}}\right)^{3},\label{eq:omega_decay_width}
\end{equation}
where the respective form of the form factor \eqref{eq:form_factor_Vwp} reads
\begin{align}
\mathcal{F}_{V\omega\pi^{0}}(0)&=-\frac{1}{F_{\omega}M_{\omega}(M_{V_{1}}^{2}-M_{V_{2}}^{2})(M_{V_{1}}^{2}-M_{V_{3}}^{2})}\nonumber\\
&\times\left(3FM_{V_{1}}^{4}+\frac{N_{c}}{8\pi^{2}F}M_{V_{1}}^{4}M_{V_{2}}^{2}\beta_{2}-\frac{N_{c}}{8\pi^{2}F}M_{V_{1}}^{2}M_{V_{2}}^{2}(M_{V_{1}}^{2}-M_{V_{3}}^{2})\right).\label{eq:omega_form_factor}
\end{align}
An expression, obtained by inserting \eqref{eq:omega_form_factor} back into \eqref{eq:omega_decay_width}, gives our prediction
\begin{equation}
\Gamma_{\omega\,\to\,\pi^{0}\gamma}^{\mathrm{theory}}=(6.73\pm 0.54 \pm 0.03)\cdot 10^{-4}\,\mathrm{GeV}\,,
\end{equation}
which can be compared with the PDG value of the decay width \cite{ParticleDataGroup:2020ssz},
\begin{equation}
\Gamma_{\omega\,\to\,\pi^{0}\gamma}^{\mathrm{PDG}}=(7.24\pm 0.25)\cdot 10^{-4}\,\mathrm{GeV}\,.
\end{equation}

\subsubsection{Consistency check: \texorpdfstring{$\beta_{6}$}{} and \texorpdfstring{$\beta_{10}$}{}}
An appropriate phenomenological manner in order to verify the values of the parameters $\beta_{6}$ and $\beta_{10}$ must include a relevant double off-shell form factor so that the said parameters are taken into account. As it turns out, one can make use of the three-body decay of the $\omega(782)$ meson, namely the decay channel $\omega\to\pi^{0}\ell^{+}\ell^{-}$ with the lepton pair $\ell^{\pm}$ in the final state. Obviously, due to the kinematics, the leptons can be either the electron and positron ($e^{\pm}$) or the muon and antimuon ($\mu^{\pm}$), respectively. According to the PDG, the measured decay widths for both cases are determined to be \cite{ParticleDataGroup:2020ssz}:
\begin{align}
\Gamma_{\omega\,\to\,\pi^{0}e^{+}e^{-}}^{\mathrm{PDG}}&=(6.7\pm 0.5)\cdot 10^{-6}\,\mathrm{GeV}\,,\label{eq:decay_width_w_to_pie+e-_value}\\
\Gamma_{\omega\,\to\,\pi^{0}\mu^{+}\mu^{-}}^{\mathrm{PDG}}&=(1.16\pm 0.16)\cdot 10^{-6}\,\mathrm{GeV}\,.\label{eq:decay_width_w_to_pimu+mu-_value}
\end{align}

As we have already mentioned, one can make use of the form factor \eqref{eq:form_factor_Vwp}, due to which the decay width can be written down in the form \cite{Husek:2015wta}
\begin{align}
&\Gamma_{\omega\,\to\,\pi^{0}\ell^{+}\ell^{-}}^{\mathrm{theory}}=\frac{\alpha^{2}}{72\pi M_{\omega}^{3}}\nonumber\\
&\hspace{20pt}\times\int_{4M_{\ell}^{2}}^{(M_{\omega}-M_{\pi})^{2}}\frac{|\mathcal{F}_{V\omega\pi^{0}}(q^{2})|^{2}}{q^{2}}\sqrt{1-\frac{4M_{\ell}^{2}}{q^{2}}}\left(1+\frac{2M_{\ell}^{2}}{q^{2}}\right)\lambda_{\mathrm{K}}^{3/2}\big(M_{\omega}^{2},M_{\pi}^{2},q^{2}\big)\,\mathrm{d}q^{2}\,,\label{eq:decay_width_w_to_pil+l-}
\end{align}
with the integration over the allowed values of the dilepton invariant mass and where
\begin{equation}
\lambda_{\mathrm{K}}(x,y,z)=x^{2}+y^{2}+z^{2}-2xy-2xz-2yz
\end{equation}
stands for the K\"{a}ll\'{e}n's (triangle) function and $\alpha$ is the fine-structure constant.

\begin{figure}[t!]
\small
\centering
\includegraphics[width=0.8\textwidth]{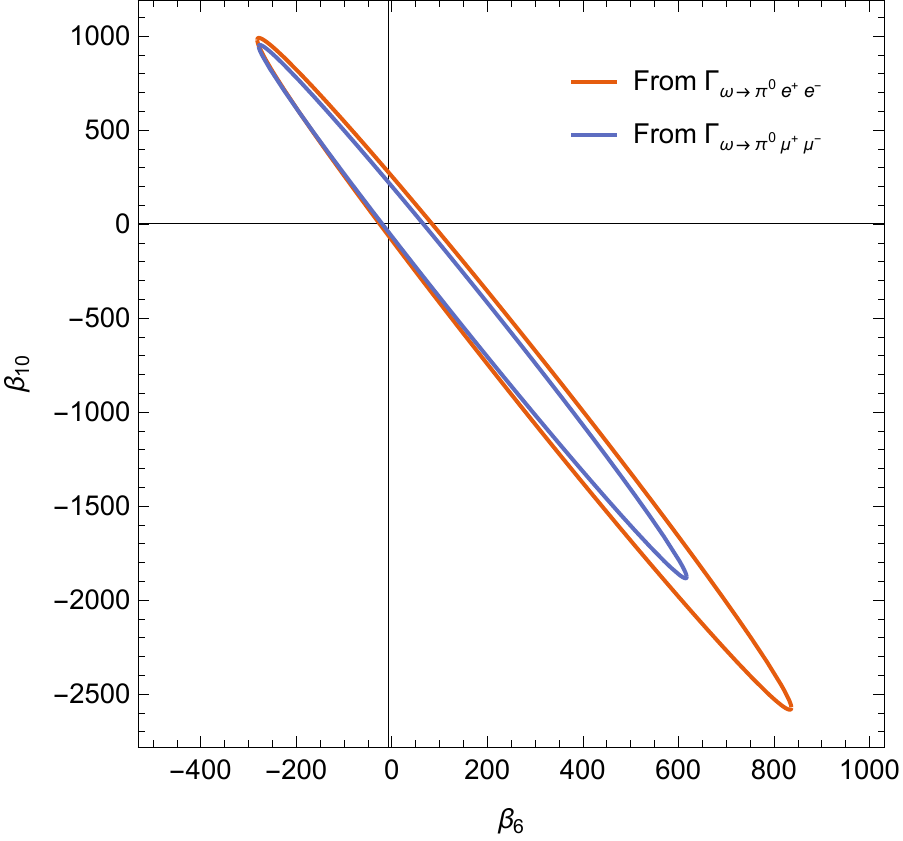}
\caption{A tentative plot of the mutual dependence of the values $\beta_{6}$ and $\beta_{10}$ obtained from eq.~\eqref{eq:decay_width_w_to_pil+l-}. The vertical and horizontal lines correspond to the central values of these parameters according to \eqref{eq:fit_result}. The error bands are omitted in all cases for simplicity.}
\label{fig:B6_B10_plot}
\end{figure}

Having the formula \eqref{eq:decay_width_w_to_pil+l-} at our disposal, one can simply substitute for the form factor \eqref{eq:form_factor_Vwp}, the respective lepton masses and the values of the parameters $\beta_{2}$, $\beta_{6}$ and $\beta_{10}$, according to \eqref{eq:fit_result}. Performing the integration is then a trivial task, so we present the final obtained results:
\begin{equation}
\Gamma_{\omega\,\to\,\pi^{0}e^{+}e^{-}}^{\mathrm{theory}}=(6.22\pm 0.53\pm 0.03)\cdot 10^{-6}\,\mathrm{GeV}
\end{equation}
and
\begin{equation}
\Gamma_{\omega\,\to\,\pi^{0}\mu^{+}\mu^{-}}^{\mathrm{theory}}=(8.38\pm 0.84\pm 0.09)\cdot 10^{-7}\,\mathrm{GeV}\,.
\end{equation}
While the first prediction is well-compatible with the experimental value, the compatibility of the second one is rather worse. 
Let us also discuss the following aspect. Suppose that we turn about the above consideration and instead of predicting the decay rates for $\omega\to\pi^{0}\ell^{+}\ell^{-}$, we  try to solve for the parameters $\beta_{6}$ and $\beta_{10}$ directly from the experimental values (\ref{eq:decay_width_w_to_pie+e-_value}), (\ref{eq:decay_width_w_to_pimu+mu-_value}) and the analytical expressions obtained from \eqref{eq:decay_width_w_to_pil+l-}. Such an approach is, however, quite sensitive to all the present experimental uncertainties and one obtains a rather wide variety of solutions covering a wide range in the $\beta_{6}$--$\beta_{10}$ plane --- many of them not being the relevant ones and far from our lattice fit. As an illustration, we show a graph at fig.~\ref{fig:B6_B10_plot}, where the mutual dependence of the values $\beta_{6}$ and $\beta_{10}$ is depicted. One can clearly see that a large stretch of both ellipses is close to each other and even, in two places, cross each other. Such proximity eventually leads to numerical uncertainties that are difficult to keep under control, and thus it prevents determining $\beta_{6}$ and $\beta_{10}$ from these two decays with satisfactory precision. 

\subsection{The parameter \texorpdfstring{$\chi^{(\mathrm{r})}$}{} and \texorpdfstring{$g-2$}{}}
As the last objective, we turn our attention to two phenomenologically important issues related to the pion transition form factor, namely the pion-pole contribution to the muon $g-2$ factor and the rare decay $\pi^{0}\to e^{+}e^{-}$.

\paragraph{The decay $\bm{\pi^{0}\to e^{+}e^{-}}$ and $\bm{\chi^{(\mathrm{r})}}$.}
The $\pi^{0}\to e^{+}e^{-}$ decay can be described at the leading order $\mathcal{O}(\alpha^2)$
by the triangle graph corresponding to the two-photon exchange.
This graph is UV finite since it includes a well-behaved nonlocal vertex given by the doubly off-shell pion transition form factor.
Alternatively, at the same order, we can calculate the amplitude of this process within ChPT, where the nonlocal vertex is replaced with the leading term of its  chiral expansion.
In this case, however, the loop integral is modified in the UV region and becomes divergent.
The finite part of the corresponding counterterm ${\chi^{(\mathrm{r})}}$ takes into account effectively the UV contribution of the original triangle graph with nonlocal vertex.
Its approximate form up to corrections proportional to $M_{e}/M_{\pi}$ and $M_{\pi}/\mu$ has been presented in ref.~\cite{Dorokhov:2007bd} and it reads
\begin{align}
\chi^{(\mathrm{r})}(\mu)\simeq\frac{5}{4}+\frac{3}{2}\int_{0}^{\infty}\left(\frac{\partial}{\partial t}\frac{\mathcal{F}_{\gamma^{\ast}\gamma^{\ast}\pi^{0}}(-t\mu^{2},-t\mu^{2})}{\mathcal{F}_{\gamma^{\ast}\gamma^{\ast}\pi^{0}}(0,0)}\right)\log{t}\,\mathrm{d}t\,,\label{eq:chi_r}
\end{align}
where the scale of $\mu=M_{\rho}$ is usually considered.

\begin{table}[t!]
\centering
\begin{tabular}{|C{3.3cm}|C{3.3cm}|C{3.3cm}|}
\hline
Model/data & $\chi^{(\mathrm{r})}(M_{\rho})$ & Reference \\ \hline\hline
LMD        & $2.29$                          & -        \\ \hline
VMD        & $2.87$                          & -        \\ \hline
THS        & $2.2\pm 0.2$                    & \cite{Husek:2015wta}        \\ \hline
LMD+V      & $2.5$                           & \multirow{4}{*}{\cite{Dorokhov:2007bd}} \\ \cline{1-2}
QCDsr      & $2.8\pm 0.1$                    &  \\ \cline{1-2}
CLEO+OPE   & $2.6\pm 0.3$                    &  \\ \cline{1-2}
N$\chi$QM  & $2.4\pm 0.5$                    &  \\ \hline
KTeV/$\pi^{0}\to e^{+}e^{-}$ & $4.5\pm 1.0$  & \cite{Husek:2014tna} \\ \hline
$K_{L}\to\mu^{+}\mu^{-}$ & $5.84\pm 0.20$    & \multirow{2}{*}{\cite{Cirigliano:2011ny}} \\ \cline{1-2}
$K_{L}\to\mu^{+}\mu^{-}$ & $8.07\pm 0.20$    &  \\ \hline
$\eta\to\mu^{+}\mu^{-}$ & $8.0\pm 0.9$       & \cite{GomezDumm:1998gw} \\ \hline\hline
This work  & $2.51\pm 0.03$                  & - \\ \hline
\end{tabular}
\caption{An overview of the values of the parameter $\chi^{(\mathrm{r})}(M_{\rho})$.}
\label{tab:chi_r_values}
\end{table}

Substituting for the pion transition form factor \eqref{eq:form_factor_dim16_v2} into \eqref{eq:chi_r} and performing the integration leads, unsurprisingly, to a complicated analytical expression in terms of $\beta_{2}$, $\beta_{6}$, $\beta_{10}$, $\delta^{2}$ and $M_{V_{3}}$. Nevertheless, after employing the results \eqref{eq:fit_result}  one obtains the estimate  of the value of $\chi^{(\mathrm{r})}(M_{\rho})$ --- see the table \ref{tab:chi_r_values} for the result\footnote{In table \ref{tab:chi_r_values}, we present for simplicity the result for mean values of $M_{V_3}$ and $\delta^2$. The uncertainty is almost entirely induced by  the error bars of the lattice fit of $\beta_i$ including the correlations, while the variation of the central value of $\chi^{(\mathrm{r})}(M_{\rho})$ with  $M_{V_3}$ and $\delta^2$ is tiny.} and for a comparison with other values obtained in various ways.
One can thus see that our estimate corresponds to lower values of $\chi^{(\mathrm{r})}(M_{\rho})$, with an agreement with ref.~\cite{Husek:2015wta} and unlike the conclusion made in \cite{Husek:2014tna}.

\begin{figure}[t!]
\small
\centering
\includegraphics[width=0.8\textwidth]{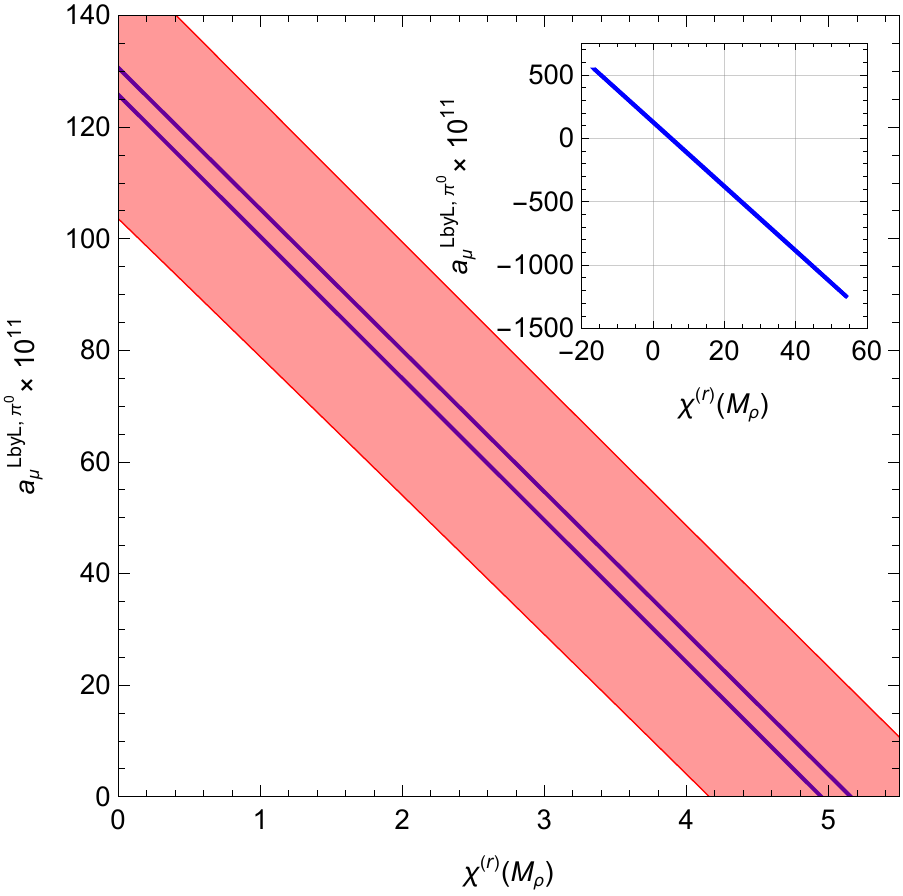}
\caption{A plot of the relation between the value of the pion-pole contribution to the muon $g-2$ on the effective parameter $\chi^{(\mathrm{r})}(M_{\rho})$ for $\beta_{6}$ and $\beta_{10}$ constrained by \eqref{eq:decay_width_w_to_pil+l-}. The graph over the whole range is given by a very narrow ellipse (the smaller plot), from which we have zoomed on the relevant area, where both $\chi^{(\mathrm{r})}(M_{\rho})$ and $g-2$ are positive. The central values of the $g-2$ are depicted in blue, while the error range is given by the red band.}
\label{fig:chiamu}
\end{figure}

\paragraph{The $\bm{g-2}$ factor.}
It is straightforward to calculate the contribution of the pion pole to the muon anomalous magnetic moment using the standard integral formula~\cite{Jegerlehner:2009ry}, which can be done for our model analytically. 
This allows us also to study the sensitivity of the result on the cutoff which determines the relevant integration range of the spacelike virtualities. From Tab.~\ref{tab:g2} follows that the contribution stabilized numerically around the cutoff 10~GeV.
The final number, combining all sources of uncertainty (i.e.~the statistical error, stemming from the fitted values of the parameters $\beta_{i}$, as well as the variation of the values $\delta^{2}$ and $M_{V_{3}}$) in quadrature is\footnote{Again, the variation with $M_{V_3}$ and $\delta^2$ is negligible.}
\begin{equation}
a_{\mu}^{\mathrm{LbyL},\,\pi^{0}} = (63.5\pm 0.8)\cdot 10^{-11}\,.\label{eq:g2_result}
\end{equation}
The estimate above should be compared with the lattice-based analysis in ref.~\cite{Gerardin:2019vio}. 
The authors provided therein two final results for the pion-pole contribution to the muon anomalous magnetic moment, that is $(59.7\pm 3.6)\cdot 10^{-11}$ and $(62.3\pm 2.3)\cdot 10^{-11}$. The first value corresponds to the model-independent lattice estimate, whilst the second one is based on the lattice and the normalization from PrimEx experiment \cite{PrimEx:2010fvg}. 
Note also that the state-of-the-art analysis based on the dispersive calculation \cite{Hoferichter:2018kwz} gives $62.6_{-2.5}^{+3.0}\cdot 10^{-11}$. 
We can see that the prediction of our model is compatible with all values summarized above. Note also that the final error is rather small. However, one should bear in mind that it does not include systematic errors which are difficult to be estimated, since the calculation is based solely on our three-multiplet-resonance model which can be biased. Therefore the error can be underestimated. Let us note that if no lattice data were used, the phenomenological input based on (\ref{eq:decay_width_w_to_pie+e-_value}), (\ref{eq:decay_width_w_to_pimu+mu-_value}) within our model can provide only the allowed region in the $a_{\mu}^{\mathrm{LbyL},\,\pi^{0}}$--$\chi_{\rm {r}}(M_\rho)$ plane --- see fig.~\ref{fig:chiamu}.
\begin{table}[t!]
\centering
\begin{tabular}{|C{2.5cm}|C{2.5cm}|}
\hline
cutoff $\mathrm{[GeV]}$  & $a_{\mu}^{\mathrm{LbyL},\,\pi^{0}}\times 10^{11}$ \\ \hline\hline
$1.5$    & $59.4$                          \\ \hline
$2$      & $61.1$                          \\ \hline
$3$      & $62.4$                          \\ \hline
$4$      & $62.9$                    \\ \hline
$5$      & $63.1$                  \\ \hline
$10$     & $63.4$                  \\ \hline
$100$ & $63.5$                  \\ \hline
$\infty$ & $63.5$                  \\ \hline
\end{tabular}
\caption{An overview of the central values of the pion-pole contribution to the muon anomalous magnetic moment as a function of the cutoff.}
\label{tab:g2}
\end{table}
%


\section{Summary}
We have investigated a specific class of the three-point Green functions of the odd-intrinsic parity sector of QCD within the Resonance chiral theory. We have focused on the order parameters of the chiral symmetry breaking in the chiral limit, namely the $\langle VVP\rangle$, $\langle VAS\rangle$ and $\langle AAP\rangle$. The other part of the odd-intrinsic parity sector consists of two anomalous three-point Green functions, the $\langle VVA\rangle$ and $\langle AAA\rangle$. Although their high-energy behaviour is available \cite{Kadavy:2020hox}, the analysis is more complicated due to the presence of the perturbative contributions. These include the logarithmic terms that cannot be treated within RChT with a finite number of resonances. We leave this problem and their study for future work.

For selected correlators, we have required their high-energy behaviour to match OPE with both the contributions of the quark and quark-gluon condensates. Based on the algebraical properties of correlators calculated within RChT, we have established the need to duplicate all the lowest vector, axial-vector, scalar and pseudoscalar resonance multiplets in the corresponding Lagrangian. Also, higher derivative operators have to be added.  Modifying the RChT in the mentioned way, the correlators were then evaluated and matching onto OPE was performed. This procedure eventually leads us to the constraints for the respective coupling constants of the resonance Lagrangians in the form of ``sum rules'' --- see the attached \mathematica notebook in \cite{KadavyThesis:2022}.

In order to satisfy additional high energy constraints on the top of the OPE mentioned above, we have investigated the $\langle VVP\rangle$ Green function in a particular resonance model when three exemplars of vector and pseudoscalar resonance multiplets are taken into account. Due to the enormous number of monomials and possible combinations, this has been done purely on an algebraic and not on a Lagrangian level. As a result, we have constructed the relevant double off-shell pion transition form factor $\mathcal{F}_{\gamma^{\ast}\gamma^{\ast}\pi^{0}}(p^{2},q^{2})$ that depends on three parameters (if one of the momenta is on-shell, only one parameter remains). Their values have been
fitted using the lattice data provided in ref.~\cite{Gerardin:2019vio}.
The consistency check of such a fit was performed by means of comparing the predictions of our model with the experimental data for the semi-on-shell form factor $Q^{2}\mathcal{F}_{\gamma^{\ast}\gamma^{\ast}\pi^{0}}(0,-Q^{2})$, or the  rates for the decays $\rho^{+}\to\pi^{+}\gamma$, $\omega\to\pi^{0}\gamma$, $\omega\to\pi^{0}e^{+}e^{-}$ and $\omega\to\pi^{0}\mu^{+}\mu^{-}$. With the exception of the latter process (where the deviation is 1.8$\sigma$), the data are well-compatible.
Then, we have obtained the predictions for the pion pole contribution to the
 muon $g-2$ factor (including the dependence on the cutoff) and the effective parameter $\chi^{(\mathrm{r})}$ related to the $\pi^0 \to e^+ e^-$ decay. The result for $g-2$  is compatible with previous state-of-the-art values \cite{Gerardin:2019vio} and \cite{Hoferichter:2018kwz}, while our prediction of $\chi^{(\mathrm{r})}$ supports rather lower values  \cite{Husek:2015wta} and {\cite{Dorokhov:2007bd}} then the higher ones \cite{Husek:2014tna}, {\cite{Cirigliano:2011ny}} and \cite{GomezDumm:1998gw}.


\acknowledgments
Symbolic computations have been performed with the use of \mathematica and \feyncalc \cite{Mertig:1990an,Shtabovenko:2016sxi}. The Feynman diagrams have been drawn using \jaxodraw \cite{Binosi:2008ig}.
We thank Tom\'a\v{s} Husek for useful discussions and comments.
We would like to also thank an anonymous referee for suggesting valuable improvements.
Our work was financially supported by The Czech Science Foundation (project GA\v{C}R no.~21-26574S).


\appendix


\section{On general structure of resonance contributions}\label{sec:struktura}
Here we gather some details on the discussion connected with taking into account two resonance multiplets of each kind, as we have briefly assessed in subsection \ref{ssec:general_structure}.

\paragraph{$\bm{\langle VVP\rangle}$ Green function.}
Let us start with writing down the general structure of the resonance contributions to the $\langle VVP\rangle$ Green function, which we subsequently test on the fulfilment of the conditions \eqref{eq:vvp_ope}, \eqref{eq:vvp_ope_2-op}, \eqref{eq:form_factor_BL} and \eqref{eq:form_factor_anomaly}.

\begin{itemize}
\item The first and the most natural general form of the resonance contribution to the $\langle VVP\rangle$ Green function corresponds to the LMD+P ansatz. We choose the corresponding parametrization of the Lorentz-invariant function to be
\begin{align}
\mathcal{F}_{VVP}(p^{2},q^{2},r^{2})=\frac{\mathcal{P}_{VVP}(p^{2},q^{2},r^{2})}{(p^{2}-M_{V}^{2})(q^{2}-M_{V}^{2})}\frac{B_{0}F^{2}}{r^{2}(r^{2}-M_{P}^{2})}\,,\label{eq:vvp_correlator_v0}
\end{align}
with the most general polynomial of the mass dimension $4$, that is Bose-symmetrical in the first two arguments, given as
\begin{align}
\mathcal{P}_{VVP}(p^{2},q^{2},r^{2})&=a_{1}+a_{2}(p^{2}+q^{2})+a_{3}r^{2}+a_{4}(p^{4}+q^{4})\nonumber\\
&+a_{5}(p^{2}+q^{2})r^{2}+a_{6}p^{2}q^{2}+a_{7}r^{4}\,.\label{eq:vvp_polynom_v0}
\end{align}
On the Lagrangian level within the context of the ref.~\cite{Kampf:2011ty}, ansatz \eqref{eq:vvp_correlator_v0} corresponds to the resonance contribution in eq.~(34) at page no.~13 therein. At this point, we are then familiar with the fact that \eqref{eq:vvp_correlator_v0} satisfies the matching onto the OPE with only the contribution of the quark condensate taken into account (see subsection 4.1 in \cite{Kampf:2011ty} or section \ref{sec:lowest_resonances} here). Nevertheless, including the contribution of the quark-gluon condensate, one easily finds out that \eqref{eq:vvp_correlator_v0} does not satisfy the matching onto the OPE \eqref{eq:vvp_ope}.
\item The next possibility is to include the second multiplet of vector resonances. Therefore, we thus denote $V_{1}\equiv V$, add the higher resonance state $V_{2}$ and write down the general ansatz as
\begin{align}
\mathcal{F}_{VVP}(p^{2},q^{2},r^{2})&=\frac{\mathcal{Q}_{VVP}(p^{2},q^{2},r^{2})}{(p^{2}-M_{V_{1}}^{2})(p^{2}-M_{V_{2}}^{2})(q^{2}-M_{V_{1}}^{2})(q^{2}-M_{V_{2}}^{2})}\nonumber\\
&\times\frac{B_{0}F^{2}}{r^{2}(r^{2}-M_{P}^{2})}\,,\label{eq:ths_correlator}
\end{align}
which corresponds to the two-hadron saturation (THS), as introduced in ref.~\cite{Husek:2015wta}. Here, the respective polynomial is of mass dimension $8$ and given by 22 terms. Its form reads
\begin{align}
\mathcal{Q}_{VVP}(p^{2},q^{2},r^{2})&=b_{1}+b_{2}(p^{2}+q^{2})+b_{3}r^{2}+b_{4}(p^{4}+q^{4})+b_{5}(p^{2}+q^{2})r^{2}+b_{6}p^{2}q^{2}\nonumber\\
&\hspace{-30pt}+b_{7}r^{4}+b_{8}(p^{6}+q^{6})+b_{9}(p^{4}+q^{4})r^{2}+b_{10}(p^{2}+q^{2})p^{2}q^{2}+b_{11}(p^{2}+q^{2})r^{4}\nonumber\\
&\hspace{-30pt}+b_{12}p^{2}q^{2}r^{2}+b_{13}r^{6}+b_{14}(p^{8}+q^{8})+b_{15}(p^{6}+q^{6})r^{2}+b_{16}(p^{4}+q^{4})p^{2}q^{2}\nonumber\\
&\hspace{-30pt}+b_{17}(p^{4}+q^{4})r^{4}+b_{18}(p^{2}+q^{2})p^{2}q^{2}r^{2}+b_{19}(p^{2}+q^{2})r^{6}+b_{20}p^{4}q^{4}\nonumber\\
&\hspace{-30pt}+b_{21}p^{2}q^{2}r^{4}+b_{22}r^{8}\,.\label{eq:ths_polynom}
\end{align}
We will now go through the required properties and find out whether these are satisfied.
\begin{itemize}
\item[1)] According to \cite{Husek:2015wta}, the THS ansatz \eqref{eq:ths_correlator} satisfies the matching onto the contribution of the quark condensate to the $\langle VVP\rangle$ correlator. Including the contribution of the quark-gluon condensate, we find out that the matching of \eqref{eq:ths_correlator} onto \eqref{eq:vvp_ope} can be fulfilled and the comparison leads to the following constraints for some of the unknown parameters:\footnote{We note that using the formula \eqref{eq:quark_condensate_v2}, the ratio $\langle\overline{q}q\rangle/B_{0}F^{2} = -3$.}
\begin{align}
b_{8}=\frac{3}{4}b_{13}=&-\frac{g_{s}\langle\overline{q}\sigma\hspace{-1pt}\cdot\hspace{-1pt}Gq\rangle}{24B_{0}F^{2}}\,,\notag\\
b_{9}=&-\frac{\langle\overline{q}q\rangle}{6B_{0}F^{2}}(M_{V_{1}}^{2}+M_{V_{2}}^{2})-\frac{g_{s}\langle\overline{q}\sigma\hspace{-1pt}\cdot\hspace{-1pt}Gq\rangle}{72B_{0}F^{2}}\,,\notag\\
b_{10}=&-\frac{\langle\overline{q}q\rangle}{6B_{0}F^{2}}M_{P}^{2}+\frac{g_{s}\langle\overline{q}\sigma\hspace{-1pt}\cdot\hspace{-1pt}Gq\rangle}{24B_{0}F^{2}}\,,\notag\\
b_{11}=&-\frac{\langle\overline{q}q\rangle}{6B_{0}F^{2}}(M_{V_{1}}^{2}+M_{V_{2}}^{2})\,,\notag\\
b_{12}=&-\frac{\langle\overline{q}q\rangle}{6B_{0}F^{2}}(2M_{V_{1}}^{2}+2M_{V_{2}}^{2}+M_{P}^{2})\,,\notag\\
b_{18}=b_{21}=&\quad\,\frac{\langle\overline{q}q\rangle}{6B_{0}F^{2}}\label{eq:bs}
\end{align}
and
\begin{equation}
b_{14}=b_{15}=b_{16}=b_{17}=b_{19}=b_{20}=b_{22}=0\,.
\end{equation}
The matching in question thus allowed us to restrict the number of unknown parameters from 22 to only 7. These are $b_{1},\ldots,b_{7}$, which are contained in the subleading terms with respect to such a matching and need to be obtained by other means.
\item[2)] Next, we insert the constraints \eqref{eq:bs} into \eqref{eq:ths_correlator} and scale the momenta according to \eqref{eq:vvp_ope_2-op}. We find out that the THS ansatz satisfies the OPE \eqref{eq:vvp_ope_2-op} automatically and no other relations for the unknown parameters are thus yet obtained.
\item[3)] The pion transition form factor \eqref{eq:form_factor}, within the THS ansatz \eqref{eq:ths_correlator}, takes a bit complicated form,
\begin{align}
\mathcal{F}_{\gamma^{\ast}\gamma^{\ast}\pi^{0}}(p^{2},q^{2})&=-\frac{1}{36B_{0}F(p^{2}-M_{V_{1}}^{2})(q^{2}-M_{V_{1}}^{2})(p^{2}-M_{V_{2}}^{2})(q^{2}-M_{V_{2}}^{2})M_{P}^{2}}\nonumber\\
&\hspace{-60pt}\times\Big[-g_{s}\langle\overline{q}\sigma\hspace{-1pt}\cdot\hspace{-1pt}Gq\rangle(p^{6}+q^{6})+\Big(g_{s}\langle\overline{q}\sigma\hspace{-1pt}\cdot\hspace{-1pt}Gq\rangle-4M_{P}^{2}\langle\overline{q}q\rangle\Big)p^{2}q^{2}(p^{2}+q^{2})\nonumber\\
&\hspace{-42pt}+24B_{0}F^{2}\Big(b_{1}+b_{2}(p^{2}+q^{2})+b_{4}(p^{4}+q^{4})+b_{6}p^{2}q^{2}\Big)\Big]\,,\label{eq:THS_form_factor}
\end{align}
which, however, simplifies a lot after inserting the respective kinematics:
\begin{align}
\mathcal{F}_{\gamma^{\ast}\gamma^{\ast}\pi^{0}}(0,-Q^{2})&=-\frac{24B_{0}F^{2}(b_{1}-b_{2}Q^{2}+b_{4}Q^{4})+g_{s}\langle\overline{q}\sigma\hspace{-1pt}\cdot\hspace{-1pt}Gq\rangle Q^{6}}{36B_{0}FM_{V_{1}}^{2}M_{V_{2}}^{2}M_{P}^{2}(Q^{2}+M_{V_{1}}^{2})(Q^{2}+M_{V_{2}}^{2})}\,.\label{eq:THS_form_factor_v2}
\end{align}
For $Q^{2}\rightarrow\infty$, the form factor \eqref{eq:THS_form_factor_v2} behaves as
\begin{align}
\mathcal{F}_{\gamma^{\ast}\gamma^{\ast}\pi^{0}}(0,-Q^{2})=&-\frac{24B_{0}F^{2}b_{4}-g_{s}\langle\overline{q}\sigma\hspace{-1pt}\cdot\hspace{-1pt}Gq\rangle(M_{V_{1}}^{2}+M_{V_{2}}^{2})}{36B_{0}F M_{V_{1}}^{2} M_{V_{2}}^{2}M_{P}^{2}}\nonumber\\
&-\frac{g_{s}\langle\overline{q}\sigma\hspace{-1pt}\cdot\hspace{-1pt}Gq\rangle}{36B_{0}F M_{V_{1}}^{2}M_{V_{2}}^{2}M_{P}^{2}}Q^{2}+\mathcal{O}\bigg(\frac{1}{Q^{2}}\bigg)\,.\label{eq:THS_form_factor_v3}
\end{align}
The Brodsky--Lepage behaviour \eqref{eq:form_factor_BL} of the form factor, however, requires the polynomial structure in front of the residual part proportional to $1/Q^{2}$ to vanish. This can not be achieved without setting the value of the quark-gluon condensate to zero. We thus conclude that the THS ansatz \eqref{eq:ths_correlator} is not a suitable choice for our case.
\end{itemize}
\item The natural extension of the THS ansatz is an addition of another multiplet of pseudoscalar resonances. Similarly, as in the previous case, we denote $P_{1}\equiv P$, and the higher resonance state $P_{2}$. We further suggest calling such an ansatz simply as THS+P, which then reads
\begin{align}
\mathcal{F}_{VVP}(p^{2},q^{2},r^{2})&=\frac{\mathcal{R}_{VVP}(p^{2},q^{2},r^{2})}{(p^{2}-M_{V_{1}}^{2})(p^{2}-M_{V_{2}}^{2})(q^{2}-M_{V_{1}}^{2})(q^{2}-M_{V_{2}}^{2})}\nonumber\\
&\times\frac{B_{0}F^{2}}{r^{2}(r^{2}-M_{P_{1}}^{2})(r^{2}-M_{P_{2}}^{2})}\,,\label{eq:vvp_correlator_v2}
\end{align}
with the respective polynomial of mass dimension $10$, given by 34 individual Bose-symmetrical terms as
\begin{align}
\mathcal{R}_{VVP}(p^{2},q^{2},r^{2})&=c_{1}+c_{2}(p^{2}+q^{2})+c_{3}r^{2}+c_{4}(p^{4}+q^{4})+c_{5}(p^{2}+q^{2})r^{2}+c_{6}p^{2}q^{2}\nonumber\\
&\hspace{-30pt}+c_{7}r^{4}+c_{8}(p^{6}+q^{6})+c_{9}(p^{4}+q^{4})r^{2}+c_{10}(p^{2}+q^{2})p^{2}q^{2}+c_{11}(p^{2}+q^{2})r^{4}\nonumber\\
&\hspace{-30pt}+c_{12}p^{2}q^{2}r^{2}+c_{13}r^{6}+c_{14}(p^{8}+q^{8})+c_{15}(p^{6}+q^{6})r^{2}+c_{16}(p^{4}+q^{4})p^{2}q^{2}\nonumber\\
&\hspace{-30pt}+c_{17}(p^{4}+q^{4})r^{4}+c_{18}(p^{2}+q^{2})p^{2}q^{2}r^{2}+c_{19}(p^{2}+q^{2})r^{6}+c_{20}p^{4}q^{4}\nonumber\\
&\hspace{-30pt}+c_{21}p^{2}q^{2}r^{4}+c_{22}r^{8}+c_{23}(p^{10}+q^{10})+c_{24}(p^{8}+q^{8})r^{2}+c_{25}(p^{6}+q^{6})p^{2}q^{2}\nonumber\\
&\hspace{-30pt}+c_{26}(p^{6}+q^{6})r^{4}+c_{27}(p^{4}+q^{4})p^{2}q^{2}r^{2}+c_{28}(p^{4}+q^{4})r^{6}+c_{29}(p^{2}+q^{2})p^{4}q^{4}\nonumber\\
&\hspace{-30pt}+c_{30}(p^{2}+q^{2})p^{2}q^{2}r^{4}+c_{31}(p^{2}+q^{2})r^{8}+c_{32}p^{4}q^{4}r^{2}+c_{33}p^{2}q^{2}r^{6}\nonumber\\
&\hspace{-30pt}+c_{34}r^{10}\,.\label{eq:vvp_polynom}
\end{align}
\begin{itemize}
\item[1)] Matching of \eqref{eq:vvp_correlator_v2} onto \eqref{eq:vvp_ope} is indeed possible and the comparison leads to the following matching conditions:
\begin{align}
\frac{4}{3}c_{15}=c_{22}=&-\frac{g_{s} \langle\overline{q}\sigma\hspace{-1pt}\cdot\hspace{-1pt}Gq\rangle}{18B_{0}F^{2}}\,,\notag\\
c_{17}=&-\frac{\langle\overline{q}q\rangle}{6B_{0}F^{2}}(M_{V_{1}}^{2}+M_{V_{2}}^{2})-\frac{g_{s} \langle\overline{q}\sigma\hspace{-1pt}\cdot\hspace{-1pt}Gq\rangle}{72B_{0}F^{2}}\,,\notag\\
c_{18}=&-\frac{\langle\overline{q}q\rangle}{6B_{0}F^{2}}(M_{P_{1}}^{2}+M_{P_{2}}^{2})+\frac{g_{s} \langle\overline{q}\sigma\hspace{-1pt}\cdot\hspace{-1pt}Gq\rangle}{24B_{0}F^{2}}\,,\notag\\
c_{19}=&-\frac{\langle\overline{q}q\rangle}{6B_{0}F^{2}}(M_{V_{1}}^{2}+M_{V_{2}}^{2})\,,\notag\\
c_{21}=&-\frac{\langle\overline{q}q\rangle}{6B_{0}F^{2}}(2M_{V_{1}}^{2}+2M_{V_{2}}^{2}+M_{P_{1}}^{2}+M_{P_{2}}^{2})\,,\notag\\
c_{30}=c_{33}=&\quad\,\frac{\langle\overline{q}q\rangle}{6B_{0}F^{2}}\label{eq:vvp_polynom_constraints}
\end{align}
and
\begin{equation}
c_{14}=c_{16}=c_{20}=c_{23}=c_{24}=c_{25}=c_{26}=c_{27}=c_{28}=c_{29}=c_{31}=c_{32}=c_{34}=0\,.
\end{equation}
Once again, we see that from 34 unknown parameters, only 13 are left undetermined. These are $c_{1},\ldots,c_{13}$ and they will be subjected to other means of restrictions.
\item[2)] We now insert the obtained constraints \eqref{eq:vvp_polynom_constraints} into \eqref{eq:vvp_correlator_v2} and try to recover the OPE \eqref{eq:vvp_ope_2-op} by simply scaling the ansatz \eqref{eq:vvp_correlator_v2} accordingly. A straightforward calculation leads to
\begin{align}
\mathcal{F}_{VVP}\Big((\lambda p)^{2}&,(q-\lambda p)^{2},q^{2}\Big)=-\frac{\langle\overline{q}q\rangle}{3\lambda^{2}}\frac{1}{p^{2}q^{2}}\nonumber\\
&\times\frac{2(c_{8}+c_{10})+(M_{P_{1}}^{2}+M_{P_{2}}^{2})q^{2}-q^{4}}{(q^{2}-M_{P_{1}}^{2})(q^{2}-M_{P_{2}}^{2})}+\mathcal{O}\bigg(\frac{1}{\lambda^{3}}\bigg)\,.\label{eq:vvp_ope_2-op_THSP}
\end{align}
Once again, to be able to match this onto \eqref{eq:vvp_ope_2-op}, the remainder of a division of the numerator of \eqref{eq:vvp_ope_2-op_THSP} by $(q^{2}-M_{P_{1}}^{2})(q^{2}-M_{P_{2}}^{2})$, treated as polynomial in $q^{2}$, must be zero. This then easily leads to
\begin{equation}
c_{10}=-c_{8}-\frac{M_{P_{1}}^{2}M_{P_{2}}^{2}}{2}\,,\label{eq:vvp_ope_2-op_THSP_v2}
\end{equation}
which further reduces the number of unknown parameters to 12.
\item[3)] The third constraint is the expected behaviour of the pion transition form factor. According to \eqref{eq:form_factor}, it takes the form
\begin{align}
\mathcal{F}_{\gamma^{\ast}\gamma^{\ast}\pi^{0}}(p^{2},q^{2})&=\frac{2F}{3M_{P_{1}}^{2}M_{P_{2}}^{2}}\label{eq:THSP_form_factor}\\
&\hspace{-65pt}\times\frac{c_{1}+c_{2}(p^{2}+q^{2})+c_{4}(p^{4}+q^{4})+c_{6}p^{2}q^{2}+c_{8}(p^{6}+q^{6})+c_{10}p^{2}q^{2}(p^{2}+q^{2})}{(p^{2}-M_{V_{1}}^{2})(p^{2}-M_{V_{2}}^{2})(q^{2}-M_{V_{1}}^{2})(q^{2}-M_{V_{2}}^{2})}\,,\nonumber
\end{align}
where we have not yet intentionally inserted the constraint \eqref{eq:vvp_ope_2-op_THSP_v2}, which will become clear soon. Employing the desired kinematical variables, we get
\begin{align}
\mathcal{F}_{\gamma^{\ast}\gamma^{\ast}\pi^{0}}(0,-Q^{2})&=\frac{2F(c_{1}-c_{2}Q^{2}+c_{4}Q^{4}-c_{8}Q^{6})}{3M_{V_{1}}^{2}M_{V_{2}}^{2}M_{P_{1}}^{2}M_{P_{2}}^{2}(Q^{2}+M_{V_{1}}^{2})(Q^{2}+M_{V_{2}}^{2})}\nonumber\\
&=\frac{2F}{3}\frac{c_{4}+c_{8}(M_{V_{1}}^{2}+M_{V_{2}}^{2}-Q^{2})}{M_{V_{1}}^{2}M_{V_{2}}^{2}M_{P_{1}}^{2}M_{P_{2}}^{2}}+\mathcal{O}\bigg(\frac{1}{Q^{2}}\bigg)\label{eq:THSP_form_factor_v2}
\end{align}
for $Q^{2}\rightarrow\infty$. Naturally, the polynomial part in front of the part proportional to $1/Q^{2}$ must vanish so that
\begin{equation}
c_{4}=c_{8}=0\,.\label{eq:vvp_ope_2-op_THSP_v3}
\end{equation}
Employing \eqref{eq:vvp_ope_2-op_THSP_v3} into \eqref{eq:THSP_form_factor_v2}, the form factor reduces to
\begin{align}
\mathcal{F}_{\gamma^{\ast}\gamma^{\ast}\pi^{0}}(0,-Q^{2})&=\frac{2F(c_{1}-c_{2}Q^{2})}{3M_{V_{1}}^{2}M_{V_{2}}^{2}M_{P_{1}}^{2}M_{P_{2}}^{2}(Q^{2}+M_{V_{1}}^{2})(Q^{2}+M_{V_{2}}^{2})}\,,\label{eq:THSP_form_factor_v3}
\end{align}
i.e.
\begin{equation}
\lim_{Q^{2}\,\rightarrow\,\infty}Q^{2}\mathcal{F}_{\gamma^{\ast}\gamma^{\ast}\pi^{0}}(0,-Q^{2})=-\frac{2Fc_{2}}{3M_{V_{1}}^{2}M_{V_{2}}^{2}M_{P_{1}}^{2}M_{P_{2}}^{2}}\,.\label{eq:THSP_form_factor_v4}
\end{equation}
Comparison of the right-hand sides of \eqref{eq:THSP_form_factor_v4} and \eqref{eq:form_factor_BL} thus gives us
\begin{equation}
c_{2}=-3M_{V_{1}}^{2}M_{V_{2}}^{2}M_{P_{1}}^{2}M_{P_{2}}^{2}\,.\label{eq:vvp_ope_2-op_THSP_v4}
\end{equation}
At this point, we are left with 9 unknown constants.
\item[4)] The last condition the form factor is obliged to satisfy is the recovery of the anomaly for vanishing momenta. We get
\begin{align}
\mathcal{F}_{\gamma^{\ast}\gamma^{\ast}\pi^{0}}(0,0)&=\frac{2Fc_{1}}{3M_{V_{1}}^{4}M_{V_{2}}^{4}M_{P_{1}}^{2}M_{P_{2}}^{2}}\,,\label{eq:THSP_form_factor_v5}
\end{align}
which, after comparison with \eqref{eq:form_factor_anomaly}, gives
\begin{equation}
c_{1}=\frac{N_{c}}{8\pi^{2}F^{2}}M_{V_{1}}^{4}M_{V_{2}}^{4}M_{P_{1}}^{2}M_{P_{2}}^{2}\,,\label{eq:vvp_ope_2-op_THSP_v6}
\end{equation}
which leaves us with 8 undetermined parameters that the THS+P ansatz \eqref{eq:vvp_correlator_v2} eventually depends on.
\end{itemize}
\end{itemize}

The summary of the discussion above is thus that in order to satisfy the theoretical and phenomenological requirements \eqref{eq:vvp_ope}, \eqref{eq:vvp_ope_2-op}, \eqref{eq:form_factor_BL} and \eqref{eq:form_factor_anomaly} of the $\langle VVP\rangle$ Green function, it is demanded to duplicate both the vector and pseudoscalar resonance multiplets. Nevertheless, the situation is not quite in favour of finding genuinely new prospects. To this end, we may substitute the obtained constraints on the parameters $c_{i}$ back into the pion transition form factor \eqref{eq:THSP_form_factor}. We then get
\begin{align}
&\mathcal{F}_{\gamma^{\ast}\gamma^{\ast}\pi^{0}}(p^{2},q^{2})=\frac{N_{c}}{12\pi^{2}F}\frac{M_{V_{1}}^{4}M_{V_{2}}^{4}}{(p^{2}-M_{V_{1}}^{2})(p^{2}-M_{V_{2}}^{2})(q^{2}-M_{V_{1}}^{2})(q^{2}-M_{V_{2}}^{2})}\nonumber\\
&\hspace{-5pt}\times\left[1-\frac{4\pi^{2}F^{2}}{N_{c}}\frac{p^{2}+q^{2}}{M_{V_{1}}^{2}M_{V_{2}}^{2}}\left(6+\frac{p^{2}q^{2}}{M_{V_{1}}^{2}M_{V_{2}}^{2}}\right)+\frac{8\pi^{2}F^{2}}{N_{c}}\frac{c_{6}p^{2}q^{2}}{M_{V_{1}}^{4}M_{V_{2}}^{4}M_{P_{1}}^{2}M_{P_{2}}^{2}}\right],\label{eq:THSP_form_factor_final}
\end{align}
which coincides with the THS form factor, see eq.~(17) in \cite{Husek:2015wta}, upon the conversion relation
\begin{equation}
c_{6}=\frac{M_{V_{1}}^{4}M_{V_{2}}^{4}M_{P_{1}}^{2}M_{P_{2}}^{2}}{(4\pi F)^{6}}\kappa
\end{equation}
is utilized, with $\kappa$ being a dimensionless parameter used in ref.~\cite{Husek:2015wta}.\footnote{In detail, the authors of ref.~\cite{Husek:2015wta} have obtained its value to be $\kappa=21\pm 3$, see eq.~(31) therein} This fact is a fairly remarkable feature of our THS+P parametrization --- although we have taken into account two multiplets of both vector and pseudoscalar resonances at once, one ultimately recovers exactly the THS form factor, where the second pseudoscalar multiplet of resonances is absent from the beginning.

\paragraph{$\bm{\langle VAS\rangle}$ Green function.}
Here, we will construct a rational ansatz being able to satisfy the OPE \eqref{eq:vas_ope}.

\begin{itemize}
\item We start with the easiest case of the $\langle VAS\rangle$ correlator, given by an ansatz
\begin{align}
\mathcal{F}_{VAS}(p^{2},q^{2},r^{2})=\frac{\mathcal{P}_{VAS}(p^{2},q^{2},r^{2})}{(p^{2}-M_{V}^{2})(q^{2}-M_{A}^{2})}\frac{B_{0}F^{2}}{(r^{2}-M_{S}^{2})}\,,\label{eq:vas_correlator_v0}
\end{align}
where the most general polynomial of mass dimension $2$ is simply
\begin{align}
\mathcal{P}_{VAS}(p^{2},q^{2},r^{2})=d_{1}+d_{2}p^{2}+d_{3}q^{2}+d_{4}r^{2}\,.\label{eq:vas_polynom_v0}
\end{align}
Since \eqref{eq:vas_correlator_v0} essentially corresponds to the resonance contribution as presented in ref.~\cite{Kampf:2011ty}, we know that it satisfies the OPE contribution given by the quark condensate (see eqs.~(81) and (82) at page no.~21 therein or appendix \ref{sec:lowest_resonances} here). Nevertheless, including the contribution of the quark-gluon condensate, one ascertains that \eqref{eq:vas_correlator_v0} does not satisfy the matching onto the OPE \eqref{eq:vas_ope}, similarly as in the case of the $\langle VVP\rangle$ correlator.
\item The obvious possibility, based on the approach chosen in the previous case, is to include other multiplets of resonances. There are several possibilities of such an extension for the $\langle VAS\rangle$ correlator. We thus denote $V_{1}\equiv V$, $A_{1}\equiv A$, $S_{1}\equiv S$ and try to add different combinations of higher resonances. In this step we will consider only two possibilities of duplicating the resonances. We can either duplicate only one type of resonances at once or we can take into account such a duplication for two different types of resonance multiplets. Examples of these possibilities can be schematically written down as
\begin{align}
\mathcal{F}_{VAS}(p^{2},q^{2},r^{2})\sim\frac{1}{(p^{2}-M_{V_{1}}^{2})(p^{2}-M_{V_{2}}^{2})(q^{2}-M_{A_{1}}^{2})(r^{2}-M_{S_{1}}^{2})}\label{eq:vas_correlator_v1}
\end{align}
and
\begin{align}
\mathcal{F}_{VAS}(p^{2},q^{2},r^{2})\sim\frac{1}{(p^{2}-M_{V_{1}}^{2})(p^{2}-M_{V_{2}}^{2})(q^{2}-M_{A_{1}}^{2})(q^{2}-M_{A_{2}}^{2})(r^{2}-M_{S_{1}}^{2})}\,,\label{eq:vas_correlator_v2}
\end{align}
with the corresponding polynomial $\mathcal{Q}_{VAS}(p^{2},q^{2},r^{2})$ having the mass dimension of $4$ for \eqref{eq:vas_correlator_v1} and $6$ for \eqref{eq:vas_correlator_v2}, that would consist of 10 and 20 terms, respectively, with corresponding parameters $e_{i}$. It is important to mention that the parametrizations \eqref{eq:vas_correlator_v1} and \eqref{eq:vas_correlator_v2} have three combinations each, depending on which resonances are being duplicated. The procedure of finding out which combination of parametrizations satisfies the OPE \eqref{eq:vas_ope} is quite lengthy and the results is that none of them is a suitable choice.
\item The natural advance is thus to take into account duplication of all three types of resonance multiplets and write down the ansatz in the form
\begin{align}
\mathcal{F}_{VAS}(p^{2},q^{2},r^{2})&=\frac{\mathcal{R}_{VAS}(p^{2},q^{2},r^{2})}{(p^{2}-M_{V_{1}}^{2})(p^{2}-M_{V_{2}}^{2})(q^{2}-M_{A_{1}}^{2})(q^{2}-M_{A_{2}}^{2})}\nonumber\\
&\times\frac{B_{0}F^{2}}{(r^{2}-M_{S_{1}}^{2})(r^{2}-M_{S_{2}}^{2})}\,,\label{eq:vas_correlator}
\end{align}
with the polynomial consisting of 35 terms and given as
\begin{align}
\mathcal{R}_{VAS}(p^{2},q^{2},r^{2})&=f_{1}+f_{2}p^{2}+f_{3}q^{2}+f_{4}r^{2}+f_{5}p^{4}+f_{6}p^{2} q^{2}+f_{7}p^{2}r^{2}+f_{8}q^{4}+f_{9}q^{2}r^{2}\nonumber\\
&\hspace{-50pt}+f_{10}r^{4}+f_{11}p^{6}+f_{12}p^{4}q^{2}+f_{13}p^{4}r^{2}+f_{14}p^{2}q^{4}+f_{15}p^{2}q^{2}r^{2}+f_{16}p^{2}r^{4}+f_{17}q^{6}\nonumber\\
&\hspace{-50pt}+f_{18}q^{4}r^{2}+f_{19}q^{2}r^{4}+f_{20}r^{6}+f_{21}p^{8}+f_{22}p^{6}q^{2}+f_{23}p^{6}r^{2}+f_{24}p^{4}q^{4}+f_{25}p^{4}q^{2}r^{2}\nonumber\\
&\hspace{-50pt}+f_{26}p^{4}r^{4}+f_{27}p^{2}q^{6}+f_{28}p^{2}q^{4}r^{2}+f_{29}p^{2}q^{2}r^{4}+f_{30}p^{2}r^{6}+f_{31}q^{8}+f_{32}q^{6}r^{2}\nonumber\\
&\hspace{-50pt}+f_{33}q^{4}r^{4}+f_{34}q^{2}r^{6}+f_{35}r^{8}\,.\label{eq:vas_polynom}
\end{align}
Such a choice is finally the one that fulfils the matching onto \eqref{eq:vas_ope}, which gives us the following constraints on the unknown parameters:
\begin{align}
-f_{11}=f_{17}=\frac{3}{4}f_{20}&=\frac{g_{s}\langle\overline{q}\sigma\hspace{-1pt}\cdot\hspace{-1pt}Gq\rangle}{24B_{0}F^{2}}\,,\notag\\
-f_{12}=f_{14}&=\frac{\langle\overline{q}q\rangle}{6B_{0}F^{2}}(M_{S_{1}}^{2}+M_{S_{2}}^{2})-\frac{g_{s}\langle\overline{q}\sigma\hspace{-1pt}\cdot\hspace{-1pt}Gq\rangle}{24B_{0}F^{2}}\,,\notag\\
-f_{13}&=\frac{\langle\overline{q}q\rangle}{6B_{0}F^{2}}(M_{A_{1}}^{2}+M_{A_{2}}^{2})+\frac{g_{s}\langle\overline{q}\sigma\hspace{-1pt}\cdot\hspace{-1pt}Gq\rangle}{72B_{0}F^{2}}\,,\notag\\
-f_{15}&=\frac{\langle\overline{q}q\rangle}{6B_{0}F^{2}}(M_{V_{1}}^{2}+M_{V_{2}}^{2}-M_{A_{1}}^{2}-M_{A_{2}}^{2}-M_{S_{1}}^{2}-M_{S_{2}}^{2})\,,\notag\\
f_{16}&=\frac{\langle\overline{q}q\rangle}{6B_{0}F^{2}}(M_{A_{1}}^{2}+M_{A_{2}}^{2})\,,\notag\\
f_{18}&=\frac{\langle\overline{q}q\rangle}{6B_{0}F^{2}}(M_{V_{1}}^{2}+M_{V_{2}}^{2})+\frac{g_{s}\langle\overline{q}\sigma\hspace{-1pt}\cdot\hspace{-1pt}Gq\rangle}{72B_{0}F^{2}}\,,\notag\\
f_{19}&=\frac{\langle\overline{q}q\rangle}{6B_{0}F^{2}}(M_{V_{1}}^{2}+M_{V_{2}}^{2})\,,\notag\\
f_{25}=-f_{28}=-f_{29}&=\frac{\langle\overline{q}q\rangle}{6B_{0}F^{2}}\label{eq:vas_polynom_constraints}
\end{align}
and
\begin{equation}
f_{21}=f_{22}=f_{23}=f_{24}=f_{26}=f_{27}=f_{30}=f_{31}=f_{32}=f_{33}=f_{34}=f_{35}=0\,.
\end{equation}
\end{itemize}

To conclude, similarly as in the previous case, the matching of the resonance contribution to the $\langle VAS\rangle$ onto the OPE \eqref{eq:vas_ope} can be performed if all the resonance multiplets are at least duplicated.

\paragraph{$\bm{\langle AAP\rangle}$ Green function.}
Let us propose the respective ans\"{a}tze and find out whether the two conditions \eqref{eq:aap_ope}  and \eqref{eq:aap_ope_2-op} are met. This can be seen as unnecessary since, from the Lagrangian point of view, we are now aware of the fact that the duplication of both the axial-vector and pseudoscalar resonances is inevitable. In spite of this, we provide a shortened discussion, similarly to the cases of previous correlators.

\begin{itemize}
\item Due to the apparent resemblance to the $\langle VVP\rangle$, it is most likely not surprising that the parametrization of the $\langle AAP\rangle$ Green function, equivalent to the ansatz \eqref{eq:vvp_correlator_v0}, can not be matched onto \eqref{eq:aap_ope}. Similarly as in the previous cases, such a matching would be feasible if the contribution of the quark-gluon condensate is omitted (see appendix \ref{sec:lowest_resonances} for details). We thus consider right away the duplication of the axial-vector resonances and move on to the parametrization of the type \eqref{eq:ths_correlator}. Once again, it reads
\begin{align}
\mathcal{F}_{AAP}(p^{2},q^{2},r^{2})&=\frac{\mathcal{Q}_{AAP}(p^{2},q^{2},r^{2})}{(p^{2}-M_{A_{1}}^{2})(p^{2}-M_{A_{2}}^{2})(q^{2}-M_{A_{1}}^{2})(q^{2}-M_{A_{2}}^{2})}\nonumber\\
&\times\frac{B_{0}F^{2}}{r^{2}(r^{2}-M_{P}^{2})}\,,\label{eq:ths_correlator_aap}
\end{align}
with the polynomial $\mathcal{Q}_{AAP}(p^{2},q^{2},r^{2})$ given by the same form as \eqref{eq:ths_polynom}, where the respective constants $b_{i}$ are changed for $g_{i}$ due to consistency of the notation.
\begin{itemize}
\item[1)] The general resonance contribution \eqref{eq:ths_correlator_aap} can be matched onto \eqref{eq:aap_ope} and the following constraints on the respective constants can be obtained:
\begin{align}
g_{8}=-\frac{3}{4}g_{13}=&-\frac{g_{s}\langle\overline{q}\sigma\hspace{-1pt}\cdot\hspace{-1pt}Gq\rangle}{24B_{0}F^{2}}\,,\notag\\
g_{9}=&-\frac{\langle\overline{q}q\rangle}{6B_{0}F^{2}}(M_{A_{1}}^{2}+M_{A_{2}}^{2})-\frac{g_{s}\langle\overline{q}\sigma\hspace{-1pt}\cdot\hspace{-1pt}Gq\rangle}{72B_{0}F^{2}}\,,\notag\\
g_{10}=&-\frac{\langle\overline{q}q\rangle}{6B_{0}F^{2}}M_{P}^{2}+\frac{g_{s}\langle\overline{q}\sigma\hspace{-1pt}\cdot\hspace{-1pt}Gq\rangle}{24B_{0}F^{2}}\,,\notag\\
g_{11}=&\quad\,\frac{\langle\overline{q}q\rangle}{6B_{0}F^{2}}(M_{A_{1}}^{2}+M_{A_{2}}^{2})\,,\notag\\
g_{12}=&-\frac{\langle\overline{q}q\rangle}{6B_{0}F^{2}}(2M_{A_{1}}^{2}+2M_{A_{2}}^{2}-M_{P}^{2})\,,\notag\\
g_{18}=-g_{21}=&\quad\,\frac{\langle\overline{q}q\rangle}{6B_{0}F^{2}}\label{eq:AAP_gs}
\end{align}
and
\begin{equation}
g_{14}=g_{15}=g_{16}=g_{17}=g_{19}=g_{20}=g_{22}=0\,,
\end{equation}
which leaves us with 7 unknown parameters $g_{1},\ldots,g_{7}$.
\item[2)] Inserting the expressions \eqref{eq:AAP_gs} into \eqref{eq:ths_correlator_aap}, we find out that the OPE \eqref{eq:aap_ope_2-op} is fulfilled automatically, similarly as in the respective case of the $\langle VVP\rangle$ correlator.
\end{itemize}
\item Let us now finally discuss the case of the general ansatz of the resonance contribution to the $\langle AAP\rangle$ Green function with both the axial-vector and pseudoscalar resonances duplicated. Its structure is equivalent to the ansatz \eqref{eq:vvp_correlator_v2} and reads
\begin{align}
\mathcal{F}_{AAP}(p^{2},q^{2},r^{2})&=\frac{\mathcal{R}_{AAP}(p^{2},q^{2},r^{2})}{(p^{2}-M_{A_{1}}^{2})(p^{2}-M_{A_{2}}^{2})(q^{2}-M_{A_{1}}^{2})(q^{2}-M_{A_{2}}^{2})}\nonumber\\
&\times\frac{B_{0}F^{2}}{r^{2}(r^{2}-M_{P_{1}}^{2})(r^{2}-M_{P_{2}}^{2})}\,,\label{eq:aap_correlator}
\end{align}
where the polynomial $\mathcal{R}_{AAP}(p^{2},q^{2},r^{2})$ is given by \eqref{eq:vvp_polynom}, with the respective constants $c_{i}$ changed for $h_{i}$.
\begin{itemize}
\item[1)] The matching of the structure \eqref{eq:aap_correlator} onto \eqref{eq:aap_ope} can be performed easily and the following conditions can be found:
\begin{align}
h_{15}=-\frac{3}{4}h_{22}=&-\frac{g_{s}\langle\overline{q}\sigma\hspace{-1pt}\cdot\hspace{-1pt}Gq\rangle}{24B_{0}F^{2}}\,,\notag\\
h_{17}=&-\frac{\langle\overline{q}q\rangle}{6B_{0}F^{2}}(M_{A_{1}}^{2}+M_{A_{2}}^{2})-\frac{g_{s} \langle\overline{q}\sigma\hspace{-1pt}\cdot\hspace{-1pt}Gq\rangle}{72B_{0}F^{2}}\,,\notag\\
h_{18}=&-\frac{\langle\overline{q}q\rangle}{6B_{0}F^{2}}(M_{P_{1}}^{2}+M_{P_{2}}^{2})+\frac{g_{s}\langle\overline{q}\sigma\hspace{-1pt}\cdot\hspace{-1pt}Gq\rangle}{24B_{0}F^{2}}\,,\notag\\
h_{19}=&\quad\,\frac{\langle\overline{q}q\rangle}{6B_{0}F^{2}}(M_{A_{1}}^{2}+M_{A_{2}}^{2})\,,\notag\\
h_{21}=&-\frac{\langle\overline{q}q\rangle}{6B_{0}F^{2}}(2 M_{A_{1}}^{2}+2 M_{A_{2}}^{2}-M_{P_{1}}^{2}-M_{P_{2}}^{2})\,,\notag\\
h_{30}=-h_{33}=&\quad\,\frac{\langle\overline{q}q\rangle}{6B_{0}F^{2}}\label{eq:AAP_hs}
\end{align}
and
\begin{align}
h_{14}&=h_{16}=h_{20}=h_{23}=h_{24}=h_{25}=h_{26}=\nonumber\\
&=h_{27}=h_{28}=h_{29}=h_{31}=h_{32}=h_{34}=0\,.
\end{align}
The matching above thus allowed us to identify many of the 34 parameters, out of which only 13 remain unknown.
\item[2)] An additional constraint can be obtained by taken into account the OPE \eqref{eq:aap_ope_2-op}. Inserting the obtained constraints \eqref{eq:AAP_hs} into \eqref{eq:aap_correlator} and scaling it accordingly, we get
\begin{align}
\mathcal{F}_{AAP}^{\mathrm{OPE}}\Big((\lambda p)^{2}&,(q-\lambda p)^{2},q^{2}\Big)=-\frac{\langle\overline{q}q\rangle}{3\lambda^{2}}\frac{1}{p^{2}q^{2}}\nonumber\\
&\times\frac{2(h_{8}+h_{10})+(M_{P_{1}}^{2}+M_{P_{2}}^{2})q^{2}-q^{4}}{(q^{2}-M_{P_{1}}^{2})(q^{2}-M_{P_{2}}^{2})}+\mathcal{O}\bigg(\frac{1}{\lambda^{3}}\bigg)\,.\label{eq:aap_ope_2-op_v2}
\end{align}
To be able to match \eqref{eq:aap_ope_2-op_v2} onto \eqref{eq:aap_ope_2-op}, the remainder of a division of the numerator of \eqref{eq:aap_ope_2-op_v2} by $(q^{2}-M_{P_{1}}^{2})(q^{2}-M_{P_{2}}^{2})$, treated as polynomial in $q^{2}$, must be zero. We thus obtain
\begin{equation}
h_{10}=-h_{8}-\frac{M_{P_{1}}^{2}M_{P_{2}}^{2}}{2}\,,\label{eq:aap_ope_2-op_v3}
\end{equation}
which reduces the number of unknown parameters from 13 to 12.
\end{itemize}
\end{itemize}


\section{Resonance saturation}\label{sec:resonance_saturation}
The results of the resonance saturation, i.e.~matching the ChPT contributions \eqref{eq:VVP_ChPT}, \eqref{eq:VAS_ChPT} and \eqref{eq:AAP_ChPT} onto the RChT contributions \eqref{eq:VVP_RChT_tensor_final}, \eqref{eq:VAS_RChT_tensor_final} and \eqref{eq:AAP_RChT_tensor_final}, respectively, are quite lengthy --- although still presentable quite comfortably. They are as follows:\footnote{In the expressions below, we have highlighted in bold such contributions that one would have obtained if only the lowest resonance multiplets would have been taken into account. These can be compared also with eqs.~(39) and (87) in ref.~\cite{Kampf:2011ty} and with eq.~\eqref{eq:C9C23_v2} in appendix \ref{sec:lowest_resonances}.}
\begin{subequations}
\begin{align}
C_{7}^{W}=&\bm{-\frac{F_{V_{1}} \kappa_{12}^{V}}{2 \sqrt{2} M_{V_{1}}^{2}}-\frac{\sqrt{2} F_{V_{1}} \kappa_{14}^{V}}{M_{V_{1}}^{2}}-\frac{F_{V_{1}} \kappa_{16}^{V}}{4 \sqrt{2} M_{V_{1}}^{2}}+\frac{2 d_{m}^{(1)} \kappa_{5}^{P}}{M_{P_{1}}^{2}}+\frac{F_{V_{1}}^{2} \kappa_{2}^{VV}}{M_{V_{1}}^{4}}}\nonumber\\
&\bm{-\frac{F_{V_{1}}^{2} \kappa_{3}^{VV}}{8 M_{V_{1}}^{4}}-\frac{\sqrt{2} F_{V_{1}}d_{m}^{(1)} \kappa_{3}^{PV}}{M_{V_{1}}^{2}M_{P_{1}}^{2}}+\frac{F_{V_{1}}^{2}d_{m}^{(1)} \kappa^{VVP}}{2M_{V_{1}}^{4}M_{P_{1}}^{2}}}-\frac{F_{V_{2}} \lambda_{12}^{V}}{2 \sqrt{2} M_{V_{2}}^{2}}\nonumber\\
&-\frac{\sqrt{2} F_{V_{2}} \lambda_{14}^{V}}{M_{V_{2}}^{2}}-\frac{F_{V_{2}} \lambda_{16}^{V}}{4 \sqrt{2} M_{V_{2}}^{2}}+\frac{2 d_{m}^{(2)} \lambda_{5}^{P}}{M_{P_{2}}^{2}}+\frac{F_{V_{1}} F_{V_{2}} \lambda_{21}^{VV}}{M_{V_{1}}^{2} M_{V_{2}}^{2}}+\frac{F_{V_{2}}^{2} \lambda_{22}^{VV}}{M_{V_{2}}^{4}}\nonumber\\
&-\frac{F_{V_{1}} F_{V_{2}} \lambda_{31}^{VV}}{8 M_{V_{1}}^{2} M_{V_{2}}^{2}}-\frac{F_{V_{1}} F_{V_{2}} \lambda_{32}^{VV}}{4 M_{V_{1}}^{2} M_{V_{2}}^{2}}-\frac{F_{V_{1}} F_{V_{2}} \lambda_{33}^{VV}}{8 M_{V_{1}}^{2} M_{V_{2}}^{2}}-\frac{F_{V_{2}}^{2} \lambda_{34}^{VV}}{8 M_{V_{2}}^{4}}\nonumber\\
&-\frac{\sqrt{2} F_{V_{2}}d_{m}^{(1)} \lambda_{31}^{PV}}{M_{V_{2}}^{2}M_{P_{1}}^{2}}-\frac{\sqrt{2}F_{V_{1}}d_{m}^{(2)} \lambda_{32}^{PV}}{M_{V_{1}}^{2}M_{P_{2}}^{2}}-\frac{\sqrt{2}F_{V_{2}}d_{m}^{(2)} \lambda_{33}^{PV}}{M_{V_{2}}^{2}M_{P_{2}}^{2}}\nonumber\\
&+\frac{F_{V_{1}} F_{V_{2}}d_{m}^{(1)} \lambda_{1}^{VVP}}{M_{V_{1}}^{2} M_{V_{2}}^{2}M_{P_{1}}^{2}}+\frac{F_{V_{1}}^{2}d_{m}^{(2)} \lambda_{2}^{VVP}}{2M_{V_{1}}^{4}M_{P_{2}}^{2}}+\frac{F_{V_{2}}^{2}d_{m}^{(1)} \lambda_{3}^{VVP}}{2M_{V_{2}}^{4}M_{P_{1}}^{2}}\nonumber\\
&+\frac{F_{V_{1}} F_{V_{2}}d_{m}^{(2)} \lambda_{4}^{VVP}}{M_{V_{1}}^{2} M_{V_{2}}^{2}M_{P_{2}}^{2}}+\frac{F_{V_{2}}^{2}d_{m}^{(2)} \lambda_{5}^{VVP}}{2M_{V_{2}}^{4}M_{P_{2}}^{2}}\,,\\
C_{9}^{W}=&\bm{-\frac{F_{A_{1}} \kappa_{3}^{A}}{4 \sqrt{2} M_{A_{1}}^{2}}-\frac{F_{A_{1}} \kappa_{8}^{A}}{2 \sqrt{2} M_{A_{1}}^{2}}-\frac{\sqrt{2} F_{A_{1}} \kappa_{11}^{A}}{M_{A_{1}}^{2}}-\frac{F_{A_{1}} \kappa_{12}^{A}}{\sqrt{2} M_{A_{1}}^{2}}-\frac{F_{A_{1}} \kappa_{15}^{A}}{4 \sqrt{2} M_{A_{1}}^{2}}}\nonumber\\
&\bm{+\frac{2 d_{m}^{(1)} \kappa_{1}^{P}}{M_{P_{1}}^{2}}+\frac{F_{A_{1}}^{2} \kappa_{2}^{AA}}{M_{A_{1}}^{4}}-\frac{F_{A_{1}}^{2} \kappa_{3}^{AA}}{8 M_{A_{1}}^{4}}-\frac{\sqrt{2} F_{A_{1}}d_{m}^{(1)} \kappa_{1}^{PA}}{M_{A_{1}}^{2} M_{P_{1}}^{2}}-\frac{F_{A_{1}}d_{m}^{(1)} \kappa_{2}^{PA}}{\sqrt{2} M_{A_{1}}^{2} M_{P_{1}}^{2}}}\nonumber\\
&\bm{+\frac{F_{A_{1}}^{2}d_{m}^{(1)} \kappa^{AAP}}{2 M_{A_{1}}^{4} M_{P_{1}}^{2}}}-\frac{F_{A_{2}} \lambda_{3}^{A}}{4 \sqrt{2} M_{A_{2}}^{2}}-\frac{F_{A_{2}} \lambda_{8}^{A}}{2 \sqrt{2} M_{A_{2}}^{2}}-\frac{\sqrt{2} F_{A_{2}} \lambda_{11}^{A}}{M_{A_{2}}^{2}}-\frac{F_{A_{2}} \lambda_{12}^{A}}{\sqrt{2} M_{A_{2}}^{2}}\nonumber\\
&-\frac{F_{A_{2}} \lambda_{15}^{A}}{4 \sqrt{2} M_{A_{2}}^{2}}+\frac{2 d_{m}^{(2)} \lambda_{1}^{P}}{M_{P_{2}}^{2}}+\frac{F_{A_{1}} F_{A_{2}} \lambda_{21}^{AA}}{M_{A_{1}}^{2} M_{A_{2}}^{2}}+\frac{F_{A_{2}}^{2} \lambda_{22}^{AA}}{M_{A_{2}}^{4}}-\frac{F_{A_{1}} F_{A_{2}} \lambda_{31}^{AA}}{8 M_{A_{1}}^{2} M_{A_{2}}^{2}}\nonumber\\
&-\frac{F_{A_{1}} F_{A_{2}} \lambda_{32}^{AA}}{4 M_{A_{1}}^{2} M_{A_{2}}^{2}}-\frac{F_{A_{1}} F_{A_{2}} \lambda_{33}^{AA}}{8 M_{A_{1}}^{2} M_{A_{2}}^{2}}-\frac{F_{A_{2}}^{2} \lambda_{34}^{AA}}{8 M_{A_{2}}^{4}}-\frac{\sqrt{2} F_{A_{2}}d_{m}^{(1)} \lambda_{11}^{PA}}{M_{A_{2}}^{2} M_{P_{1}}^{2}}\nonumber\\
&-\frac{\sqrt{2} F_{A_{1}}d_{m}^{(2)} \lambda_{12}^{PA}}{M_{A_{1}}^{2} M_{P_{2}}^{2}}-\frac{\sqrt{2} F_{A_{2}}d_{m}^{(2)} \lambda_{13}^{PA}}{M_{A_{2}}^{2} M_{P_{2}}^{2}}-\frac{F_{A_{2}}d_{m}^{(1)} \lambda_{21}^{PA}}{\sqrt{2} M_{A_{2}}^{2} M_{P_{1}}^{2}}-\frac{F_{A_{1}}d_{m}^{(2)} \lambda_{22}^{PA}}{\sqrt{2} M_{A_{1}}^{2} M_{P_{2}}^{2}}\nonumber\\
&-\frac{F_{A_{2}}d_{m}^{(2)} \lambda_{23}^{PA}}{\sqrt{2} M_{A_{2}}^{2} M_{P_{2}}^{2}}+\frac{F_{A_{1}}F_{A_{2}}d_{m}^{(1)}\lambda_{1}^{AAP}}{M_{A_{1}}^{2} M_{A_{2}}^{2} M_{P_{1}}^{2}}+\frac{F_{A_{1}}^{2}d_{m}^{(2)} \lambda_{2}^{AAP}}{2 M_{A_{1}}^{4} M_{P_{2}}^{2}}+\frac{F_{A_{2}}^{2}d_{m}^{(1)} \lambda_{3}^{AAP}}{2 M_{A_{2}}^{4} M_{P_{1}}^{2}}\nonumber\\
&+\frac{F_{A_{1}} F_{A_{2}}d_{m}^{(2)} \lambda_{4}^{AAP}}{M_{A_{1}}^{2} M_{A_{2}}^{2} M_{P_{2}}^{2}}+\frac{F_{A_{2}}^{2}d_{m}^{(2)} \lambda_{5}^{AAP}}{2 M_{A_{2}}^{4} M_{P_{2}}^{2}}\,,\\
C_{11}^{W}=&\quad\,\bm{\frac{F_{V_{1}} \kappa_{4}^{V}}{2 \sqrt{2} M_{V_{1}}^{2}}-\frac{F_{V_{1}} \kappa_{15}^{V}}{\sqrt{2} M_{V_{1}}^{2}}+\frac{F_{A_{1}} \kappa_{14}^{A}}{\sqrt{2} M_{A_{1}}^{2}}+\frac{c_{m}^{(1)} \kappa_{2}^{S}}{M_{S_{1}}^{2}}-\frac{F_{A_{1}}c_{m}^{(1)} \kappa_{1}^{SA}}{\sqrt{2} M_{A_{1}}^{2} M_{S_{1}}^{2}}}\nonumber\\
&\bm{+\frac{F_{V_{1}}c_{m}^{(1)} \kappa_{1}^{SV}}{\sqrt{2} M_{V_{1}}^{2}M_{S_{1}}^{2}}+\frac{F_{V_{1}}c_{m}^{(1)} \kappa_{2}^{SV}}{2 \sqrt{2} M_{V_{1}}^{2}M_{S_{1}}^{2}}+\frac{F_{V_{1}}F_{A_{1}} \kappa_{6}^{VA}}{2 M_{V_{1}}^{2}M_{A_{1}}^{2}}+\frac{F_{V_{1}}F_{A_{1}}c_{m}^{(1)} \kappa^{VAS}}{2 M_{V_{1}}^{2}M_{A_{1}}^{2} M_{S_{1}}^{2}}}\nonumber\\
&+\frac{F_{V_{2}} \lambda_{4}^{V}}{2 \sqrt{2} M_{V_{2}}^{2}}-\frac{F_{V_{2}} \lambda_{15}^{V}}{\sqrt{2} M_{V_{2}}^{2}}+\frac{F_{A_{2}} \lambda_{14}^{A}}{\sqrt{2} M_{A_{2}}^{2}}+\frac{c_{m}^{(2)} \lambda_{2}^{S}}{M_{S_{2}}^{2}}-\frac{F_{A_{2}}c_{m}^{(1)} \lambda_{11}^{SA}}{\sqrt{2} M_{A_{2}}^{2} M_{S_{1}}^{2}}\nonumber\\
&-\frac{F_{A_{1}}c_{m}^{(2)} \lambda_{12}^{SA}}{\sqrt{2} M_{A_{1}}^{2} M_{S_{2}}^{2}}-\frac{F_{A_{2}}c_{m}^{(2)} \lambda_{13}^{SA}}{\sqrt{2} M_{A_{2}}^{2} M_{S_{2}}^{2}}+\frac{F_{V_{2}}c_{m}^{(1)} \lambda_{11}^{SV}}{\sqrt{2} M_{V_{2}}^{2}M_{S_{1}}^{2}}+\frac{F_{V_{1}}c_{m}^{(2)} \lambda_{12}^{SV}}{\sqrt{2} M_{V_{1}}^{2}M_{S_{2}}^{2}}\nonumber\\
&+\frac{F_{V_{2}}c_{m}^{(2)} \lambda_{13}^{SV}}{\sqrt{2} M_{V_{2}}^{2}M_{S_{2}}^{2}}+\frac{F_{V_{2}}c_{m}^{(1)} \lambda_{21}^{SV}}{2 \sqrt{2} M_{V_{2}}^{2}M_{S_{1}}^{2}}+\frac{F_{V_{1}}c_{m}^{(2)} \lambda_{22}^{SV}}{2 \sqrt{2} M_{V_{1}}^{2}M_{S_{2}}^{2}}+\frac{F_{V_{2}}c_{m}^{(2)} \lambda_{23}^{SV}}{2 \sqrt{2} M_{V_{2}}^{2}M_{S_{2}}^{2}}\nonumber\\
&+\frac{F_{V_{2}}F_{A_{1}} \lambda_{61}^{VA}}{2 M_{V_{2}}^{2}M_{A_{1}}^{2}}+\frac{F_{V_{1}}F_{A_{2}} \lambda_{62}^{VA}}{2 M_{V_{1}}^{2}M_{A_{2}}^{2}}+\frac{F_{V_{2}}F_{A_{2}} \lambda_{63}^{VA}}{2 M_{V_{2}}^{2}M_{A_{2}}^{2}}+\frac{F_{V_{2}}F_{A_{1}}c_{m}^{(1)} \lambda_{1}^{VAS}}{2 M_{V_{2}}^{2}M_{A_{1}}^{2} M_{S_{1}}^{2}}\nonumber\\
&+\frac{F_{V_{1}}F_{A_{2}}c_{m}^{(1)} \lambda_{2}^{VAS}}{2 M_{V_{1}}^{2}M_{A_{2}}^{2} M_{S_{1}}^{2}}+\frac{F_{V_{1}}F_{A_{1}}c_{m}^{(2)} \lambda_{3}^{VAS}}{2 M_{V_{1}}^{2}M_{A_{1}}^{2} M_{S_{2}}^{2}}+\frac{F_{V_{2}}F_{A_{2}}c_{m}^{(1)} \lambda_{4}^{VAS}}{2 M_{V_{2}}^{2}M_{A_{2}}^{2} M_{S_{1}}^{2}}\nonumber\\
&+\frac{F_{V_{2}}F_{A_{1}}c_{m}^{(2)} \lambda_{5}^{VAS}}{2 M_{V_{2}}^{2}M_{A_{1}}^{2} M_{S_{2}}^{2}}+\frac{F_{V_{1}} F_{A_{2}}c_{m}^{(2)} \lambda_{6}^{VAS}}{2 M_{V_{1}}^{2}M_{A_{2}}^{2} M_{S_{2}}^{2}}+\frac{F_{V_{2}} F_{A_{2}}c_{m}^{(2)} \lambda_{7}^{VAS}}{2 M_{V_{2}}^{2}M_{A_{2}}^{2} M_{S_{2}}^{2}}\,,\\
C_{22}^{W}=&\bm{-\frac{F_{V_{1}}\kappa_{17}^{V}}{\sqrt{2}M_{V_{1}}^{2}}}\bm{-\frac{F_{V_{1}}^{2}\kappa_{3}^{VV}}{2M_{V_{1}}^{4}}}-\frac{F_{V_{2}}\lambda_{17}^{V}}{\sqrt{2}M_{V_{2}}^{2}}-\frac{F_{V_{1}}F_{V_{2}}\lambda_{31}^{VV}}{2M_{V_{1}}^{2}M_{V_{2}}^{2}}-\frac{F_{V_{1}}F_{V_{2}}\lambda_{33}^{VV}}{2M_{V_{1}}^{2}M_{V_{2}}^{2}}-\frac{F_{V_{2}}^{2}\lambda_{34}^{VV}}{2M_{V_{2}}^{4}}\,,\\
C_{23}^{W}=&\bm{-\frac{F_{A_{1}}\kappa_{16}^{A}}{\sqrt{2}M_{A_{1}}^{2}}-\frac{F_{A_{1}}^{2}\kappa_{3}^{AA}}{2 M_{A_{1}}^{4}}}-\frac{F_{A_{2}}\lambda_{16}^{A}}{\sqrt{2}M_{A_{2}}^{2}}-\frac{F_{A_{1}}F_{A_{2}}\lambda_{31}^{AA}}{2M_{A_{1}}^{2}M_{A_{2}}^{2}}-\frac{F_{A_{1}}F_{A_{2}}\lambda_{33}^{AA}}{2M_{A_{1}}^{2}M_{A_{2}}^{2}}-\frac{F_{A_{2}}^{2}\lambda_{34}^{AA}}{2M_{A_{2}}^{4}}\,.
\end{align}
\end{subequations}

Upon applying the constraints obtained from the RChT--OPE matching, the expressions above can be rewritten as follows:
\begin{subequations}
\begin{align}
C_{7}^{W}=&\quad\,\frac{1}{M_{V_{1}}^{2}+M_{V_{2}}^{2}}\bigg(\frac{F_{V_{1}}^{2} \kappa_{3}^{VV} M_{V_{1}}^{4}}{4M_{V_{2}}^{4} M_{P_{2}}^{2} }+\frac{F_{V_{1}} F_{V_{2}} \lambda_{32}^{VV} M_{V_{1}}^{2}}{4 M_{V_{2}}^{4} }+\frac{F_{V_{2}}^{2}d_{m}^{(1)} \lambda_{3}^{VVP} M_{V_{1}}^{2}}{2 M_{V_{2}}^{4}M_{P_{1}}^{2} }\nonumber\\
&+\frac{\sqrt{2}F_{V_{1}}d_{m}^{(1)} \kappa_{3}^{PV} M_{V_{1}}^{2}}{M_{V_{2}}^{2}M_{P_{1}}^{2} }-\frac{\sqrt{2} F_{V_{1}}d_{m}^{(1)} \kappa_{3}^{PV} M_{V_{1}}^{2}}{M_{V_{2}}^{2}M_{P_{2}}^{2} }-\frac{F_{V_{1}}^{2} \kappa_{3}^{VV} M_{V_{1}}^{2}}{4 M_{V_{2}}^{2}M_{P_{2}}^{2} }\nonumber\\
&-\frac{3 F^{2} M_{V_{1}}^{2}}{16 M_{V_{2}}^{2}M_{P_{2}}^{2} }-\frac{F_{V_{1}} F_{V_{2}} \lambda_{21}^{VV} M_{V_{1}}^{2}}{2 M_{V_{2}}^{4} }-\frac{F_{V_{1}} F_{V_{2}}d_{m}^{(1)} \lambda_{1}^{VVP} M_{V_{1}}^{2}}{M_{V_{2}}^{4}M_{P_{2}}^{2} }\nonumber\\
&-\frac{F_{V_{1}} F_{V_{2}}d_{m}^{(2)} \lambda_{4}^{VVP} M_{V_{1}}^{2}}{M_{V_{2}}^{4}M_{P_{2}}^{2} }-\frac{F_{V_{1}}^{2}d_{m}^{(1)} \kappa^{VVP} M_{V_{1}}^{2}}{2 M_{V_{2}}^{4} M_{P_{2}}^{2}}-\frac{F_{V_{1}}^{2} d_{m}^{(2)}\lambda_{2}^{VVP} M_{V_{1}}^{2}}{2 M_{V_{2}}^{4}M_{P_{2}}^{2} }\nonumber\\
&-\frac{F_{V_{2}}^{2}d_{m}^{(1)} \lambda_{3}^{VVP} M_{V_{1}}^{2}}{2 M_{V_{2}}^{4}M_{P_{2}}^{2} }+\frac{N_{c} M_{V_{1}}^{2}}{256\pi^{2} M_{P_{2}}^{2}}-\frac{\langle\overline{q}q\rangle M_{V_{1}}^{2}}{192B_{0}M_{V_{2}}^{4}}-\frac{\langle\overline{q}q\rangle M_{V_{1}}^{2}}{48B_{0}M_{V_{2}}^{2}M_{P_{2}}^{2}}\nonumber\\
&+\frac{g_{s}\langle\overline{q}\sigma\hspace{-1pt}\cdot\hspace{-1pt}Gq\rangle M_{V_{1}}^{2}}{192B_{0}M_{V_{2}}^{2}M_{P_{1}}^{2}M_{P_{2}}^{2}}-\frac{\langle\overline{q}q\rangle}{32B_{0}M_{P_{2}}^{2}}-\frac{\langle\overline{q}q\rangle}{64B_{0}M_{V_{2}}^{2}}-\frac{\langle\overline{q}q\rangle}{192B_{0}M_{V_{1}}^{2}}\nonumber\\
&+\frac{g_{s}\langle\overline{q}\sigma\hspace{-1pt}\cdot\hspace{-1pt}Gq\rangle}{192B_{0}M_{P_{1}}^{2}M_{P_{2}}^{2}}-\frac{g_{s}\langle\overline{q}\sigma\hspace{-1pt}\cdot\hspace{-1pt}Gq\rangle}{1152B_{0}M_{V_{2}}^{2}M_{P_{2}}^{2}}-\frac{\langle\overline{q}q\rangle M_{V_{2}}^{2}}{96B_{0}M_{V_{1}}^{2}M_{P_{2}}^{2}}\nonumber\\
&-\frac{g_{s}\langle\overline{q}\sigma\hspace{-1pt}\cdot\hspace{-1pt}Gq\rangle}{1152B_{0}M_{V_{1}}^{2}M_{P_{2}}^{2}}-\frac{\langle\overline{q}q\rangle M_{V_{2}}^{2}}{192B_{0}M_{V_{1}}^{4}}+\frac{F_{V_{1}}^{2} M_{V_{2}}^{2} \kappa_{3}^{VV}}{4M_{V_{1}}^{2} M_{P_{2}}^{2}}+\frac{F_{V_{1}}^{2} \kappa_{3}^{VV}}{2  M_{V_{1}}^{2}}\nonumber\\
&+\frac{F_{V_{1}}^{2}d_{m}^{(1)} M_{V_{2}}^{2} \kappa^{VVP}}{2 M_{V_{1}}^{4}M_{P_{1}}^{2}}+\frac{F_{V_{1}}^{2}d_{m}^{(1)} \kappa^{VVP}}{2M_{V_{1}}^{2} M_{P_{1}}^{2}}+\frac{F_{V_{1}} F_{V_{2}}d_{m}^{(1)} \lambda_{1}^{VVP}}{M_{V_{1}}^{2}M_{P_{1}}^{2}}\nonumber\\
&+\frac{F_{V_{1}} F_{V_{2}}d_{m}^{(1)} \lambda_{1}^{VVP}}{M_{V_{2}}^{2}M_{P_{1}}^{2} }+\frac{F_{V_{1}} F_{V_{2}} \lambda_{21}^{VV}}{2 M_{V_{2}}^{2} }+\frac{F_{V_{1}}^{2}d_{m}^{(2)} M_{V_{2}}^{2} \lambda_{2}^{VVP}}{2 M_{V_{1}}^{4}M_{P_{2}}^{2}}+\frac{N_{c}}{256  \pi^{2}}\nonumber\\
&+\frac{F_{V_{1}} F_{V_{2}} \lambda_{21}^{VV}}{2  M_{V_{1}}^{2}}+\frac{F_{V_{1}}^{2} d_{m}^{(2)}\lambda_{2}^{VVP}}{2 M_{V_{1}}^{2}M_{P_{2}}^{2}}+\frac{F_{V_{1}} F_{V_{2}} \lambda_{32}^{VV}}{4  M_{V_{1}}^{2}}+\frac{F_{V_{2}}^{2}d_{m}^{(1)} \lambda_{3}^{VVP}}{2 M_{V_{2}}^{2}M_{P_{1}}^{2} }\nonumber\\
&+\frac{F_{V_{1}} F_{V_{2}}d_{m}^{(2)} \lambda_{4}^{VVP}}{M_{V_{1}}^{2}M_{P_{2}}^{2} }+\frac{\sqrt{2}F_{V_{1}}d_{m}^{(1)} M_{V_{2}}^{2} \kappa_{3}^{PV}}{M_{V_{1}}^{2}M_{P_{2}}^{2}}+\frac{3 F^{2}}{32  M_{V_{1}}^{2}}-\frac{N_{c}M_{V_{2}}^{2}}{256  \pi^{2} M_{V_{1}}^{2}}\nonumber\\
&-\frac{\sqrt{2} F_{V_{1}}d_{m}^{(1)} M_{V_{2}}^{2} \kappa_{3}^{PV}}{M_{V_{1}}^{2}M_{P_{1}}^{2}}-\frac{F_{V_{1}} F_{V_{2}} M_{V_{2}}^{2} \lambda_{21}^{VV}}{2  M_{V_{1}}^{4}}-\frac{F_{V_{1}}^{2} M_{V_{2}}^{2} \kappa_{3}^{VV}}{4  M_{V_{1}}^{4}}-\frac{F_{V_{1}}^{2} \kappa_{3}^{VV}}{4 M_{P_{2}}^{2} }\nonumber\\
&-\frac{F_{V_{1}} F_{V_{2}} \lambda_{32}^{VV}}{2 M_{V_{2}}^{2} }-\frac{F_{V_{1}}^{2} \kappa_{3}^{VV}}{4 M_{V_{2}}^{2} }-\frac{3 F^{2}}{32 M_{V_{2}}^{2} }+\frac{N_{c}M_{V_{2}}^{2}}{256\pi^{2} M_{P_{2}}^{2}}-\frac{F_{V_{1}} F_{V_{2}}d_{m}^{(1)} \lambda_{1}^{VVP}}{M_{V_{2}}^{2}M_{P_{2}}^{2} }\nonumber\\
&-\frac{3 F^{2}}{16 M_{P_{2}}^{2} }-\frac{F_{V_{1}}^{2}d_{m}^{(1)} \kappa^{VVP}}{2 M_{P_{2}}^{2} M_{V_{2}}^{2} }-\frac{F_{V_{1}}^{2}d_{m}^{(2)} \lambda_{2}^{VVP}}{2 M_{V_{2}}^{2}M_{P_{2}}^{2} }-\frac{F_{V_{2}}^{2} d_{m}^{(1)}\lambda_{3}^{VVP}}{2 M_{V_{2}}^{2}M_{P_{2}}^{2} }\bigg)\,,\\
C_{9}^{W}=&\quad\,\frac{g_{s} M_{A_{1}}^{2} \langle\overline{q}\sigma\hspace{-1pt}\cdot\hspace{-1pt}Gq\rangle}{384 B_{0} M_{A_{2}}^{4} M_{P_{1}}^{2} M_{P_{2}}^{2}}+\frac{g_{s} \langle\overline{q}\sigma\hspace{-1pt}\cdot\hspace{-1pt}Gq\rangle}{128 B_{0}  M_{A_{2}}^{2} M_{P_{1}}^{2} M_{P_{2}}^{2}}-\frac{g_{s}\langle\overline{q}\sigma\hspace{-1pt}\cdot\hspace{-1pt}Gq\rangle}{192B_{0} M_{A_{2}}^{2} M_{P_{1}}^{4}}\nonumber\\
&-\frac{g_{s}\langle\overline{q}\sigma\hspace{-1pt}\cdot\hspace{-1pt}Gq\rangle}{1152 B_{0} M_{A_{1}}^{2} M_{A_{2}}^{2}M_{P_{2}}^{2}}-\frac{\langle\overline{q}q\rangle}{96 B_{0} M_{A_{2}}^{2}M_{P_{2}}^{2}}+\frac{\langle\overline{q}q\rangle}{192 B_{0} M_{A_{2}}^{4}}\nonumber\\
&-\frac{\langle\overline{q}q\rangle}{96 B_{0} M_{A_{1}}^{2}M_{P_{2}}^{2}}+\frac{\langle\overline{q}q\rangle}{192 B_{0} M_{A_{1}}^{2} M_{A_{2}}^{2}}+\frac{\langle\overline{q}q\rangle}{192 B_{0} M_{A_{1}}^{4}}-\frac{N_{c}}{768 \pi ^{2} M_{P_{2}}^{2}}\nonumber\\
&+\frac{d_{m}^{(1)} F_{A_{1}}^{2} \kappa^{AAP}}{2 M_{A_{1}}^{4} M_{P_{1}}^{2}}-\frac{d_{m}^{(1)} F_{A_{1}}^{2} \kappa^{AAP}}{2 M_{A_{2}}^{4} M_{P_{2}}^{2}}+\frac{d_{m}^{(1)} F_{A_{1}} F_{A_{2}} \lambda_{1}^{AAP}}{M_{A_{1}}^{2} M_{A_{2}}^{2} M_{P_{1}}^{2}}-\frac{d_{m}^{(1)} F_{A_{1}} F_{A_{2}} \lambda_{1}^{AAP}}{M_{A_{2}}^{4} M_{P_{2}}^{2}}\nonumber\\
&-\frac{\sqrt{2} d_{m}^{(1)} F_{A_{1}} \kappa_{1}^{PA}}{M_{A_{1}}^{2} M_{P_{1}}^{2}}-\frac{d_{m}^{(1)} F_{A_{1}} \kappa_{2}^{PA}}{\sqrt{2} M_{A_{1}}^{2} M_{P_{1}}^{2}}+\frac{\sqrt{2} d_{m}^{(1)} F_{A_{1}} \kappa_{1}^{PA}}{M_{A_{1}}^{2} M_{P_{2}}^{2}}+\frac{d_{m}^{(1)} F_{A_{1}} \kappa_{2}^{PA}}{\sqrt{2} M_{A_{1}}^{2} M_{P_{2}}^{2}}\nonumber\\
&+\frac{\sqrt{2} d_{m}^{(1)} F_{A_{1}} \kappa_{1}^{PA}}{M_{A_{2}}^{2} M_{P_{1}}^{2}}+\frac{d_{m}^{(1)} F_{A_{1}} \kappa_{2}^{PA}}{\sqrt{2} M_{A_{2}}^{2} M_{P_{1}}^{2}}-\frac{\sqrt{2} d_{m}^{(1)} F_{A_{1}} \kappa_{1}^{PA}}{M_{A_{2}}^{2} M_{P_{2}}^{2}}-\frac{d_{m}^{(1)} F_{A_{1}} \kappa_{2}^{PA}}{\sqrt{2} M_{A_{2}}^{2} M_{P_{2}}^{2}}\nonumber\\
&+\frac{d_{m}^{(1)} F_{A_{2}}^{2} \lambda_{3}^{AAP}}{2 M_{A_{2}}^{4} M_{P_{1}}^{2}}-\frac{d_{m}^{(1)} F_{A_{2}}^{2} \lambda_{3}^{AAP}}{2 M_{A_{2}}^{4} M_{P_{2}}^{2}}+\frac{d_{m}^{(2)} F_{A_{1}}^{2} \lambda_{2}^{AAP}}{2 M_{A_{1}}^{4} M_{P_{2}}^{2}}-\frac{d_{m}^{(2)} F_{A_{1}}^{2} \lambda_{2}^{AAP}}{2 M_{A_{2}}^{4} M_{P_{2}}^{2}}\nonumber\\
&+\frac{d_{m}^{(2)} F_{A_{1}} F_{A_{2}} \lambda_{4}^{AAP}}{M_{A_{1}}^{2} M_{A_{2}}^{2} M_{P_{2}}^{2}}-\frac{d_{m}^{(2)} F_{A_{1}} F_{A_{2}} \lambda_{4}^{AAP}}{M_{A_{2}}^{4} M_{P_{2}}^{2}}+\frac{F_{A_{1}}^{2} \kappa_{3}^{AA} M_{A_{1}}^{2}}{4 M_{A_{2}}^{4} M_{P_{2}}^{2}}-\frac{F_{A_{1}}^{2} \kappa_{3}^{AA}}{4 M_{A_{1}}^{2} M_{P_{2}}^{2}}\nonumber\\
&-\frac{F_{A_{1}} F_{A_{2}} \lambda_{21}^{AA}}{2 M_{A_{1}}^{4}}+\frac{F_{A_{1}} F_{A_{2}} \lambda_{31}^{AA}}{16 M_{A_{1}}^{4}}+\frac{F_{A_{1}} F_{A_{2}} \lambda_{32}^{AA}}{8 M_{A_{1}}^{4}}+\frac{F_{A_{1}} F_{A_{2}} \lambda_{33}^{AA}}{16 M_{A_{1}}^{4}}\nonumber\\
&+\frac{F_{A_{1}} F_{A_{2}} \lambda_{31}^{AA} M_{A_{1}}^{2}}{8 M_{A_{2}}^{4} M_{P_{2}}^{2}}-\frac{F_{A_{1}} F_{A_{2}} \lambda_{32}^{AA} M_{A_{1}}^{2}}{4 M_{A_{2}}^{4} M_{P_{2}}^{2}}+\frac{F_{A_{1}} F_{A_{2}} \lambda_{33}^{AA} M_{A_{1}}^{2}}{8 M_{A_{2}}^{4} M_{P_{2}}^{2}}\nonumber\\
&+\frac{F_{A_{1}} F_{A_{2}} \lambda_{21}^{AA}}{M_{A_{1}}^{2} M_{A_{2}}^{2}}-\frac{F_{A_{1}} F_{A_{2}} \lambda_{31}^{AA}}{8 M_{A_{1}}^{2} M_{A_{2}}^{2}}-\frac{F_{A_{1}} F_{A_{2}} \lambda_{32}^{AA}}{4 M_{A_{1}}^{2} M_{A_{2}}^{2}}-\frac{F_{A_{1}} F_{A_{2}} \lambda_{33}^{AA}}{8 M_{A_{1}}^{2} M_{A_{2}}^{2}}\nonumber\\
&-\frac{F_{A_{1}} F_{A_{2}} \lambda_{31}^{AA}}{8 M_{A_{1}}^{2} M_{P_{2}}^{2}}-\frac{F_{A_{1}} F_{A_{2}} \lambda_{32}^{AA}}{4 M_{A_{1}}^{2} M_{P_{2}}^{2}}-\frac{F_{A_{1}} F_{A_{2}} \lambda_{33}^{AA}}{8 M_{A_{1}}^{2} M_{P_{2}}^{2}}-\frac{F_{A_{1}} F_{A_{2}} \lambda_{21}^{AA}}{2 M_{A_{2}}^{4}}\nonumber\\
&+\frac{F_{A_{1}} F_{A_{2}} \lambda_{31}^{AA}}{16 M_{A_{2}}^{4}}+\frac{F_{A_{1}} F_{A_{2}} \lambda_{32}^{AA}}{8 M_{A_{2}}^{4}}+\frac{F_{A_{1}} F_{A_{2}} \lambda_{33}^{AA}}{16 M_{A_{2}}^{4}}+\frac{F_{A_{1}} F_{A_{2}} \lambda_{32}^{AA}}{2 M_{A_{2}}^{2} M_{P_{2}}^{2}}\,,\\
C_{11}^{W}=&-\frac{F_{A_{1}} c_{m}^{(1)}\kappa_{1}^{SA}}{\sqrt{2} M_{A_{1}}^{2} M_{S_{1}}^{2}}+\frac{F_{A_{1}}c_{m}^{(1)} \kappa_{1}^{SA}}{\sqrt{2} M_{A_{1}}^{2} M_{S_{2}}^{2}}+\frac{F_{A_{1}}c_{m}^{(1)} \kappa_{1}^{SA}}{\sqrt{2} M_{A_{2}}^{2} M_{S_{1}}^{2}}-\frac{F_{A_{1}}c_{m}^{(1)} \kappa_{1}^{SA}}{\sqrt{2} M_{A_{2}}^{2} M_{S_{2}}^{2}}\nonumber\\
&+\frac{F_{V_{1}}c_{m}^{(1)} \kappa_{1}^{SV}}{\sqrt{2} M_{V_{1}}^{2}M_{S_{1}}^{2}}-\frac{F_{V_{1}}c_{m}^{(1)} \kappa_{1}^{SV}}{\sqrt{2} M_{V_{1}}^{2}M_{S_{2}}^{2}}+\frac{F_{V_{1}}c_{m}^{(1)} \kappa_{1}^{SV}}{\sqrt{2} M_{V_{2}}^{2}M_{S_{2}}^{2}}-\frac{F_{V_{1}}c_{m}^{(1)} \kappa_{1}^{SV}}{\sqrt{2} M_{V_{2}}^{2}M_{S_{1}}^{2}}\nonumber\\
&+\frac{F_{V_{1}}c_{m}^{(1)} \kappa_{2}^{SV}}{2 \sqrt{2} M_{V_{1}}^{2}M_{S_{1}}^{2}}-\frac{F_{V_{1}}c_{m}^{(1)} \kappa_{2}^{SV}}{2 \sqrt{2} M_{V_{2}}^{2}M_{S_{1}}^{2} }-\frac{F_{V_{1}} c_{m}^{(1)}\kappa_{2}^{SV}}{2 \sqrt{2} M_{V_{1}}^{2}M_{S_{2}}^{2}}+\frac{F_{V_{1}}c_{m}^{(1)} \kappa_{2}^{SV}}{2 \sqrt{2} M_{V_{2}}^{2}M_{S_{2}}^{2}}\nonumber\\
&+\frac{F_{V_{1}}F_{A_{1}} \kappa_{6}^{VA}}{2 M_{V_{1}}^{2}M_{A_{1}}^{2}}-\frac{F_{V_{1}}F_{A_{1}} \kappa_{6}^{VA}}{2 M_{V_{2}}^{2}M_{A_{1}}^{2}}-\frac{F_{V_{1}}F_{A_{1}} \kappa_{6}^{VA}}{2 M_{V_{1}}^{2}M_{A_{2}}^{2}}+\frac{F_{V_{1}}F_{A_{1}} \kappa_{6}^{VA}}{2M_{V_{2}}^{2}M_{A_{2}}^{2}}\nonumber\\
&+\frac{F_{V_{1}}F_{A_{1}}c_{m}^{(1)} \kappa^{VAS}}{2 M_{V_{1}}^{2}M_{A_{1}}^{2} M_{S_{1}}^{2} }-\frac{F_{V_{1}}F_{A_{1}} c_{m}^{(1)} \kappa^{VAS}}{2 M_{V_{2}}^{2}M_{A_{2}}^{2} M_{S_{2}}^{2} }+\frac{F_{V_{2}}F_{A_{1}}c_{m}^{(1)} \lambda_{1}^{VAS}}{2 M_{V_{2}}^{2}M_{A_{1}}^{2} M_{S_{1}}^{2} }\nonumber\\
&-\frac{F_{V_{2}}F_{A_{1}} c_{m}^{(1)} \lambda_{1}^{VAS}}{2 M_{V_{2}}^{2}M_{A_{2}}^{2} M_{S_{2}}^{2}}+\frac{F_{V_{1}}F_{A_{2}}c_{m}^{(1)} \lambda_{2}^{VAS}}{2 M_{V_{1}}^{2}M_{A_{2}}^{2} M_{S_{1}}^{2} }-\frac{F_{V_{1}}F_{A_{2}}c_{m}^{(1)} \lambda_{2}^{VAS}}{2 M_{V_{2}}^{2}M_{A_{2}}^{2} M_{S_{2}}^{2}}\nonumber\\
&+\frac{F_{V_{1}}F_{A_{1}} c_{m}^{(2)} \lambda_{3}^{VAS}}{2 M_{V_{1}}^{2}M_{A_{1}}^{2} M_{S_{2}}^{2}}-\frac{F_{V_{1}}F_{A_{1}} c_{m}^{(2)} \lambda_{3}^{VAS}}{2 M_{V_{2}}^{2}M_{A_{2}}^{2} M_{S_{2}}^{2} }+\frac{F_{V_{2}}F_{A_{2}} c_{m}^{(1)}\lambda_{4}^{VAS}}{2 M_{V_{2}}^{2}M_{A_{2}}^{2} M_{S_{1}}^{2} }\nonumber\\
&-\frac{F_{V_{2}}F_{A_{2}}c_{m}^{(1)} \lambda_{4}^{VAS}}{2M_{V_{2}}^{2} M_{A_{2}}^{2} M_{S_{2}}^{2}}+\frac{F_{V_{2}}F_{A_{1}}c_{m}^{(2)}\lambda_{5}^{VAS}}{2 M_{V_{2}}^{2}M_{A_{1}}^{2} M_{S_{2}}^{2}}-\frac{F_{V_{2}}F_{A_{1}}c_{m}^{(2)} \lambda_{5}^{VAS}}{2M_{V_{2}}^{2} M_{A_{2}}^{2} M_{S_{2}}^{2}}\nonumber\\
&+\frac{F_{V_{1}}F_{A_{2}}c_{m}^{(2)} \lambda_{6}^{VAS}}{2 M_{V_{1}}^{2}M_{A_{2}}^{2} M_{S_{2}}^{2}}-\frac{F_{V_{1}}F_{A_{2}}c_{m}^{(2)} \lambda_{6}^{VAS}}{2 M_{V_{2}}^{2}M_{A_{2}}^{2} M_{S_{2}}^{2}}-\frac{\langle\overline{q}q\rangle}{192B_{0}M_{V_{1}}^{2}M_{A_{2}}^{2}}\nonumber\\
&-\frac{\langle\overline{q}q\rangle}{192B_{0}M_{V_{1}}^{2}M_{S_{2}}^{2}}-\frac{\langle\overline{q}q\rangle}{192B_{0}M_{V_{2}}^{2}M_{S_{1}}^{2}}-\frac{\langle\overline{q}q\rangle}{192B_{0}M_{V_{2}}^{2}M_{A_{2}}^{2}}\nonumber\\
&+\frac{\langle\overline{q}q\rangle}{192B_{0}M_{A_{2}}^{2}M_{S_{2}}^{2}}+\frac{\langle\overline{q}q\rangle}{192B_{0}M_{A_{2}}^{2} M_{S_{1}}^{2}}-\frac{\langle\overline{q}q\rangle}{192B_{0}M_{A_{1}}^{2}M_{V_{2}}^{2}}\nonumber\\
&+\frac{\langle\overline{q}q\rangle}{192B_{0}M_{A_{1}}^{2}M_{S_{2}}^{2}}-\frac{g_{s}\langle\overline{q}\sigma\hspace{-1pt}\cdot\hspace{-1pt}Gq\rangle}{2304B_{0}M_{V_{1}}^{2}M_{V_{2}}^{2}M_{S_{2}}^{2}}+\frac{g_{s}\langle\overline{q}\sigma\hspace{-1pt}\cdot\hspace{-1pt}Gq\rangle}{768B_{0}M_{V_{2}}^{2}M_{S_{1}}^{2}M_{S_{2}}^{2}}\nonumber\\
&-\frac{g_{s}\langle\overline{q}\sigma\hspace{-1pt}\cdot\hspace{-1pt}Gq\rangle}{768B_{0}M_{A_{2}}^{2}M_{S_{1}}^{2}M_{S_{2}}^{2}}+\frac{g_{s}\langle\overline{q}\sigma\hspace{-1pt}\cdot\hspace{-1pt}Gq\rangle}{2304B_{0}M_{A_{1}}^{2}M_{A_{2}}^{2}M_{S_{2}}^{2}}-\frac{\langle\overline{q}q\rangle}{192B_{0}M_{V_{2}}^{2}M_{S_{2}}^{2}}\,,\\
C_{22}^{W}=&-\frac{3F^{2}}{8M_{V_{1}}^{2}M_{V_{2}}^{2}}+\frac{N_{c}}{64\pi^{2}M_{V_{1}}^{2}}+\frac{N_{c}}{64\pi^{2}M_{V_{2}}^{2}}\,,\label{eq:C22W_v2}\\
C_{23}^{W}=&-\frac{F_{A_{1}} \kappa_{3}^{A}}{2 \sqrt{2} M_{A_{1}}^{2}}+\frac{F_{A_{1}} \kappa_{3}^{A}}{\sqrt{2} M_{A_{2}}^{2}}-\frac{F_{A_{1}} \kappa_{3}^{A} M_{A_{1}}^{2}}{2 \sqrt{2} M_{A_{2}}^{4}}-\frac{F_{A_{1}} \kappa_{8}^{A}}{\sqrt{2} M_{A_{1}}^{2}}+\frac{\sqrt{2} F_{A_{1}} \kappa_{8}^{A}}{M_{A_{2}}^{2}}\nonumber\\
&-\frac{F_{A_{1}} \kappa_{8}^{A} M_{A_{1}}^{2}}{\sqrt{2} M_{A_{2}}^{4}}-\frac{F_{A_{1}} \kappa_{15}^{A}}{2 \sqrt{2} M_{A_{1}}^{2}}+\frac{F_{A_{1}} \kappa_{15}^{A}}{\sqrt{2} M_{A_{2}}^{2}}-\frac{F_{A_{1}} \kappa_{15}^{A} M_{A_{1}}^{2}}{2 \sqrt{2} M_{A_{2}}^{4}}-\frac{F_{A_{1}}^{2} \kappa_{3}^{AA}}{2 M_{A_{1}}^{4}}\nonumber\\
&+\frac{F_{A_{1}}^{2} \kappa_{3}^{AA}}{2 M_{A_{2}}^{4}}-\frac{F_{A_{1}} F_{A_{2}} \lambda_{31}^{AA}}{2 M_{A_{1}}^{2} M_{A_{2}}^{2}}+\frac{F_{A_{1}} F_{A_{2}} \lambda_{31}^{AA}}{2 M_{A_{2}}^{4}}-\frac{F_{A_{1}} F_{A_{2}} \lambda_{33}^{AA}}{2 M_{A_{1}}^{2} M_{A_{2}}^{2}}+\frac{F_{A_{1}} F_{A_{2}} \lambda_{33}^{AA}}{2 M_{A_{2}}^{4}}\nonumber\\
&+\frac{N_{c}}{192 \pi ^{2} M_{A_{2}}^{2}}+\frac{\langle\overline{q}q\rangle}{48 B_{0} M_{A_{2}}^{4}}-\frac{g_{s}\langle\overline{q}\sigma\hspace{-1pt}\cdot\hspace{-1pt}Gq\rangle}{192 B_{0} M_{A_{2}}^{4} M_{P_{1}}^{2}}+\frac{g_{s}\langle\overline{q}\sigma\hspace{-1pt}\cdot\hspace{-1pt}Gq\rangle}{192 B_{0} M_{A_{1}}^{2} M_{A_{2}}^{2} M_{P_{1}}^{2}}\,.
\end{align}
\end{subequations}

Interestingly enough, we see that we have obtained a simple relation \eqref{eq:C22W_v2} for $C_{22}^{W}$, in which all the parameters are known. This thus allows us to extract its numerical value:
\begin{equation}
C_{22}^{W}\doteq 7.63\cdot 10^{-3}\,\mathrm{GeV}^{-2}
\end{equation}


\section{Contribution of lowest resonances only: \texorpdfstring{$\bm{\langle AAP\rangle}$}{}}\label{sec:lowest_resonances}
In the section \ref{sec:two_multiplets}, we have mentioned that the results of the RChT-OPE matching and the resonance saturation for the $\langle VVP\rangle$, $\langle VAS\rangle$ and $\langle AAP\rangle$ Green functions are somewhat complicated. On the other hand, let us recall that such a matching for the special case of only the lightest resonance multiplets taken into account has already been published in ref.~\cite{Kampf:2011ty}, and that is only for the $\langle VVP\rangle$ and $\langle VAS\rangle$ correlators.

In what follows, we thus present the results of the RChT-OPE matching and the resonance saturation for the $\langle AAP\rangle$ Green function, where only the lightest axial-vector and pseudoscalar resonance multiplets are considered --- the respective resonance contribution is then given by eq.~\eqref{eq:AAP_RChT_tensor}. We emphasize that since such a contribution can not satisfy the OPE \eqref{eq:aap_ope} with the quark-gluon condensate contribution included, the results presented below have been obtained with the OPE given by the contribution of the quark condensate only.

Performing the matching of the resonance contribution onto the OPE, we obtain the following constraints for the coupling constants of the resonance Lagrangian \eqref{eq:odd6-lagrangian}:
\begin{align}
-(\kappa_{3}^{A}+2\kappa_{8}^{A}+\kappa_{15}^{A})=4(2\kappa_{11}^{A}+\kappa_{12}^{A})=-2\kappa_{16}^{A}&=\frac{N_{c}}{96\sqrt{2}\pi^{2}F_{A_{1}}}\,,\notag\\
\kappa_{1}^{P}&=0\,,\notag\\
8\kappa_{2}^{AA}-\kappa_{3}^{AA}&=\frac{\langle\overline{q}q\rangle}{24B_{0}F_{A_{1}}^{2}}\,,\label{eq:AAP_OPE_5}\\
2\kappa_{1}^{PA}+\kappa_{2}^{PA}-\frac{F_{A_{1}}\kappa_{3}^{AA}}{2\sqrt{2}d_{m}^{(1)}}&=\frac{\langle\overline{q}q\rangle}{48\sqrt{2}B_{0}F_{A_{1}}d_{m}^{(1)}}+\frac{N_{c}M_{A_{1}}^{2}}{384\sqrt{2}\pi^{2}F_{A_{1}}d_{m}^{(1)}}\,.\nonumber
\end{align}
Applying the constraints above into the resonance contribution \eqref{eq:AAP_RChT_tensor} leads to
\begin{align}
\mathcal{F}_{AAP}^{\mathrm{RChT}}(p^{2},q^{2},r^{2})&\nonumber\\
&\hspace{-70pt}=\frac{B_{0}N_{c}M_{A_{1}}^{2}M_{P_{1}}^{2}(p^{2}+q^{2}-2M_{A_{1}}^{2})}{48\pi^{2}(p^{2}-M_{A_{1}}^{2})(q^{2}-M_{A_{1}}^{2})(r^{2}-M_{P_{1}}^{2})r^{2}}-\frac{4B_{0}F_{A_{1}}^{2}\kappa_{3}^{AA}\left(2M_{A_{1}}^{2}r^{2}-M_{P_{1}}^{2}(p^{2}+q^{2})\right)}{(p^{2}-M_{A_{1}}^{2})(q^{2}-M_{A_{1}}^{2})(r^{2}-M_{P_{1}}^{2})r^{2}}\nonumber\\
&\hspace{-70pt}+\frac{16B_{0}F_{A_{1}}^{2}d_{m}^{(1)}\kappa^{AAP}}{(p^{2}-M_{A_{1}}^{2})(q^{2}-M_{A_{1}}^{2})(r^{2}-M_{P_{1}}^{2})}+\frac{\langle\overline{q}q\rangle(p^{2}+q^{2}-r^{2}-2M_{A_{1}}^{2}+M_{P_{1}}^{2})}{6(p^{2}-M_{A_{1}}^{2})(q^{2}-M_{A_{1}}^{2})(r^{2}-M_{P_{1}}^{2})}\,,
\end{align}
from which we see that the $\langle AAP\rangle$ Green function is thus given by two unknown constants, $\kappa_{3}^{AA}$ and $\kappa^{AAP}$.

Matching the ChPT result \eqref{eq:AAP_ChPT} of the $\langle AAP\rangle$ correlator onto its RChT contribution \eqref{eq:AAP_RChT_tensor} gives us the relation for the respective low-energy constants in the form
\begin{align}
C_{9}^{W}=&-\frac{F_{A_{1}}(\kappa_{3}^{A}+2\kappa_{8}^{A}+8\kappa_{11}^{A}+4\kappa_{12}^{A}+\kappa_{15}^{A})}{4\sqrt{2}M_{A_{1}}^{2}}+\frac{2d_{m}^{(1)}\kappa_{1}^{P}}{M_{P_{1}}^{2}}+\frac{F_{A_{1}}^{2}(8\kappa_{2}^{AA}-\kappa_{3}^{AA})}{8M_{A_{1}}^{4}}\nonumber\\
&-\frac{F_{A_{1}}d_{m}^{(1)}(2\kappa_{1}^{PA}+\kappa_{2}^{PA})}{\sqrt{2}M_{A_{1}}^{2}M_{P_{1}}^{2}}+\frac{F_{A_{1}}^{2}d_{m}^{(1)}\kappa^{AAP}}{2M_{A_{1}}^{4}M_{P_{1}}^{2}}\,,\notag\\
C_{23}^{W}=&-\frac{F_{A_{1}}\kappa_{16}^{A}}{\sqrt{2}M_{A_{1}}^{2}}-\frac{F_{A_{1}}^{2}\kappa_{3}^{AA}}{2M_{A_{1}}^{4}}\,,\label{eq:C9C23_v2}
\end{align}
that can be further simplified by applying the relations \eqref{eq:AAP_OPE_5}, i.e.
\begin{align}
C_{9}^{W}=&-\frac{N_{c}}{768\pi^{2}M_{P_{1}}^{2}}+\frac{\langle\overline{q}q\rangle}{192B_{0}M_{A_{1}}^{2}}\bigg(\frac{1}{M_{A_{1}}^{2}}-\frac{2}{M_{P_{1}}^{2}}\bigg)-\frac{F_{A_{1}}^{2}\kappa_{3}^{AA}}{4M_{A_{1}}^{2}M_{P_{1}}^{2}}+\frac{F_{A_{1}}^{2}d_{m}^{(1)}\kappa^{AAP}}{2M_{A_{1}}^{4}M_{P_{1}}^{2}}\,,\notag\\
C_{23}^{W}=&\quad\,\frac{N_{c}}{384\pi^{2}M_{A_{1}}^{2}}-\frac{F_{A_{1}}^{2}\kappa_{3}^{AA}}{2M_{A_{1}}^{4}}\,.
\end{align}
%


\end{document}